\documentclass[12pt,english]{article}
\usepackage[T1]{fontenc}
\usepackage[latin9]{inputenc}
\usepackage{geometry}
\usepackage{color}
\usepackage{babel}
\usepackage{array}
\usepackage{mathrsfs}
\usepackage{multirow}
\usepackage{dsfont}
\usepackage{amsmath}
\usepackage{amsthm}
\usepackage{amssymb}
\usepackage{graphicx}
\usepackage{setspace}
\usepackage[authoryear,round]{natbib}
\usepackage[linesnumbered,ruled,vlined]{algorithm2e}

\usepackage{xspace}
\usepackage{chngcntr}
\usepackage{apptools}
\AtAppendix{\counterwithin{lemma}{section}}

\usepackage[unicode=true,pdfusetitle,
 bookmarks=true,bookmarksnumbered=false,bookmarksopen=false,
 breaklinks=false,pdfborder={0 0 1},backref=false,colorlinks=true]
 {hyperref}
\hypersetup{
 linkcolor=blue,citecolor=blue}

\makeatletter

\providecommand{\tabularnewline}{\\}

\theoremstyle{plain}
\newtheorem{assumption}{\protect\assumptionname}
\theoremstyle{remark}
\newtheorem{remark}{\protect\remarkname}
\theoremstyle{plain}

\theoremstyle{plain}
\newtheorem{lemma}{\protect\lemmaname}
\theoremstyle{plain}
\newtheorem{theorem}{\protect\theoremname}

\makeatother

\providecommand{\assumptionname}{Assumption}
\providecommand{\lemmaname}{Lemma}
\providecommand{\propositionname}{Proposition}
\providecommand{\remarkname}{Remark}
\providecommand{\theoremname}{Theorem}

\usepackage{babel}

\makeatother

\providecommand{\lemmaname}{Lemma}
\providecommand{\propositionname}{Proposition}
\providecommand{\remarkname}{Remark}
\providecommand{\theoremname}{Theorem}

\begin{document}
\title{\textbf{Stochastic Learning of Semiparametric Monotone Index
Models with Large Sample Size} }

\begin{spacing}{1}
\author{Qingsong Yao\footnote{
Department of Economics, Boston College. Email: yaoq@bc.edu. I sincerely thank my advisors Shakeeb Khan, Zhijie Xiao and Arthur Lewbel for their continuous guidance and help  during my PhD studies. I also thank my previous advisor, Guoqing Zhao, for his support. I thank Shengtao Dai, David Hughes, and participants at BU-BC econometric workshop for their constructive comments on my paper and my presentation. All remaining errors are my own.}
\\
\\
This Version: Oct 27, 2023/First Version: Sep 12, 2023}

\date{}

\maketitle

\begin{abstract}
I study the estimation of semiparametric monotone index models in the scenario where the number of observation points $n$ is extremely large and conventional approaches fail to work due to heavy computational burdens. Motivated by the mini-batch gradient descent algorithm (MBGD) that is widely used as a stochastic optimization tool in the machine learning field, I proposes a novel subsample- and iteration-based  estimation procedure. In particular, starting from any initial guess of the true parameter, I progressively update the parameter using a sequence of subsamples randomly drawn from the data set whose sample size is much smaller than $n$. The update is  based on  the  gradient of some well-chosen loss function, where the nonparametric component is replaced with its Nadaraya-Watson kernel estimator based on subsamples.  My proposed algorithm essentially  generalizes   MBGD algorithm to the semiparametric setup. Compared with full-sample-based method, the new method reduces the computational time by roughly $n$ times if the subsample size and the kernel function are chosen properly, so  can be easily applied when the sample size $n$ is large.  Moreover, I show that if I further conduct averages across the estimators produced during iterations, the difference between the average estimator and full-sample-based estimator will be $1/\sqrt{n}$-trivial. Consequently,  the average estimator is $1/\sqrt{n}$-consistent and asymptotically normally distributed.  In other words, the new estimator substantially improves the computational speed, while at the same time maintains the estimation accuracy.

\end{abstract}
\end{spacing}
\newpage

\section{Introduction}\label{section1}

With the rapid development of  technology in data collection and data storage,  it's becoming more and more common nowadays for data analysts to deal with data set with extraordinary amount of observations. This offers the researchers unprecedented  opportunities to more precisely understand the potential mechanism lurking behind the data, while on the same time brings about a series of new challenges. Among others, the key challenge is the  heavy computational burdens that make the existing statistical methods numerically prohibitive. For example, when estimating a model using gradient-based iterative optimization procedure, the gradient of some  objective function is repeatedly evaluated at a sequence of candidate parameters so that the optimal point can be numerically found. When the sample size is extremely large, even a single evaluation of the gradient  would cost a huge amount of computation time, let along evaluating repeatedly at many points, making model estimation practically infeasible. Consequently, it's more urgent than ever before to study estimation methods that is applicable in the big-data era.

This paper studies \textit{semiparametric} estimation of monotone index models in a large $n$ scenario.  To fix idea, throughout this paper I will focus on the following 
binary choice model 
\begin{equation}
y=\mathds{1}\left(X_{0}\beta_{0}^{\star}+\mathbf{X}^{\mathrm{T}}\boldsymbol{\beta}^{\star}-u>0\right),\label{eq:model}
\end{equation}
where $\mathds{1}\left(\cdot\right)$ is indicator function, $\mathbf{X}_{e}=\left(X_{0},\mathbf{X}^{\mathrm{T}}\right)^{\mathrm{T}}=\left(X_{0},X_{1},\cdots,X_{p}\right)^{\mathrm{T}}\in\mathcal{X}_{e}$
is $(p+1)\times 1$  covariate vector, $\boldsymbol{\beta}_{e}^{\star}=\left(\beta_{0}^{\star},\boldsymbol{\beta}^{\star\mathrm{T}}\right)^{\mathrm{T}}=\left(\beta_{0}^{\star},\beta_{1}^{\star},\cdots,\beta_{p}^{\star}\right)^{\mathrm{T}}\in\mathcal{B}_{e}$
is the unknown true parameter vector, and $u$ is the unobserved individual
shock with CDF  $G\left(\cdot\right)$. Binary choice model    is a leading example of the class of monotone index models, which has a wide range of applications in many areas such as economics, business,  and biostatistics. I also point out that all of the conclusions obtained under such setup can be trivially extended to more general class of monotone index models.

When the CDF $G(\cdot)$ in (\ref{eq:model}) is known, parametric estimation method such as maximum likelihood estimation can be applied. However, as I have discussed before, even under such setup estimation  can be computationally costly  when the data size is massive. To deal with the ``large $n$'' issue, subsample-based estimation strategy are widely applied. For example, when applying the \textit{gradient descent} algorithm to iteratively search for the maxima of the log-likelihood function, instead of using the full sample, it's generally proposed to use a random subsample whose sample size is much smaller than $n$ to perform the update, which is known as the \textit{mini-batch gradient descent algorithm} (MBGD, see \citet{bottou2018optimization,ruder2016overview}). The batch size can be chosen as small as 1, in which case the algorithm is known as the \textit{stochastic gradient descent} \citep{toulis2017asymptotic}. For another example, \citet{forneron2022estimation} studies stochastic optimization based on Newton-Raphson and quasi Newton iterations for a general class of parametric objective functions, and proposes subsample-based estimation and inference procedure for the unknown parameters.

In this paper, I focus on the  semiparametric estimation of $\boldsymbol{\beta}_{e}^{\star}$. In other words,  I seek to estimate $\boldsymbol{\beta}_{e}^{\star}$
 without specifying the functional form of  $G\left(\cdot\right)$. The main advantages of semiparametric specification are model flexibility as well as tractability. 
In the existing literature, semiparametric estimation for monotone index models and binary choice model in particular has been extensively studied. The  methods    can be roughly classified into two categories: M-estimation approach and direct construction approach. 
For the  first category, the estimator is obtained by optimizing
some objective functions. The standing estimators include maximum
score estimator \citep{manski1975maximum,manski1985semiparametric,horowitz1992smoothed},
maximum rank correlation estimator \citep{han1987non,sherman1993limiting,cavanagh1998rank,fan2020rank},
semiparametric least squares estimator \citep{hardle1993optimal,ichimura1993semiparametric}
and semiparametric maximum likelihood estimator \citep{cosslett1983distribution,klein1993efficient}.
Apart from M-estimation, the second class of estimation methods features direct construction of the estimators, which includes average derivative estimator \citep{stoker1986consistent,powell1989semiparametric,horowitz1996direct,hristache2001direct}, special regressor approach \citep{lewbel2000semiparametric} and eigenvalue
approach \citep{ahn2018simple}.

The key feature that distinguishes my paper from the existing literature is that I try to estimate the model in a scenario where the sample size $n$ is extremely large. Large sample size $n$ imposes computational challenges to model estimation even in the parametric setup, and such issue turns out to be far more serious in the semiparametric setup. In his famous paper,  \citet{ichimura1993semiparametric} pointed out that for semiparametric least square estimator, ``the
computation time is roughly $n$ times more than with smooth parametric nonlinear
regression estimation''. So if I estimate the semiparametric model based on a data set of millions of observations, the estimation time would be roughly millions of  times longer than  parametric estimation, say, Logit or Probit regression. This makes semiparametric  estimation  almost computationally infeasible when $n$ is extremely large. Indeed, for many semiparametric M-estimators such as \citet{ichimura1993semiparametric}'s semiparametric least squares estimator and \citet{klein1993efficient}'s semiparametric maximum likelihood estimator, the unknown CDF (or monotonic link function for more general monotone index models) $G(\cdot)$ in the objective function is replaced with its Nadaraya-Watson kernel estimator. So evaluating the objective function (or its gradient) generally involves calculating kernel estimators (or their gradients) at $n$ points. Since each kernel estimator (or its gradient) requires computational complexity of order $O(n)$, a single assessment of the objective function (or its gradient) requires computational time of order $O(n^2)$, which increases fast with the sample size $n$. This makes the conventional semiparametric estimation method not applicable even for data set with only tens of thousands of observation points.
Apart from intensive computational burdens, there are many other crucial limitations that prohibit the use of existing semiparametric estimation methods\footnote{For  M-estimation approach, the objective
functions involved are usually heavily discontinuous and/or non-convex
with respect to the parameter. In this case, even looking for a local optimum is generally NP-Hard \citep{murty1987some},
let alone the global optimum.  This makes the optimization procedure computationally
infeasible. On the other side, the direct construction approach generally
imposes more structure on the covariates. For example, the average
derivative approach requires that the covariates are all continuous, so can not
be directly applied to discrete covariates such as dummy variables. 
Moreover, the application of such method
usually involves nonparametric estimation of the density functions  or their partial derivative of some random variables conditional on the covariates. Such estimation becomes an intractable problem even when the number of covariates is modest. Although there have been some attempts to reduce the dimensionality of conditional density estimation (e.g., \citet{hall2004cross}), the methods are still computationally-intensive, which may not be applicable in a data-rich environment, see \citet{ouyang2023} and references therein.}.

In this paper, I propose a novel semiparametric estimation procedure for (\ref{eq:model}) that can be easily implemented with very fast speed even on a regular laptop when the sample size $n$ is extremely large. My method is motivated by the MBGD algorithm. For any random variable $Z$, parameter $\theta$, and loss function $L(Z,\theta)$, given a sequence of realizations $Z_1, \cdots, Z_n$ of $Z$, to search for the optimal point $\theta^{\star}$  that minimizes the population loss function $\mathbb{E}_Z\left(L(Z, \theta)\right)$,  MBGD conducts the following iteration, 
\begin{align}\label{MBGD}
\theta_{k+1} = \theta_k - \frac{\delta_k}{\left|\mathfrak{I}_{k}\right|}\sum_{i\in \mathfrak{I}_{k}} \frac{\partial L(Z_i,\theta_k)}{\partial \theta},
\end{align}
where $\theta_1$ is some initial guess, $\delta_k >0$ is the learning rate, and $\mathfrak{I}_{k}$ is the subsample used in the $k$-th round of iteration. In other words, the MBGD algorithm updates the parameter based on the gradient of the loss function at observation points that fall into the subsample $\mathfrak{I}_k$. Compared with the full-sample-based \textit{batch gradient descent} (BGD) that uses gradient at all the data points to perform the update,  MBGD update is less accurate\footnote{When using the full sample to conduct update, the gradient of the  empirical loss function $L_n(\theta) = \frac{1}{n}\sum_{i=1}^nL(Z_i, \theta)$ is accurately evaluated at each candidate parameter $\theta$ because the gradient of the loss function at each data point $Z_i$ is evaluated.  While when using subsample-based update, the gradient of the empirical loss function is only approximated by the gradients at a subsample of observations.} but significantly alleviates the computational burden when $\left|\mathfrak{I}_k\right|\ll n$. 
Typically, the   MBGD algorithm applies only to the parametric setup where the loss function $L\left(\cdot, \cdot\right)$ is fully known. While when  estimating the binary choice model (\ref{eq:model}),  the loss function generally has form  $L\left(\left. \cdot,\cdot\right|G\right)$, so depends on the link function $G\left(\cdot\right)$\footnote{For example, the quadratic loss function is given by  $L(\mathbf{X},y,\boldsymbol{\beta}|G) = (y - G(\mathbf{X}^{\mathrm{T}}\boldsymbol{\beta}))^2$ and the  log-likelihood loss function is given by  $L(\mathbf{X},y,\boldsymbol{\beta}|G) = -(y\log(G(\mathbf{X}^{\mathrm{T}}\boldsymbol{\beta})) + (1-y)\log(1-G(\mathbf{X}^{\mathrm{T}}\boldsymbol{\beta}))) $. }. In the semiparametric setup  where $G$ is unknown, $L\left(\left. \cdot,\cdot\right|G\right)$ is then not fully specified, which makes the above MBGD update no longer feasible.

To make (\ref{MBGD}) feasible, I consider a two-step updating procedure. In the $k$-th round of update, I first nonparametrically estimate the unknown function $G\left(\cdot\right)$, whose estimator is denoted as $\widehat{G}_k$. Then in the second step, I plug the first-step estimator $\widehat{G}_k$ into the loss function $L\left(\left.Z,\theta\right|G\right)$ and perform the update (\ref{MBGD}) based on the estimated loss function $L\left(\left.Z,\theta\right|\widehat{G}\right)$ as if it were the true loss function. The key  difficulty of such two-step update in the large  $n$ scenario lies in the heavy computational burden caused by nonparametric estimation of $G\left(\cdot\right)$. Indeed, conventional nonparametric estimator such as Nadaraya-Watson kernel estimator requires computational complexity of order $O(n)$ to evaluate  $\widehat G_k$ at a single point. So if I use a subsample of size $B$ to perform the update, I need to  evaluate $\widehat G_k $ at a total of $B$ points, and the computational burden of each single update is of order $O(Bn)$. This is  too large to be practical if I choose $B \gg  1/\sqrt{n}$\footnote{Indeed, this is required if I pursue $1/\sqrt{n}$-consistency and asymptotic normality of the estimator, see  \autoref{thm8}.} and update hundreds of thousands of times. The main novelty of this paper is that instead of using conventional nonparametric estimator based on the full sample, I propose to use subsample to construct the Nadaraya-Watson kernel estimator, so that the above two-step update is fully subsample-based. The idea behind such subsample-based nonparametric estimation is intuitive: if I believe that using subsample for iteration leads to relatively accurate update, then the subsample-based    nonparametric estimator should also be reasonably close to the one based on the full sample.  When the subsample size is $B$, evaluating $B$ subsample-based Nadaraya-Watson kernel estimators requires computational complexity $O(B^2)$. This will be much smaller than $O(n^2)$ if I choose $B\ll n$. Indeed, I will show that as long as I properly choose the kernel function, $B$ can be chosen sufficiently close to $1/\sqrt{n}$, so the computational burden of update can be made close to $O(n)$, which is almost linear in $n$. This makes semiparametric estimation of monotone index models practically feasible when the sample size $n$ is  large. 

\citet{khan2022} (KLTY hereafter) also consider a similar two-step updating procedure. While the main difference between my method and theirs lies in that in KLTY, both the first-step nonparametric estimation and the second-step update are based on the full sample. Full-sample-based update increases the update accuracy, but as I discussed before, it leads to heavy computational burdens so is only applicable when the sample size is modest. Comparatively, the main novelty of my method lies in that  I propose a fully subsample-based update which substantially improves the computation speed and can be easily applied when the sample size is extremely large. Roughly speaking, the relationship between my method and KLTY's method is similar to that between  mini-batch gradient descent and batch gradient descent. Finally, similar to KLTY's method, 
my proposed method also overcomes the optimization issue of the M-estimator, see KLTY for more discussion.

I also develop the statistical properties of the above fully subsample-based two-step updating algorithm. Under some regularity conditions, I show that the proposed alogithm yields an asymptotically consistent estimator. However, its guaranteed convergence rate is slower than the parametric rate $1/\sqrt{n}$ if I choose $B\ll n$ to improve computational speed. Indeed, the guaranteed convergence rate will be even slower than rate $1/\sqrt{B}$, which is the convergence rate of conventional MBGD estimators.   Such slower convergence rate is mainly caused by subsample-based nonparametric estimation in the first step. The subsample-based nonparametric estimator is no longer an unbiased estimator for the one based on the full sample, and such bias dampens the $1/\sqrt{B}$-convergence. I then decompose the bias. I find that the first-order bias have $1/\sqrt{n}$-trivial conditional mean (conditioned on the subsamples in the previous updates and the data set), while the second-order bias are uniformly $1/\sqrt{n}$-trivial as long as I update sufficiently many times. This motivates me to  follow \citet{polyak1992acceleration} and use average to eliminate the first-order bias and accelerate the convergence rate. In particular, after some burn-in rounds of updates, all the  estimators produced during the following updates are averaged. I show that as long as the numbers of burn-in and follow-up updates are both large enough, the averaged estimator will converge at $1/\sqrt{n}$ rate and is asymptotically normally distributed.  Such a result demonstrates that our subsample-based method not only improves the computational speed, it also maintains the estimation accuracy on the same time.

Since the subsample-based estimator is asymptotically normally distributed after averaging, inference on the true parameter can be conducted if some consistent estimator of the asymptotic covariance matrix is available. Unfortunately, when sample size $n$ is extremely large, estimating the covariance matrix based on the full sample also requires large amount of time because it involves evaluating a large number of nonparametric estimators. To faciliate the inference, I also propose a subsample-based estimator of the covariance matrix, which subtantially improves the computation speed.   I show that the subsample-based estimator is a consistent estimator of the unknown covariance matrix, so the inference using such subsample-based estimator will be asymptotically valid. 

The main contribution of this paper to the econometric literature is that I propose a computationally friendly algorithm that can be used to semiparametrically estimate the monotone index models when the sample size $n$ is extremely large. My new algorithm essentially generalizes the mini-batch estimation method to the semiparametric setup. It can be easily applied when there are hundreds of covariates and hundreds of thousands of or even millions of data points. Essentially, it bridges the gap between semiparametric estimation theories and empirical applications in the data-rich environment. 

As an empirical illustration of my new method, I revisit the empirical results in \citet{helpman2008estimating}. In their paper,  \citet{helpman2008estimating} use a parametric Probit model to study how the conditional probability of one country exporting to another is affected by a set of coutry-pair factors, and such estimation results are further embedded into a second-step estimation of the gravity equation. The full data set they use contains a total of 248060 observation points and 337 covariates including large number of country and year fixed effects, which features both large $n$ and $p$.  Given that Probit estimation assumes that the random shock in the binary choice model has tail that decays at a fast speed,  the estimation results could be biased if the true random shock has heavier tails, and in that case, the subsequent inference of the true parameter will also be invalid. Above discussion motivates   semiparametric estimation, but given the size of the data set, the conventional semiparametric estimation are practically infeasible. In this paper I apply the proposed KMBGD estimation procedure to revisit the estimation results. The estimation and inference based on my method take around 8 hours and 0.8 hours respectively, which is practically feasible. Interestingly, compared with Probit distribution, I find that semiparametric estimation results are more in favor of a Logit distributed random shock in the sense that the KMBGD estimator is close to Logit estimator while differs significantly from Probit estimator. Such a result also highlights the use of semiparametric estimation as opposed to parametric estimation in applications.

The remainder of the paper is arranged as follows. In  \autoref{section2}, I formally introduce the two-step fully subsample-based updating algorithm. In \autoref{section3}, I develop the asymptotic properties of the proposed algorithm. 
Then in \autoref{section4}, I propose a subsample-based inference procedure. In \autoref{section5}, I study the finite-sample performance of the proposed algorithm by conducting some Monte Carlo simulations. In \autoref{section6}, I
apply my new algorithm to revisit \citet{helpman2008estimating}'s Probit estimation results. Finally, \autoref{section7} concludes. All the proofs of the lemmas and theorems are arranged to the Appendix.

\subsection{Notations}

For any real sequences $\left\{ a_{n}\right\} _{n=1}^{\infty}$ and
$\left\{ b_{n}\right\} _{n=1}^{\infty}$, I write $a_{n}=o\left(b_{n}\right)$
if $\lim\sup_{n\rightarrow\infty}\left|a_{n}/b_{n}\right|=0$, $a_{n}=O\left(b_{n}\right)$
if $\lim\sup_{n\rightarrow\infty}\left|a_{n}/b_{n}\right|<\infty$,
and $a_{n}\sim b_{n}$ if both $a_{n}=O\left(b_{n}\right)$ and $b_{n}=O\left(a_{n}\right)$.
For any random sequences $\left\{ a_{n}\right\} _{n=1}^{\infty}$
and $\left\{ b_{n}\right\} _{n=1}^{\infty}$, I write $a_{n}=O_{p}\left(b_{n}\right)$
if for any $0<\tau<1$ there exist $N$ and $C>0$ such that $P\left\{ \left|a_{n}/b_{n}\right|>C\right\} <\tau$
holds for all $n\geq N$, I write $a_{n}=o_{p}\left(b_{n}\right)$
if for any $C>0$, $P\left( \left|a_{n}/b_{n}\right|>C\right)\rightarrow0$.
For any Borel set $A\subseteq\mathbb{R}^{k}$, denote its Lebesgue
measure as $m\left(A\right)$. Denote $I_p$ as the $p$-dimensional identity matrix. For any symmetric matrix $A$, we write
$A\succ0$ if $A$ is positive definite, and $A\succeq0$ if $A$
is positive semi-definite. For any symmetric matrices $A$ and $B$,
I write $A\succ B$ if $A-B\succ0$ and $A\succeq B$ if $A-B\succeq0$.
For any matrix $A$, I denote $\sigma\left(A\right)$ as its singular
value, and denote $\overline{\sigma}\left(A\right)$ and $\underline{\sigma}\left(A\right)$
as its largest and smallest singular value. For any symmetric matrix
$A$, I denote $\lambda\left(A\right)$ as its eigenvalue, and denote
$\overline{\lambda}\left(A\right)$ and $\underline{\lambda}\left(A\right)$
as its largest and smallest eigenvalue. For any vector $\boldsymbol{x}=\left(x_{1},\cdots,x_{p}\right)^{\mathrm{T}}$,
I denote its Euclidean norm as $\left\Vert \boldsymbol{x}\right\Vert =\sqrt{\sum_{i=1}^{p}x_{i}^{2}}$.
For any matrices $A=\left(a_{ij}\right)_{n\times m}$, I denote $\left\Vert A\right\Vert =\sqrt{\sum_{i=1}^{n}\sum_{j=1}^{m}a_{ij}^{2}}$.

\section{The KMBGD Algorithm \label{section2}}

This section formally introduces the subsample-based learning algorithm for binary choice models. To make my illustration more intuitive, I will start with a special case where the CDF function $G(\cdot)$ is known. Given any loss function $L\left(\mathbf{X}_{e},y,\boldsymbol{\beta}_{e}|G\right)$
that depends on $G(\cdot)$ and is  differentiable with respect to $\boldsymbol{\beta}_{e}\in\mathcal{B}_{e}$,
the conventional MBGD estimator of $\boldsymbol{\beta}_{e}^{\star}$ is constructed based on
the following iteration \citep{bottou2018optimization,ruder2016overview},
\begin{equation}
	\boldsymbol{\beta}_{e,k+1}=\boldsymbol{\beta}_{e,k}-\frac{\delta_{k}}{B}\sum_{i\in \mathfrak{I}_{B,k}}\partial L\left(\mathbf{X}_{e,i},y_{i},\boldsymbol{\beta}_{e,k}|G\right)/\partial\boldsymbol{\beta}_{e},\label{known_loss}
\end{equation}
where $\boldsymbol{\beta}_1$ is given, $B$ is a positive integer and is the sbusample size. For each $k$,  $\delta_{k}>0$  is the learning rate, 
and 
\begin{align}
	\label{subsample} \mathfrak{I}_{B,k}=\{i_{k,1},i_{k,2},\cdots,i_{k,B}\}
\end{align}
is an
index set that is randomly drawn from $\{1,2,\cdots,n\}$ with replacement
and is independent over $k$. In other words, under MBGD algorithm,  in each iteration I randomly draw a subset of size $B$, and then update the estimator based on such subsample. 

Given a choice of the subsample size $B$,   to apply the MBGD algorithm  (\ref{known_loss}) to estimate $\boldsymbol{\beta}^{\star}$, it remains to choose the loss function. Following \citet{agarwal2014least} and \citet{khan2022}, I consider the loss function 
\begin{equation}
	L\left(\mathbf{X}_{e},y,\boldsymbol{\beta}_{e}|G\right)=\int_{-A}^{\mathbf{X}_{e}^{\mathrm{T}}\boldsymbol{\beta}_{e}}G\left(z\right)dz-y\mathbf{X}_{e}^{\mathrm{T}}\boldsymbol{\beta}_{e},\label{loss_function}
\end{equation}
for some sufficiently large positive constant $A$. \citet{khan2022} show that  loss function (\ref{loss_function})   has many properties such as global minimization at true parameter $\boldsymbol{\beta}^{\star}$ and positive definite Hessian matrix with respect to $\boldsymbol{\beta}_e$. 
Based on the MBGD updating rule (\ref{known_loss}) and loss function (\ref{loss_function}), the MBGD estimator of $\boldsymbol{\beta}_{e}^{\star}$ is
constructed based on the following iteration procedure:
\begin{equation}
	\boldsymbol{\beta}_{e,k+1}=\boldsymbol{\beta}_{e,k}-\frac{\delta_{k}}{B}\sum_{i\in\mathfrak{I}_{B,k}}\left(G\left(\mathbf{X}_{e,i}^{\mathrm{T}}\boldsymbol{\beta}_{e,k}\right)-y_{i}\right)\mathbf{X}_{e,i}.\label{known_G}
\end{equation}

Now I turn to the case of semiparametric estimation, which is the main focus of this paper. To ensure identification,  I set $\beta_{0}^{\star}$  
to be 1, so the estimation target now is $\boldsymbol{\beta}^{\star}$. To simplify notation,  denote the space of $\mathbf{X}$ as
$\mathcal{X}$, and the corresponding parameter space of $\boldsymbol{\beta}$
as $\mathcal{B}$. 

\begin{remark}
   Here I provide some discussion on the choice of the normalized covariate. The covariate whose coefficient is normalized to 1 must have nonzero and positive true coefficient. Since the true coefficient is unknown, I recommend choosing the covariate  based on  economic theories.   However, there could be scenarios where the (unknown) actual coefficient has the opposite sign as to that implied by economic theories. So it's also recommend to conduct a preliminary estimation based of Logit or Probit to provide some additional insights. In particular, it's suggested to choose covariate whose coefficient is significantly different from zero. If the estimated coefficient is negative, then use the negative value of such covariate for estimation. Finally, it's also recommended using continuous variable as the normalized covariate. 
\end{remark}

Note that the MBGD algorithm (\ref{known_G}) relies on the nonparametric component $G\left(\cdot\right)$ as a key input, which is unavailable in the current semiparametric setup. So the conventional MBGD algorithm is infeasible. To make the update feasible, a natural idea is to replace the unknown component with its nonparametric estimator. Intuitively,
suppose  that in the $k$-th round of iteration, the starting point $\boldsymbol{\beta}_{k}$
is close to the unknown true parameter $\boldsymbol{\beta}^{\star}$,
then there holds 
\[G\left(z\right) =\mathbb{E}\left(\left.y\right|X_{0}+\boldsymbol{\mathbf{X}}^{\mathrm{T}}\boldsymbol{\beta}^{\star}=z\right)\approx\mathbb{E}\left(\left.y\right|X_{0}+\boldsymbol{\mathbf{X}}^{\mathrm{T}}\boldsymbol{\beta}_{k}=z\right),\]
for any $z$. 
This immediately motivates the following Nadaraya-Watson kernel
estimator for $G\left(\cdot\right)$, 
\begin{equation}
	\widehat{G}\left(\left.z\right|\boldsymbol{\beta}_{k}\right)=\frac{\sum_{j=1}^{n}K_{h_{n}}\left(z-X_{0,j}-\boldsymbol{\mathbf{X}}_{j}^{\mathrm{T}}\boldsymbol{\beta}_{k}\right)y_{j}}{\sum_{j=1}^{n}K_{h_{n}}\left(z-X_{0,j}-\boldsymbol{\mathbf{X}}_{j}^{\mathrm{T}}\boldsymbol{\beta}_{k}\right)},z\in R,\label{kernel_estimator}
\end{equation}
where $K_{h}\left(\cdot\right)=h^{-1}K\left(\cdot/h\right)$, $K\left(\cdot\right)$
is kernel function, and $h_{n}$ is bandwidth parameter
depending on $n$. Given the estimated CDF $\widehat{G}\left(\left.\cdot\right|\boldsymbol{\beta}_{k}\right)$,
we can directly plug it back to (\ref{known_G}) and perform the update  as if it were the true CDF $G\left(\cdot\right)$. Note that a potential issue for   (\ref{kernel_estimator}) is that it's  based on the full data set, so evaluating its value has computational complexity of order $O(n)$ for each input $z$. If I use $B$ data points to perform the update, then a total of $B$ kernel estimators need to be evaluated in each update, which leads to computational burden of order $O(nB)$. The computational speed can be improved if I choose $B\ll n$, but note that to obtain an estimator with $1/\sqrt{n}$-consistency, it is generally required that   $B\sim \sqrt{n}$, see \citet{forneron2022estimation}. Indeed, in the current semiparametric setup, the order of $B$ has to be chosen even slightly larger,  see the following \autoref{thm8}. In this case, the computational burden will be of order at least $O(n\sqrt{n})$, which is far from being linear in $n$.

The key philosophy of my new algorithm is that, if I trust that using $B$ data points  provides relatively accurate updates, then the kernel estimation based on such $B$ points should also be reasonably close to that based on the full sample for all input $z$. Such an idea motivates me to use only the randomly-drawn subset  to construct the kernel estimator. In particular, consider the following Nadaraya-Watson
kernel estimator of $G(z)$
constructed based on the data points in subsample $\mathfrak{I}_{B,k}$,
\begin{equation}
	\widehat{G}\left(\left.z\right|\boldsymbol{\beta},\mathfrak{I}_{B,k},\underline{c}_{f}\right)=\frac{\frac{1}{B}\sum_{i\in\mathfrak{I}_{B,k}}K_{h_{n}}\left(z-X_{0,i}-\mathbf{X}_{i}^{\mathrm{T}}\boldsymbol{\beta}\right)y_{j}}{\left\{ \frac{1}{B}\sum_{i\in\mathfrak{I}_{B,k}}K_{h_{n}}\left(z-X_{0,i}-\mathbf{X}_{i}^{\mathrm{T}}\boldsymbol{\beta}\right)\right\} \lor\underline{c}_{f}},\label{subsample_NW}
\end{equation}
where $K_h$, $K$ and $h_n$ are all similarly defined as before, and   $\underline{c}_{f}>0$ is some sufficiently small constant.
Basically, the subsample-based estimator (\ref{subsample_NW})
is constructed as if I   only observe the random subsample $\left\{ \left(\mathbf{X}_{e,i},y_{i}\right)\right\} _{i\in\mathfrak{I}_{B,k}}$.
The computational complexity for evaluating $\widehat{G}\left(\left.z\right|\boldsymbol{\beta},\mathfrak{I}_{B,k},\underline{c}_{f}\right)$
is obviously of order $O\left(B\right)$. 

\begin{remark}
Note that different from $\widehat{G}\left(\left.z\right|\boldsymbol{\beta}\right)$ in (\ref{kernel_estimator}),
when using subsample $\mathfrak{I}_{B,k}$ to construct the kernel
estimator, I make truncation to the denominator so that it is
lower bounded by some positive constant $\underline{c}_{f}$.
This mainly aims to decrease the instability caused by subsampling. Note that under
truncation, I have that $\left|\widehat{G}\left(\left.z\right|\boldsymbol{\beta},\mathfrak{I}_{B,k},\underline{c}_{f}\right)\right|\leq Ch_{n}^{-1}$
for some positive constant $C>0$. Note also that although I use subsample to construct the kernel estimator, the  bandwidth parameter $h_n$ is still determined  by the full sample size $n$. This ensures that the subsample-based kernel estimator  concentrates around the  one based on the full sample.
\end{remark}

Given the subsample-based kernel estimator, I can formally illustrate my subsample-based learning algorithm. At the beginning of the $k$-th update, the initial point
$\boldsymbol{\beta}_{k}$ is given. Then using the subsample-based kernel estimator of $G\left(z\right)$ given in (\ref{subsample_NW}),
I consider the following updating algorithm, 
\begin{equation}
	\boldsymbol{\beta}_{k+1}=\boldsymbol{\beta}_{k}-\frac{\delta_{k}}{B}\sum_{i\in\mathfrak{I}_{B,k}}\left(\widehat{G}\left(\left.X_{0,i}+\mathbf{X}_{i}^{\mathrm{T}}\boldsymbol{\beta}_{k}\right|\boldsymbol{\beta}_{k},\mathfrak{I}_{B,k},\underline{c}_{f}\right)-y_{i}\right)\mathbf{X}_{i}^{\phi},\label{KcMBGD estimator}
\end{equation}
where $\mathbf{X}_{i}^{\phi} = \mathbf{X}_{i}\cdot \mathds{1}(\boldsymbol{\mathbf{X}}_{e,i}\in\mathcal{X}_{e}^{\phi})$,
and 
$
	\mathcal{X}_{e}^{\phi}=\left\{ \boldsymbol{\mathbf{X}}_{e}\in\mathcal{X}_{e}:\left|X_{j}\right|\leq1-\phi,0\leq j\leq p\right\} 
$
for some $0<\phi<1$\footnote{Such truncation is basically used to improve the uniform convergence speed of kernel estimation. Similar method is applied in many research such as \citet{ichimura1993semiparametric} and \citet{klein1993efficient}. }. 
Since the above algorithm generalizes the  conventional mini-batch gradient descent procedure to the semiparametric setup, I label the new algorithm the \textit{kernel-based mini-batch gradient descent algorithm} (KMBGD).  The algorithm is summarized in \autoref{KMBGD Estimator Algorithm}.

\IncMargin{1em}
\begin{algorithm} \SetKwData{Left}{left}\SetKwData{This}{this}\SetKwData{Up}{up} \SetKwFunction{Union}{Union}\SetKwFunction{FindCompress}{FindCompress} \SetKwInOut{Input}{input}\SetKwInOut{Output}{output}
	
	\Input{Data set $\left\{\left( \mathbf{X}_{e,i},y_i\right)\right\}_{i=1}^n$, sequence of learning rate $\left\{\delta_k\right\}_{k=1}^{\infty}$, initial guess $\boldsymbol{\beta}_{1}$, kernel function $K$, bandwidth $h_n$, subsample size $B$, number of iterations $T$, trimming parameter $\phi$ and $\underline{c}_f$} 
	\Output{The KMBGD estimator $\widehat{\boldsymbol{\beta}}$}
	\BlankLine 
	
	$k \leftarrow 1$;
	
	\While{$k\leq  T$}
	{Generate index set  $\mathfrak{I}_{B,k}$; \\
	\For{$l\leftarrow 1$ \KwTo $B$}{$\widehat{G}\left(\left.X_{0,i_{k,l}}+\mathbf{X}_{i_{k,l}}^{\mathrm{T}}\boldsymbol{\beta}_k\right|\boldsymbol{\beta}_{k}, \mathfrak{I}_{B,k}, \underline{c}_f\right)\leftarrow \frac{\frac{1}{B}\sum_{j\in\mathfrak{I}_{B,k}}K_{h_{n}}\left(X_{0,i_{k,l}}+\mathbf{X}_{i_{k,l}}^{\mathrm{T}}\boldsymbol{\beta}_k-X_{0,j}-\boldsymbol{\mathbf{X}}_{j}^{\mathrm{T}}\boldsymbol{\beta}_{k}\right)y_{j}}{\left\{\frac{1}{B}\sum_{j\in\mathfrak{I}_{B,k}}K_{h_{n}}\left(X_{0,i_{k,l}}+\mathbf{X}_{i_{k,l}}^{\mathrm{T}}\boldsymbol{\beta}_k-X_{0,j}-\boldsymbol{\mathbf{X}}_{j}^{\mathrm{T}}\boldsymbol{\beta}_{k}\right)\right\}\lor\underline{c}_f}$;}
		
		$\boldsymbol{\beta}_{k+1}\leftarrow \boldsymbol{\beta}_{k}-\frac{\delta_{k}}{B}\sum_{i\in{\mathfrak{I}_{B,k}}}\left(\widehat{G}\left(\left.X_{0,i}+\mathbf{X}_i^{\mathrm{T}}\boldsymbol{\beta}_k\right|\boldsymbol{\beta}_{k}, \mathfrak{I}_{B,k}, \underline{c}_f\right)-y_{i}\right)\mathbf{X}_{i}^{\phi}$;
		
		$k \leftarrow k+1$;}
	
	$
	\widehat{\boldsymbol{\beta}} \leftarrow \boldsymbol{\beta}_{T+1}
	$;
	
	\caption{The KMBGD Estimator}
	\label{KMBGD Estimator Algorithm} 
\end{algorithm}
\DecMargin{1em}

\begin{remark}
    I provide some comparisons between my KMBGD algorithm and the KBGD algorithm proposed in KLTY. Basically,  the KBGD algorithm is a full-sample-based algorithm; if I choose $\mathfrak{I}_{B,k} = \{1, \cdots, n\}$ for all $k$, then  KMBGD degenerates to KBGD. For computational burden, I obviously have that KBGD has computational complexity of order $O(n^2)$ in eahc update, while the update of KMBGD has complexity  of order $O\left(B^{2}\right)$. If I choose $B$ close to $1/\sqrt{n}$, the computational complexity of KMBGD will be close to $n$, which is linear in the sample size and is roughly $n$ times smaller than that of KBGD. This implies that when $n$ is extremely large, KMBGD is a better option.
\end{remark}

\begin{remark}
  Similar to the KBGD algorithm, my method is also iteration-based and  does not rely on any optimization procedure, so it can be easily implemented when the number of the covaraites $p$ is also large. In other words, the KMBGD estimator applies to the scenario where both $n$ and $p$ are large. For example, in the empirical application in \autoref{section6}, I consider semiparametric estimation of binary choice models when $p=337$ and $n=248060$. However, since in this paper I mainly focus on the scenario where the sample size $n$ is extremely large, in my following theoretical analysis I will take $p$ as being fixed. 
  \end{remark}

\section{\label{section3} Statistical Properties of KMBGD Estimator}

In this section, I formally study the statistical properties of the proposed KMBGD estimator. Under some regularity conditions, I first show that as long as I update sufficiently many times, the KMBGD estimator is consistent. However, the convergence rate is slower than $1/\sqrt{n}$ if I choose $B\ll n$. Indeed, such rate is even slower than $1/\sqrt{B}$, which is the convergence rate of general mini-batch estimators \citep{forneron2022estimation}. Then I will show that although KMBGD estimator itself converges at a slow rate, I can conduct averages across all the estimators produced during updates to accelerate the convergence rate. In particular, I show that if we properly choose subsample size, bandwidth prameter, order of kernel function, and number of iterations, the average estimator obtains $1/\sqrt{n}$-consistency.

Before I illustrate the main results, I first introduce some notations. Let $f_e\left(\mathbf{X}_e\right)$ and $f\left(\mathbf{X}\right)$ denote the joint density of $\mathbf{X}_e$ and $\mathbf{X}$\footnote{By assuming $\mathbf{X}_e$ has joint density function, we require that $\mathbf{X}_e$ is continuous, which facilitates our following discussion. However, I point out that my analysis can be trivially extended to the case where there are some discrete covariates, see KLTY. }. Define $z\left(\mathbf{X}_e, \boldsymbol{\beta}\right) = X_0 + \mathbf{X}^{\mathrm{T}}\boldsymbol{\beta}$. Let  $f_{\mathbf{X}|z}\left(\left.\mathbf{X}\right|z,\boldsymbol{\beta}\right)$
be the conditional density of $\boldsymbol{\mathbf{X}}$ given $z\left(\mathbf{X}_e, \boldsymbol{\beta}\right) =z$
and $\boldsymbol{\beta}$. Define 
\begin{align*}
	& W\left(\boldsymbol{\mathbf{X}}_{e},\widetilde{\boldsymbol{\mathbf{X}}}_{e},\boldsymbol{\beta}\right) =G^{\prime}\left(z\left(\boldsymbol{\mathbf{X}}_{e},\boldsymbol{\beta}^{\star}\right)+\left(\boldsymbol{\mathbf{X}}-\widetilde{\boldsymbol{\mathbf{X}}}\right)^{\mathrm{T}}\Delta\boldsymbol{\beta}\right)f_{\boldsymbol{X}|z}\left(\left.\widetilde{\boldsymbol{\mathbf{X}}},\right|z\left(\boldsymbol{\mathbf{X}}_{e},\boldsymbol{\beta}\right),\boldsymbol{\beta}\right),\\
	& V\left(\boldsymbol{\mathbf{X}}_{e},\widetilde{\boldsymbol{\mathbf{X}}}_{e},\boldsymbol{\beta}\right)=\left(\boldsymbol{\boldsymbol{\mathbf{X}}}\boldsymbol{\boldsymbol{\mathbf{X}}}^{\mathrm{T}}-\boldsymbol{\boldsymbol{\mathbf{X}}}\widetilde{\boldsymbol{\boldsymbol{\mathbf{X}}}}^{\mathrm{T}}\right)W\left(\boldsymbol{\mathbf{X}}_{e},\widetilde{\boldsymbol{\mathbf{X}}}_{e},\boldsymbol{\beta}\right),\\
	& \varLambda_{\phi}\left(\boldsymbol{\beta}\right)=\mathbb{E}\left[\mathds{1}_{i}^{\phi}\cdot\int_{\mathcal{X}}V\left(\boldsymbol{\mathbf{X}}_{e,i},\boldsymbol{\mathbf{X}}_{e},\boldsymbol{\beta}\right)d\boldsymbol{\mathbf{X}}\right].
\end{align*}
The following technical assumptions are imposed.

\begin{assumption}
	\label{assu1}An i.i.d. data set $\mathscr{D}_{n}=\left\{ \left(\mathbf{X}_{e,i},y_{i}\right)\right\} _{i=1}^{n}$
	of sample size $n$ is observed, where $y_{i}$ is generated by $
	y_i=\mathds{1}\left(X_{0,i}\beta_{0}^{\star}+\mathbf{X}_i^{\mathrm{T}}\boldsymbol{\beta}^{\star}-u_i>0\right)$ with unobserved shock $u_{i}$ that is independent
	of $\mathbf{X}_{e,i}$ and has CDF $G\left(\cdot\right)$.
\end{assumption}

\begin{assumption}
	\label{assump:2}(i) $\mathcal{X}_{e}=\left[-1,1\right]^{p+1}$; (ii)
	$\mathcal{B}_{e}$ is convex, and there exists some constant $B_{0}>0$
	such that for any $\boldsymbol{\beta}_{e}\in\mathcal{B}_{e}$, $\left|\beta_{j}\right|\leq B_{0}$
	for any $0\leq j\leq p$; (iii) The CDF $G$ has up to $(D+1)$-th order bounded derivatives. 
\end{assumption}

\begin{assumption}
	\label{assu:3}The kernel function $K\left(\cdot\right)$ satisfies:
	(i) $K$ is bounded and twice continuously differentiable with bounded
	first and second derivatives, and the second derivative satisfies
	Lipschitz condition on the whole real line; (ii) $\int K\left(s\right)ds=1$;
	(iii)  $\int s^{\upsilon}K\left(s\right)du=0$
	for $1\leq\upsilon\leq D-1$ and $\int u^{D}K\left(u\right)du\neq0$;
	(iv) $K\left(s\right)=0$ for $\left|s\right|>1$. 
\end{assumption}
\begin{assumption}
	\label{assu:4}(i) There exists some constant $\zeta>1$ such that
	$\zeta^{-1}\leq f_{e}\left(\mathbf{X}_{e}\right)\leq\zeta$ holds
	for all $\mathbf{X}_{e}\in\mathcal{X}_{e}$; (ii)  $f_{e}\left(\mathbf{X}_{e}\right)$
	has up to  $(D+1)$-th  order bounded derivatives.
\end{assumption}

\begin{assumption}
	\label{assu:5}There hold 
	\[
	\sup_{\boldsymbol{\beta}\in\mathcal{B}}\overline{\lambda}\left(\varLambda_0\left(\boldsymbol{\beta}\right)+\varLambda^{\mathrm{T}}_0\left(\boldsymbol{\beta}\right)\right)\leq\overline{\lambda}_{\varLambda}<\infty,
	\]
	and 
	\[
	\inf_{\boldsymbol{\beta}\in\mathcal{B}}\underline{\lambda}\left(\varLambda_0\left(\boldsymbol{\beta}\right)+\varLambda_0^{\mathrm{T}}\left(\boldsymbol{\beta}\right)\right)\geq\underline{\lambda}_{\varLambda}>0.
	\]
	
\end{assumption}

\begin{remark}
    Note that all the assumptions are also imposed in KLTY. This implies that extending KBGD to fully subsample-based algorithm does not require additional  assumptions. 
\end{remark}

Based on the above assumptions, now I formally study the statistical properties of the iterative estimator $\boldsymbol{\beta}_{k}$
based on iteration (\ref{subsample_NW}) and (\ref{KcMBGD estimator}). I first introduce some further notations. Let $P$ denote the probability measure of the data set $\mathcal{D}_n$. 
Let $\mathbb{P}^{*}$ be the probability measure corresponding to
random variables $\{\mathfrak{I}_{B,k}\}_{k=1}^{\infty}$ and
$\mathbb{P}_{k}^{*}$ be probability measure corresponding to $\{\mathfrak{I}_{B,k^{\prime}}\}_{k^{\prime}\geq k}^{\infty}$
conditional on the observation of   $\{\mathfrak{I}_{B,k^{\prime}}\}_{k^{\prime}=1}^{k-1}$
for $k\geq2$ and $\mathbb{P}_{1}^{*}=\mathbb{P}^{*}$. Let $\mathbb{E}^{*}$
and $\mathbb{E}_{k}^{*}$ be the expectation with respect to $\mathbb{P}^*$
and $\mathbb{P}_{k}^{*}$. Finally, let $\mathbb{P}$ be the probability
measure of $\{\mathscr{D}_{n}, \mathfrak{I}_{B,1}, \mathfrak{I}_{B,2}, \cdots\}$, where  $\mathscr{D}_{n}$ is the data set. 

Recall that the Nadaraya-Watson kernel estimator for $\mathbb{E}\left(\left.y\right|X_{0}+\mathbf{X}^{\mathrm{T}}\boldsymbol{\beta}=z\right)$
based on the full data is given by $\widehat{G}\left(\left.z\right|\boldsymbol{\beta}\right)$ in (\ref{kernel_estimator}).
For any $\boldsymbol{\beta}\in\mathcal{B}$, define $\Delta \boldsymbol{\beta} = \boldsymbol{\beta}- \boldsymbol{\beta}^{\star}$.   I obviously have the following decomposition for the MBGD update (\ref{KcMBGD estimator}), 
\begin{align}\label{decomposition}
	\Delta\boldsymbol{\beta}_{k+1} & =\Delta\boldsymbol{\beta}_{k}-\frac{\delta_{k}}{n}\sum_{i=1}^{n}\left(\widehat{G}\left(\left.X_{0,i}+\mathbf{X}_{i}^{\mathrm{T}}\boldsymbol{\beta}_{k}\right|\boldsymbol{\beta}_{k}\right)-y_{i}\right)\mathbf{X}_{i}^{\phi}\nonumber\\
&-\delta_k\underset{\pi_{1,n,k}}{\underbrace{\frac{1}{B}\sum_{i\in\mathfrak{I}_{B,k}}\left(\widehat{G}\left(\left.X_{0,i}+\mathbf{X}_{i}^{\mathrm{T}}\boldsymbol{\beta}_{k}\right|\boldsymbol{\beta}_{k}\right)-y_{i}\right)\mathbf{X}_{i}^{\phi}-\frac{1}{n}\sum_{i=1}^{n}\left(\widehat{G}\left(\left.X_{0,i}+\mathbf{X}_{i}^{\mathrm{T}}\boldsymbol{\beta}_{k}\right|\boldsymbol{\beta}_{k}\right)-y_{i}\right)\mathbf{X}_{i}^{\phi}}}\nonumber \\
&-\delta_k\underset{\pi_{2,n,k}}{\underbrace{\frac{1}{B}\sum_{i\in\mathfrak{I}_{B,k}}\left(\widehat{G}\left(\left.X_{0,i}+\mathbf{X}_{i}^{\mathrm{T}}\boldsymbol{\beta}_{k}\right|\boldsymbol{\beta}_{k},\mathfrak{I}_{B,k},\underline{c}_{f}\right)-\widehat{G}\left(\left.X_{0,i}+\mathbf{X}_{i}^{\mathrm{T}}\boldsymbol{\beta}_{k}\right|\boldsymbol{\beta}_{k}\right)\right)\mathbf{X}_{i}^{\phi}}}.
\end{align}
It's not difficult to see that if $\pi_{1,n,k} = \pi_{2,n,k} =0$, then (\ref{decomposition}) degenerates to the full-sample-based KBGD algorithm. Indeed, $\pi_{1,n,k}$
describes the randomness caused by updating using only a subset of the data, whereas $\pi_{2,n,k}$
describes the randomness caused by performing nonparametric kernel estimation using only
a subset of the data points. Essentially, $\pi_{1,n,k}$ is shared by all the mini-batch estimators, while $\pi_{2,n,k}$ is specific to the semiparametric setup I consider in this paper. I have the following lemma describing the properties of $\pi_{1,n,k}$ and $\pi_{2,n,k}$.

\begin{lemma}\label{lem:4.2}
	Suppose that  \autoref{assu1}--\autoref{assu:5}  
	hold with $D\geq 4$. Suppose also that  $\underline{c}_f $ is chosen such that  $\inf_{z\in\mathcal{Z}^{\phi},\boldsymbol{\beta}\in\mathcal{B}}f_{Z}\left(\left.z\right|\boldsymbol{\beta}\right)\geq3\underline{c}_{f}$. 
	 If $\boldsymbol{\beta}_k$ is update based on (\ref{subsample_NW}) and (\ref{KcMBGD estimator}), I have that
	\[
	P\left(\sup_{k\geq1}\mathbb{E}^{*}\left(\left\Vert \pi_{1,n,k}\right\Vert ^{2}\right)\leq CB^{-1}\right)\rightarrow 1,
	\]
	and
	\[
	P\left(\sup_{k\geq1}\mathbb{E}^{*}\left(\left\Vert \pi_{2,n,k}\right\Vert ^{2}\right)\leq C\log\left(Bh_{n}^{-2}\right)/Bh_{n}^{2}\right)\rightarrow1,
	\]
	for some $C$ that does not depend on $n, B, h_n$, and $k$. 
\end{lemma}

\autoref{lem:4.2} immediately yields the following result.

\begin{theorem}\label{thm7}
	Suppose that  \autoref{assu1}--\autoref{assu:5}  
	hold with $D\geq 4$. Suppose also that  $\underline{c}_f $ is chosen such that  $\inf_{z\in\mathcal{Z}^{\phi},\boldsymbol{\beta}\in\mathcal{B}}f_{Z}\left(\left.z\right|\boldsymbol{\beta}\right)\geq3\underline{c}_{f}$. 
	Suppose moreover that 
	$\delta_{k}=\delta<\min\{ 1/\left(2\underline{\lambda}_{\varLambda}\right),1/\left(4p^{2}\left\Vert G^{\prime}\right\Vert _{\infty}\right)\} $, $\phi<\delta \underline{\lambda}_{\varLambda}/\left(16p^{2}\left\Vert G^{\prime}\right\Vert _{\infty}\zeta\right)$, 
	$h_{n}$ is chosen such that $h_nn^{1/2D}\rightarrow0$ and $h_{n}n^{1/6}/\log^{1/3} \left(n\right)\rightarrow\infty$.    If $\boldsymbol{\beta}_k$ is update based on (\ref{subsample_NW}) and (\ref{KcMBGD estimator}), define 
	\[
	k_n=\left[\frac{\log\left(h_n^{2D}+\sqrt{\log\left(Bh_{n}^{-2}\right)/Bh_{n}^{2}}\right)-\log\left(\sqrt{\mathbb{E}^{*}\left\Vert \Delta\boldsymbol{\beta}_{1}\right\Vert ^{2}}\right)}{\log\left(1-\delta\underline{\lambda}_{\varLambda}/8\right)}\right],
	\]
	I have that 
	\[
	\sup_{k\geq k_{n}+1}\mathbb{E}^{*}\left(\left\Vert \Delta\boldsymbol{\beta}_{k}\right\Vert ^{2}\right)=O_{p}\left(h_{n}^{2D}+\frac{\log\left(Bh_{n}^{-2}\right)}{Bh_{n}^{2}}\right).
	\]
\end{theorem}

According to \autoref{thm7}, if I choose $B\ll n$ to improve computational speed, the upper bounded on the estimation error  $\mathbb{E}^{*}\left(\left\Vert \Delta\boldsymbol{\beta}_{k}\right\Vert \right) $ will be of rate slower than $n^{-1/2}$ even when the order of the kernel function is large. The slower convergence rate is a common feature of all the  mini-batch  estimators. Indeed, the mini-batch estimators converge at the rate $1/\sqrt{B}$ at best, see, for example, Lemma 2 in  \citet{forneron2022estimation}. However, different from the conventional mini-batch estimator, my KMBGD estimators are guaranteed to  converge no faster than  $\sqrt{\log(n)/Bh_n^2}$. If I choose $B=1/\sqrt{n}$ and $h_n = n^{-1/6}$, then the convergence rate would be $\sqrt{\log(n)}n^{-1/12}$, which is much slower than $1/\sqrt{B}=n^{-1/4}$. 

The slower convergence rate of the KMBGD estimator is mainly due to the fact that I use subsamples to construct the kernel estimator. In this case, the subsample-based gradient is no longer an unbiased estimator (conditional on the previous subsamples) of the full-sample-based gradient, that is, $\mathbb{E}^*(\pi_{2,n,k})\neq 0$. The bias makes the convergence rate of KMBGD estimator slower than $1/\sqrt{B}$.  However, surprisingly, in the following  I will show that if I appropriately choose the kernel function  and bandwidth parameter, even with $B\ll n$, I can still obtain $1/\sqrt{n}$ by following \citet{polyak1992acceleration} and conducting average across KMBGD estimators produced during iterations. 

To formally show the above results, I first further decompose  the KMBGD dynamics. To ease my following exposition, for any $z$ and $\boldsymbol{\beta}$ denote 
$A_{n,y}\left(z,\boldsymbol{\beta}\right)=\frac{1}{n}\sum_{i=1}^{n}K_{h_{n}}\left(z-X_{0,i}-\mathbf{X}_{i}^{\mathrm{T}}\boldsymbol{\beta}\right)y_{i}$,  
$A_{n,1}\left(z,\boldsymbol{\beta}\right)=\frac{1}{n}\sum_{i=1}^{n}K_{h_{n}}\left(z-X_{0,i}-\mathbf{X}_{i}^{\mathrm{T}}\boldsymbol{\beta}\right)
$, 
	$A_{n,y}\left(\left.z,\boldsymbol{\beta}\right|\mathfrak{I}_{B,k}\right)=\frac{1}{B}\sum_{i\in\mathfrak{I}_{B,k}}K_{h_{n}}\left(z-X_{0,i}-\mathbf{X}_{i}^{\mathrm{T}}\boldsymbol{\beta}\right)y_{i}$, and 
	$A_{n,1}\left(\left.z,\boldsymbol{\beta}\right|\mathfrak{I}_{B,k}\right)=\frac{1}{B}\sum_{i\in\mathfrak{I}_{B,k}}K_{h_{n}}\left(z-X_{0,i}-\mathbf{X}_{i}^{\mathrm{T}}\boldsymbol{\beta}\right)
$. 
I have the following lemma. 

\begin{lemma}\label{lem:4.3}
Suppose that all the assumptions and conditions  in  \autoref{thm7} hold.  Suppose moreover that  $B \cdot \min\{h_n^6/\log^{2}(n), h_n^2 /(\sqrt{n}\log(n))\}\rightarrow \infty$.   Define 
$
	\boldsymbol{\xi}_{n}^{\phi}=\frac{1}{n}\sum_{i=1}^{n}(\widehat{G}\left(\left.z_{i}^{\star}\right|\boldsymbol{\beta}^{\star}\right)-y_{i})\mathbf{X}_{i}^{\phi}$, where $z_i^{\star} = z\left(\mathbf{X}_{e,i}, \boldsymbol{\beta}^{\star}\right)$. Also define $z_{i,k} = z(\mathbf{X}_{e,i}, \boldsymbol{\beta}_k)$. If $\boldsymbol{\beta}_k$ is update based on (\ref{subsample_NW}) and (\ref{KcMBGD estimator}), I have that 
	\begin{align*}
		\Delta\boldsymbol{\beta}_{k+1} & =\left(I_{p}-\delta\varLambda_{\phi}\left(\boldsymbol{\beta}^{\star}\right)\right)\Delta\boldsymbol{\beta}_{k}-\delta\boldsymbol{\xi}_{n}^{\phi}+\delta\varOmega_{k}^{\phi}\\
		&-\delta\underset{\varrho_{1,n,k}}{\underbrace{\frac{1}{B}\sum_{i\in\mathfrak{I}_{B,k}}\left(\widehat{G}\left(\left.z_{i,k}\right|\boldsymbol{\beta}_{k}\right)-y_{i}\right)\mathbf{X}_{i}^{\phi}-\frac{1}{n}\sum_{i=1}^{n}\left(\widehat{G}\left(\left.z_{i,k}\right|\boldsymbol{\beta}_{k}\right)-y_{i}\right)\mathbf{X}_{i}^{\phi}}}\\
		& -\delta\underset{\varrho_{2,n,k}}{\underbrace{\frac{1}{B}\sum_{i\in\mathfrak{I}_{B,k}}\frac{\mathbf{X}_{i}^{\phi}}{A_{n,1}\left(z_{i,k},\boldsymbol{\beta}_{k}\right)}\cdot\left(A_{n,y}\left(\left.z_{i,k},\boldsymbol{\beta}_{k}\right|\mathfrak{I}_{B,k}\right)-A_{n,y}\left(z_{i,k},\boldsymbol{\beta}_{k}\right)\right)}}\\
		& +\delta\underset{\varrho_{3,n,k}}{\underbrace{\frac{1}{B}\sum_{i\in\mathfrak{I}_{B,k}}\frac{A_{n,y}\left(z_{i,k},\boldsymbol{\beta}_{k}\right)\mathbf{X}_{i}^{\phi}}{A_{n,1}^{2}\left(z_{i,k},\boldsymbol{\beta}_{k}\right)}\cdot\left(A_{n,1}\left(\left.z_{i,k},\boldsymbol{\beta}_{k}\right|\mathfrak{I}_{B,k}\right)-A_{n,1}\left(z_{i,k},\boldsymbol{\beta}_{k}\right)\right)}},
	\end{align*}
	where 
	$\sup_{k\geq k_{n}+1}\mathbb{E}^{*}\left\Vert \varOmega_{k}^{\phi}\right\Vert =o_{p}\left(n^{-1/2}\right).
	$
\end{lemma}

I now provide some discussion for \autoref{lem:4.3}. Basically, if there are no noise terms $\varrho_{1,n,k}$, $\varrho_{2,n,k}$, and $\varrho_{3,n,k}$, then the dynamics of $\Delta \boldsymbol{\beta}_k$ simply degenerate to the full-sample-based KBGD algorithm in KLTY as implied in \autoref{thm:3.3}  in Appendix. However, since I use subsamples to perform the update, additional noises due to subsampling are introduced into the update and these noises are captured by the above three  terms. Basically, $\varrho_{1,n,k}$ describes the impacts of using subsamples instead of full sample to perform the update. Such error is shared by all the mini-batch-based methods. While the remaining two terms $\varrho_{2,n,k}$ and $\varrho_{3,n,k}$ describe the impacts of using subsamples instead of full sample to construct the Nadaraya-Watson kernel estimator, so are specific to my algorithm only. 
Simple calculation leads to 
$
\mathbb{E}^* \left(\varrho_{1,n,k}\right) = 0,
$
 $\mathbb{E}^{*}\left(\varrho_{2,n,k}\right) = O_p\left(1/Bh_n\right), 
$
and
$\mathbb{E}^{*}\left(\varrho_{3,n,k}\right) = O_p\left(1/Bh_n\right) $
uniformly with respect to $k$. The above implies that for $k$ sufficiently large, the first-order difference between KBGD and KMBGD estimators almost constitute a martingale difference sequence. By ``almost'' I mean that the conditional expectation is of order $O_p(1/Bh_n)$, which can be  made   $n^{-1/2}$-trivial if I choose $B\gg n^{1/2}h_n^{-1}$. 

\autoref{lem:4.3} implies that although the KMBGD estimator itself does not obtain $1/\sqrt{n}$-consistency due to noises caused by subsample-based kernel estimation and update,  
I can  follow \citet{polyak1992acceleration} to conduct average across the estimators produced during iterations to eliminate these noises. Similar to the conventional mini-batch gradient estimator, the resulting
estimator will be $1/\sqrt{n}$-consistent as long as we choose $B$ that diverges at some rate.  In particular, let $k^*$ be the number of burn-in  iterations and  $T$ be the number of follow-up iterations. The averaged KMBGD estimator (AKMBGD) is defined as follws,
\begin{align}\label{AKMBGD}
	\overline{\boldsymbol{\beta}} = \frac{1}{T}\sum_{t=1}^{T} \boldsymbol{\beta}_{k^* + t}.
\end{align}
I summarize the
algorithm in \autoref{AKMBGD Estimator Algorithm}.

\IncMargin{1em}
\begin{algorithm} \SetKwData{Left}{left}\SetKwData{This}{this}\SetKwData{Up}{up} \SetKwFunction{Union}{Union}\SetKwFunction{FindCompress}{FindCompress} \SetKwInOut{Input}{input}\SetKwInOut{Output}{output}
	
	\Input{Data set $\left\{\left( \mathbf{X}_{e,i},y_i\right)\right\}_{i=1}^n$, sequence of learning rate $\left\{\delta_k\right\}_{k=1}^{\infty}$, initial guess $\boldsymbol{\beta}_{1}$, kernel function $K$, bandwidth $h_n$, subsample size $B$, number of burn-in iterations $k^*$, number of follow-up iterations $T$, trimming parameter $\phi$ and $\underline{c}_f$} 
	\Output{The AKMBGD estimator $\overline{\boldsymbol{\beta}}$}
	\BlankLine 
	
	$k \leftarrow 1$;
	
	\While{$k\leq k^* + T$}
	{Generate index set  $\mathfrak{I}_{B,k}$; \\
	\For{$l\leftarrow 1$ \KwTo $B$}{$\widehat{G}\left(\left.X_{0,i_{k,l}}+\mathbf{X}_{i_{k,l}}^{\mathrm{T}}\boldsymbol{\beta}_k\right|\boldsymbol{\beta}_{k}, \mathfrak{I}_{B,k}, \underline{c}_f\right)\leftarrow \frac{\frac{1}{B}\sum_{j\in\mathfrak{I}_{B,k}}K_{h_{n}}\left(X_{0,i_{k,l}}+\mathbf{X}_{i_{k,l}}^{\mathrm{T}}\boldsymbol{\beta}_k-X_{0,j}-\boldsymbol{\mathbf{X}}_{j}^{\mathrm{T}}\boldsymbol{\beta}_{k}\right)y_{j}}{\left\{\frac{1}{B}\sum_{j\in\mathfrak{I}_{B,k}}K_{h_{n}}\left(X_{0,i_{k,l}}+\mathbf{X}_{i_{k,l}}^{\mathrm{T}}\boldsymbol{\beta}_k-X_{0,j}-\boldsymbol{\mathbf{X}}_{j}^{\mathrm{T}}\boldsymbol{\beta}_{k}\right)\right\}\lor\underline{c}_f}$;}
		
		$\boldsymbol{\beta}_{k+1}\leftarrow \boldsymbol{\beta}_{k}-\frac{\delta_{k}}{B}\sum_{i\in{\mathfrak{I}_{B,k}}}\left(\widehat{G}\left(\left.X_{0,i}+\mathbf{X}_i^{\mathrm{T}}\boldsymbol{\beta}_k\right|\boldsymbol{\beta}_{k}\right)-y_{i}\right)\mathbf{X}_{i}^{\phi}$;
		
		$k \leftarrow k+1$;}
	
	$
	\overline{\boldsymbol{\beta}} \leftarrow \frac{1}{T}\sum_{t=1}^{T} \boldsymbol{\beta}_{k^* + t}
	$;
	
	\caption{The AKMBGD Estimator}
	\label{AKMBGD Estimator Algorithm} 
\end{algorithm}
\DecMargin{1em} 

Now I provide the theoretical properties of the AKMBGD estimator.

\begin{theorem}\label{thm8}
	Suppose that all the assumptions and conditions in \autoref{thm7}
	hold. Suppose moreover that  
	$B \cdot \min\{h_n^6/\log^{2}(n), h_n^2 /(n^{1/2}\log(n))\}\rightarrow \infty$.
	Let $k^* = k_n +[-\log(n)/\log(1-\delta\underline{\lambda}_{\varLambda}/8)]
	$.  If $\boldsymbol{\beta}_k$ is update based on (\ref{subsample_NW}) and (\ref{KcMBGD estimator}), for any $T\geq 1$, I have that 
	\[
	\Delta \overline{\boldsymbol{\beta}}  = -\varLambda_{\phi}^{-1}\left(\boldsymbol{\beta}^{\star}\right)\boldsymbol{\xi}_{n}^{\phi}+O_{\mathbb{P}}\left(\frac{1}{\sqrt{Bh_n^{2}T}} + \frac{\log^{1/4}(n)}{Bh_n}\right).
	\]
	If $T$ is chosen such that $Bh_n^2Tn^{-1} \rightarrow \infty$, I have that 
	\[
	 \sqrt{n}\Delta \overline{\boldsymbol{\beta}}  \rightarrow_d \mathcal{N}\left(0,\Sigma_{\boldsymbol{\beta}}^{\phi}\right),
	\]
where $\Sigma_{\boldsymbol{\beta}}^{\phi}=\varLambda_{\phi}^{-1}\left(\boldsymbol{\beta}^{\star}\right)\Sigma_{\boldsymbol{\xi}}^{\phi}\left(\varLambda_{\phi}^{-1}\left(\boldsymbol{\beta}^{\star}\right)\right)^{\mathrm{T}}$  and 
	\[
	\Sigma_{\boldsymbol{\xi}}^{\phi}=\mathbb{E}\left[\left(1-G\left(z_{i}^{\star}\right)\right)G\left(z_{i}^{\star}\right)\left(\mathbf{X}_{i}^{\phi}-\mathbb{E}\left(\left.\mathbf{X}_{i}^{\phi}\right|z_{i}^{\star}\right)\right)\left(\mathbf{X}_{i}^{\phi}-\mathbb{E}\left(\left.\mathbf{X}_{i}^{\phi}\right|z_{i}^{\star}\right)\right)^{\mathrm{T}}\right].
	\]
\end{theorem}

\autoref{thm8} is the key result of this paper. It demonstrates that even though I only use a random subsample whose size is substaintially smaller than the full  sample size to conduct kernel estimation and perform update in each round of iteration, the average of estimators produced during iterations will be equivalent to the full-sample estimator up to some small order terms. The small order terms will be uniformly $1/\sqrt{n}$-trivial as long as I choose $B\gg \max\{\log^2(n)h_n^{-6},\sqrt{n}\log(n)h_n^{-2}\}$ and $T \gg nB^{-1}h_n^{-2}$. This implies that as long as I choose kernel function properly, my KMBGD estimator will be as efficient as the one based on the full sample, dispite the fact that I only use a much smaller subsample to perform the update in each round.

\autoref{thm8} also suggests that the computational speed of each update can be improved  by appropriately choosing the kernel function. In particular, since  $h_n$ must satisfy 
$h_n \ll n^{-1/2D}$ according to the conditions required in the theorem, then $B\gg \max\{n^{3/D}\log^2(n), n^{1/2+1/D}\log (n)\}$ must hold, so 
the computational complexity will be of order at least $O(\max\{n^{6/D}\log^4(n), n^{1+2/D}\log^2 (n)\})$. Obviously, to improve the computational speed, I can choose a high-order kernel function. For example, if I choose a 8-th order kernel, the computational complexity is of order $O(n^{5/4}\log^2(n))$; if I choose a 12-th order kernel, the computational complexity is of order $O(n^{7/6}\log^2(n))$. If I can choose sufficiently large $D$, then the computational complexity is lower bounded by $n\log^2(n)$,  which is almost the linear rate $O(n)$. 

I finally discuss the total computational time of KBGD and KMBGD estimation. Suppose $k^*$ updates are necessary to eliminate the impacts of the initial guess, then the full-sample-based KBGD algorithm requires $O(k^*n^2)$ computational time in total, while the KMBGD algorithms requires $O(k^*B^2 + B^2T)$. Since \autoref{thm8} requires that $T\gg nB^{-1}h_n^-2$, then the total computational time of KMBGD will be at least $O(k^*B^2 + nBh_n^{-2})$. If I choose $B\gg \sqrt{n}h_n^{-2}\log n$ and $h_n \ll n^{-1/2D}$, then $k^*B^2 + nBh_n^{-2}\gg k^*n^{1+2/D}\log^2(n) + n^{3/2+2/D}$. So the upper bound on the ratio between the total computational time of KBGD and KMBGD is of order 
\[
n^{1-2/D}\log^{-2}(n) + k^*n^{1/2 - 2/D}.
\]
Obviously, when $D\geq6$, the above ratio diverage at rate $n^{2/3} + k^*n^{1/6}$. More crucially, the above rate will be large when $k^*$, the number of burn-in updates, is large, which will often be the case when the number of covariates is large and $\underline{\varLambda}/\overline{\varLambda}$ is small,

\begin{remark}
    All the  theories so far are developed for binary chocie models with continuous covariates, but my method can be directly applied to the case where more general monotone index models are considered and there are some discrete covariates without any modifications. See my simulation results in \autoref{section5}. 
    
\end{remark}

\begin{remark}\label{rem7}
    
    Regarding the choice of the tuning parameter, I recommend choosing $\delta_k = 1$ for all $k$ in the first place, and if the iteration diverges, then gradually shrink it towards zero. For the choice of $B$, I recommend choosing $B = \max\{1000, \sqrt{n}h_n^{-1}\log(n)\}$. For the stopping rule, I  recommend updating until the mean of the estimators produced during iterations is stable. For example, let $T$ and $gap$ be two positive integers. First update the parameter $T+gap$ rounds. Then for each $k>T+gap$,   compare two average estimators $\frac{1}{T}\sum_{j=1}^{T}\boldsymbol{\beta}_{k-j}$ and $\frac{1}{T}\sum_{j=1}^{T}\boldsymbol{\beta}_{k-j-gap}$. If the maximum distance between arguments of the above two estimators is smaller than some given tolerance $\varrho$, then stop and use the average of last $T+gap$ estimators as the final estimator. For another example, I can choose some pre-specified numbers of burn-in and follow-up updates, as long as both are sufficiently large.  
\end{remark}

\section{Inference with Large $n$}\label{section4}

In this section, I discuss the inference-related issues when the sample size $n$ is large. According to \autoref{thm8}, the AKMBGD estimator is asymptotically normally distributed, so inference on the true parameter $\boldsymbol{\beta}^{\star}$ can be  conducted  if I can consistently estimate the asymptotic covariance matrix $\Sigma_{\boldsymbol{\beta}}^{\phi}$. In their paper, KLTY provide a consistent estimator for the covariance matrix based on the full sample. 
However, to construct such  estimator, I need to construct nonparametric estimators for conditional expectation $\mathbb{E}\left(\left. \mathbf{X}_{i}^{\phi}\right|z_i^{\star}\right)$ for each $i$, which may cost large amount of time when both $n$ and $p$ are large. 

For parametric optimization,  \citet{forneron2022estimation} proposes a stochastic Newton-Raphson udpate   and use the produced estimators for inference to alleviate the computational burden of statistical inference. But his method can not be applied in the current scenario even if I can approximate the ``Hessian'' matrix\footnote{Note that in our case, the ``Hessian'' refers to the matrix $\varLambda_{\phi} (\boldsymbol{\beta}^{\star})$, which is actually not symmetric.} accurately. This is because,  apart from $\varrho_{1,n,k}$ that captures the distribution of $\boldsymbol{\xi}^{\phi}_n$,   additional subsampling errors $\varrho_{2,n,k}$ and $\varrho_{3,n,k}$ are introduced because I use subsamples to construct the nonparametric estimator. Such additional errors are at least of the same order as $\varrho_{1,n,k}$, so they dampen the bootstrap-based inference. 

To solve the above inference issue in the large $n$ scenario,   this section provides a subsample-based estimator for the covariance matrix. Let $\{\mathfrak{I}_{B,r}\}_{r=1}^R$  be a sequence of random index sets defined in (\ref{subsample}).  For each $1\leq r\leq R$, define
	\[
	\widehat{\Sigma}_{\boldsymbol{\xi}}^{\phi, r}=\frac{1}{B}\sum_{i\in \mathfrak{I}_{B,r}}\left(\widehat{G}_{i}^r\left(1-\widehat{G}_{i}^r\right)\left(\mathbf{X}_{i}^{\phi}-\widehat{\mathbb{E}}^r\left(\left.\mathbf{X}_{i}^{\phi}\right|\widehat{z}_{i}\right)\right)\left(\mathbf{X}_{i}^{\phi}-\widehat{\mathbb{E}}^r\left(\left.\mathbf{X}_{i}^{\phi}\right|\widehat{z}_{i}\right)\right)^{\mathrm{T}}\right),
	\]
	and 
	\[
	\widehat{\varLambda}^r_{\phi}\left(\overline{\boldsymbol{\beta}}\right)=\frac{1}{B}\sum_{i\in\mathfrak{I}_{B,r}}\mathbf{X}_{i}^{\phi}\frac{\partial\widehat{G}\left(\left.z\left(\mathbf{X}_{e,i},\overline{\boldsymbol{\beta}}\right)\right|\overline{\boldsymbol{\beta}}, \mathfrak{I}_{B,r}, \overline{c}_f\right)}{\partial\boldsymbol{\beta}^{\mathrm{T}}},
	\]
	where 
	\[
	\widehat{G}_{i}^r=\frac{\frac{1}{B}\sum_{j\in\mathfrak{I}_{B,r}}K_{h_{n}}\left(\widehat{z}_{i}-\widehat{z}_{j}\right)y_{j}}{\left\{\frac{1}{B}\sum_{j\in\mathfrak{I}_{B,r}}K_{h_{n}}\left(\widehat{z}_{i}-\widehat{z}_{j}\right)\right\}\lor \overline{c}_f},\ \widehat{\mathbb{E}}^r\left(\left.\mathbf{X}_{i}^{\phi}\right|\widehat{z}_{i}\right)=\frac{\frac{1}{B}\sum_{j\in\mathfrak{I}_{B,r}}K_{h_{n}}\left(\widehat{z}_{i}-\widehat{z}_{j}\right)\mathbf{X}_{j}^{\phi}}{\left\{\frac{1}{B}\sum_{j=\in\mathfrak{I}_{B,r}}K_{h_{n}}\left(\widehat{z}_{i}-\widehat{z}_{j}\right)\right\}\lor \overline{c}_f},
	\]
	and $\widehat{z}_{i}=X_{0,i}+\mathbf{X}_{i}^{\mathrm{T}}\overline{\boldsymbol{\beta}}$. Also define  
	\begin{equation}
\widetilde{\Sigma}_{\boldsymbol{\beta}}^{\phi} = 
	\left(\frac{1}{R}\sum_{i=1}^R\widehat{\varLambda}^r_{\phi}\left(\overline{\boldsymbol{\beta}}\right)\right)^{-1}\left(\frac{1}{R}\sum_{r=1}^R\widehat{\Sigma}_{\boldsymbol{\xi}}^{\phi,r}\right)\left(\frac{1}{R}\sum_{r=1}^R\widehat{\varLambda}_{\phi }^{r \mathrm{T}}\left(\overline{\boldsymbol{\beta}}\right)\right)^{-1}.
\label{variance_estimator}	\end{equation}
 Then we have the following result.
\begin{theorem}\label{thm9}
    Suppose that all the assumptions and conditions in \autoref{thm7}
	hold. If $Bh_n^{2}\rightarrow \infty$,  I have that 
    \[
        \left\Vert \mathbb{P}^*\mathrm{lim}_{R\rightarrow\infty}\widetilde{\Sigma}_{\boldsymbol{\beta}}^{\phi} - \Sigma_{\boldsymbol{\beta}} ^{\phi}\right\Vert \rightarrow_{\mathbb{P}} 0,
    \]
    where $\mathbb{P}^*$  and $\mathbb{P}$ are defined in \autoref{section3}.
    Moreover,  
    \[
    \widetilde{\Sigma}_{\boldsymbol{\beta}}^{\phi -1/2}\sqrt{n}\Delta\overline{\boldsymbol{\beta}}\rightarrow_d \mathcal{N}(0, I_p).
    \]
\end{theorem}
 
 \begin{remark}
     When using subsamples to construct the  estimators,  $\widehat{\varLambda}^r_{\phi}$ and $\widehat{\Sigma}_{\boldsymbol{\xi}}^{\phi, r}$ may largely 
 deviate from their full-sample counterparts for some subsamples due to subsampling randomness. A large $R$ is then required to offset such randomness, which increases the computational time. To control for the subsampling randomness and alleviate the computational burden, I recommend detecting outliers of among $\{\widehat{\varLambda}^r_{\phi}\}_{r=1}^R$ and $\{\widehat{\Sigma}_{\boldsymbol{\xi}}^{\phi, r}\}_{r=1}^R$, and leaving out the subsample-based estimators which are detected as outliers. Finally, the  estimator of the variance is constructed as in (\ref{variance_estimator}) based on the remaining subsamples.  We also note that the subsample size $B$ used in the calculation of the asymptotic covariance can be different from the one used in estimaton. According to my simulations, choosing $B=3000$ and $R = 200$ lead to fairly accurate estimators.  
 \end{remark}

\section{\label{section5} Monte Carlo Experiments}
This section conducts some Monte Carlo experiments to evaluate the finite-sample performance as well as the computational efficiency of the proposed KMBGD and AKMBGD estimators. Throughout this section, I consider the following data generating process 
\begin{equation}
y_i = \mathds{1}\left(X_{0,i} + \beta_1^{\star}X_{1,i} + \cdots + \beta_{9} ^{\star} X_{9, i} - u_i >0\right), 1\leq i\leq n,
\end{equation}
where $n$ is the sample size.  For all $1\leq i\leq n$,  $X_{0,i}\sim \mathcal{N}(0,1)$, $X_{1,i}\sim \text{Bernoulli}(1/2)$, $X_{2,i}\sim \text{Poisson}(2)$, and $X_{j,i}\sim (\chi^{2}(1)-1)/\sqrt{2}$ for $3\leq j \leq 9$. So I have a mixture of both continuous and discrete covariates. Moreover, $X_{j,i}$ is independent over $j$ for each $i$. $u_i$ is the random error with cumulative distribution function $G(u)$, which is independent of the covariates. $(X_{0,i}, \cdots, X_{9,i}, u_i)$ is iid over $i$. I set the true parameter vector as 
$\boldsymbol{\beta}^{\star}=\left(1,1,0.5,2,5,-0.5,-1,-2,-5\right)^{\mathrm{T}}.$ 
 I consider four setups of error distrubtion: $\text{Cauchy}$, $t(4)$, $\chi^2(3)$, and $\mathcal{N}(0,1)$.  Finally, in the following simulations, whenever I conduct the kernel estimation, I use sixth-order Epanechnikov
 kernel to construct the Nadaraya-Watson estimator, where the kernel function is given by $K(u) = \frac{525}{256}\left(1 - u^2\right)\left(1-6u^2 - \frac{33}{5}u^4\right)\mathds{1}\left(|u|\leq 1\right)$. 
 
\begin{table}
\centering{}\caption{\label{simutab1}Finite Sample Performance of Kernel-Based Estimators}
\begin{tabular}{ll>{\centering}p{1cm}>{\centering}p{1cm}>{\centering}p{1cm}>{\centering}p{1cm}>{\centering}p{1cm}>{\centering}p{1cm}>{\centering}p{1cm}>{\centering}p{1cm}>{\centering}p{1cm}}
 & \multicolumn{10}{c}{}\tabularnewline
\hline 
\hline 
 & \multicolumn{10}{c}{$u_{i}\sim\mathrm{Cauchy}$}\tabularnewline
\hline 
 &  & $\beta_{1}$ & $\beta_{2}$ & $\beta_{3}$ & $\beta_{4}$ & $\beta_{5}$ & $\beta_{6}$ & $\beta_{7}$ & $\beta_{8}$ & $\beta_{9}$\tabularnewline
\hline 
\multirow{3}{*}{$n=50000$} & Bias & 0.0051 & 0.0010 & 0.0016 & 0.0042 & 0.0088 & 0.0003 & 0.0015 & 0.0041 & 0.0100\tabularnewline
 & RMSE & 0.0533 & 0.0314 & 0.0309 & 0.0610 & 0.1305 & 0.0258 & 0.0326 & 0.0549 & 0.1222\tabularnewline
 & CR & 0.9570 & 0.9520 & 0.9490 & 0.9660 & 0.9660 & 0.9590 & 0.9580 & 0.9550 & 0.9670\tabularnewline
\hline 
\multirow{3}{*}{$n=100000$} & Bias & 0.0006 & 0.0007 & 0.0003 & 0.0004 & 0.0016 & 0.0003 & 0.0009 & 0.0012 & 0.0036\tabularnewline
 & RMSE & 0.0366 & 0.0208 & 0.0206 & 0.0425 & 0.0924 & 0.0173 & 0.0229 & 0.0379 & 0.0879\tabularnewline
 & CR & 0.9580 & 0.9590 & 0.9530 & 0.9490 & 0.9540 & 0.9640 & 0.9540 & 0.9570 & 0.9480\tabularnewline
\hline 
 & \multicolumn{10}{c}{$u_{i}\sim t(4)$}\tabularnewline
\hline 
\multirow{3}{*}{$n=50000$} & Bias & 0.0023 &0.0003  &0.0000  &0.0014  &0.0019  &0.0002  &0.0004  &0.0011  &0.0019 \tabularnewline
 & RMSE &0.0362  &0.0201  &0.0187  &0.0397  &0.0869  &0.0169  &0.0213  &0.0357  &0.0805 \tabularnewline
 & CR & 0.9420 & 0.9490 & 0.9470 & 0.9600 &0.9450  &0.9430  & 0.9520 & 0.9470 &0.9530 \tabularnewline
\hline 
\multirow{3}{*}{$n=100000$} & Bias & 0.0001 &0.0001  &0.0000  &0.0004  &0.0003  &0.0003  &0.0001  &0.0005  &0.0011 \tabularnewline
 & RMSE & 0.0245 &0.0138  &0.0135  &0.0273  &0.0588  &0.0115  &0.0148  &0.0248  &0.0559 \tabularnewline
 & CR & 0.9490 & 0.9470 & 0.9490 & 0.9470 & 0.9600 & 0.9540 & 0.9580 &0.9530 & 0.9650\tabularnewline
\hline 
 & \multicolumn{10}{c}{$u_{i}\sim\chi^{2}\left(3\right)$}\tabularnewline
\hline 
\multirow{3}{*}{$n=50000$} & Bias & 0.0018 & 0.0015 & 0.0005 & 0.0008 & 0.0033 & 0.0001 & 0.0007 & 0.0001 & 0.0038\tabularnewline
 & RMSE & 0.0429 & 0.0246 & 0.0225 & 0.0482 & 0.1076 & 0.0217 & 0.0289 & 0.0458 & 0.1077\tabularnewline
 & CR & 0.9590 & 0.9400 & 0.9490 & 0.9430 & 0.9380 & 0.9520 & 0.9450 & 0.9410 & 0.9420\tabularnewline
\hline 
\multirow{3}{*}{$n=100000$} & Bias & 0.0001 & 0.0000 & 0.0002 & 0.0008 & 0.0020 & 0.0002 & 0.0001 & 0.0004 & 0.0002\tabularnewline
 & RMSE & 0.0301 & 0.0163 & 0.0159 & 0.0322 & 0.0718 & 0.0149 & 0.0197 & 0.0300 & 0.0707\tabularnewline
 & CR & 0.9480 & 0.9540 & 0.9550 & 0.9490 & 0.9550 & 0.9620 & 0.9520 & 0.9650 & 0.9550\tabularnewline
\hline 
 & \multicolumn{10}{c}{$u_{i}\sim \mathcal{N}\left(0,1\right)$}\tabularnewline
\hline 
\multirow{3}{*}{$n=50000$} & Bias & 0.0006 &0.0001  &0.0001  &0.0004  &0.0007  &0.0004  &0.0005  &0.0006  &0.0021 \tabularnewline
 & RMSE & 0.0315 & 0.0166 &0.0167  &0.0347  &0.0762  &0.0145  &0.0182  &0.0306  &0.0712 \tabularnewline
 & CR & 0.9500 &0.9580  &0.9570  &0.9540  &0.9500  &0.9480  &0.9590  &0.9470  &0.9420 \tabularnewline
\hline 
\multirow{3}{*}{$n=100000$} & Bias & 0.0001 & 0.0003 & 0.0008 & 0.0012 &0.0007  &0.0002  &0.0002  &0.0000  &0.0000 \tabularnewline
 & RMSE & 0.0214 & 0.0120 & 0.0119 & 0.0247 & 0.0534 & 0.0104 & 0.0134 & 0.0219 & 0.0506\tabularnewline
 & CR & 0.9510 &0.9590  &0.9430  &0.9480  &0.9540  &0.9510  &  0.9410& 0.9560 &0.9590 \tabularnewline
\hline 
\hline 
 &  &  &  &  &  &  &  &  &  & \tabularnewline
\end{tabular}
\end{table}

 \subsection{Finite-Sample Performance} \label{section5.1}
In this subsection, I conduct some Monte Carlo experiments to study
the finite sample performance of our AKMBGD estimator.  I consider three setups of sample sizes: $n= 25000$, $n = 50000$, and $n = 100000$. 
    I report the bias, root mean squared error (RMSE), and coverage rate
of AKMBGD estimators for $\beta_{1}^{\star}$ to $\beta_{9}^{\star}$.
Suppose that the simulation is repeated $R$ times, in the $r$-th
round the estimator of $\beta_{j}^{\star}$ is denoted as $\widehat{\beta}_{j}^{r}$.
Then the bias and RMSE of $\beta_{j}^{\star}$ is defined by 
\[
\text{Bias}=\left|\frac{1}{R}\sum_{r=1}^{R}\widehat{\beta}_{j}^{r}-\beta_{j}^{\star}\right|,\ \ 
\text{RMSE}=\sqrt{\frac{1}{R}\sum_{r=1}^{R}\left(\widehat{\beta}_{j}^{r}-\beta_{j}^{\star}\right)^{2}}.
\]
I consider nominal coverage rate $0.95$, so the actual coverage
rate is given by 
\[
\mathrm{CR}=\frac{1}{R}\sum_{r=1}^{R}\mathds{1}\left(\widehat{\beta}_{j}^{r}-1.96\widehat{\sigma}_{j}^{r}\leq\beta_{j}^{\star}\leq\widehat{\beta}_{j}^{r}+1.96\widehat{\sigma}_{j}^{r}\right),
\]
where $\widehat{\sigma}_{j}^{r}$ is the subsample-based estimator of the variance of $\widehat{\beta}_{j}^{r}$.

The learning rate is chosen as $\gamma_k = 1$ for all $k$. The bandwidth used in the $k$-th round of update is $h_n = c_k\cdot h_n^{-1/10}$, where $c_k = \text{std}\left(z_{i,k}\right)$ and $z_{i,k} = X_{0,i}+\mathbf{X}_i^{\mathrm{T}}\boldsymbol{\beta}_k$. The initial guess is chosen as the Logit estimator. When constructing the AKMBGD estimator, I first run 2000 burn-in updates. Then the stopping rule is chosen as that in \autoref{rem7} with $T = 10000$, $gap = 1000$, and $\varrho  = 0.001$. The subsample size $B$ is chosen as 3000 for both estimation and inference. Finally, when conducting inference, i randomly draw 200 subsamples to construct the variance estimator. 

The simulation results are reported in \autoref{simutab1}. It can be seen that the AKMBGD estimators have small bias, whose RMSE decreases with sample size almost at rate $\sqrt{n}$.
Moreover, the confidence interval constructed based on the subsample-based
variance  has actual coverage rate that is
quite close to the nominal rate $0.95$. This demonstrates that the AKMBGD estimators and subsample-based variance estimator have great finite-sample performance.

\begin{table}[t]
\begin{centering}
\caption{\label{simutab2}Comparing Updating Speed}

\begin{tabular}{ll>{\raggedright}p{2.5cm}c>{\raggedright}p{1.5cm}>{\raggedright}p{1.5cm}>{\raggedright}p{1.5cm}}
 &  &  &  & \multicolumn{1}{c}{} &  & \multicolumn{1}{c}{}\tabularnewline
\hline 
\hline 
Sample Size &  & Method &  & KBGD & SBGD & KMBGD\tabularnewline
\cline{1-1} \cline{3-3} \cline{5-7} \cline{6-7} \cline{7-7} 
\multirow{2}{*}{$n=2500$} &  & Unparalleled &  & 0.0475 & 0.0003 & 0.0081\tabularnewline
 &  & Parallel &  & 0.0412 & -- & 0.0321\tabularnewline
\multirow{2}{*}{$n=5000$} &  & Unparalleled &  & 0.2009 & 0.0004 & 0.0078\tabularnewline
 &  & Parallel &  & 0.0669 & -- & 0.0292\tabularnewline
\multirow{2}{*}{$n=10000$} &  & Unparalleled &  & 0.8335 & 0.0006 & 0.0078\tabularnewline
 &  & Parallel &  & 0.1822 & -- & 0.0302\tabularnewline
\multirow{2}{*}{$n=20000$} &  & Unparalleled &  & 3.2828 & 0.0027 & 0.0075\tabularnewline
 &  & Parallel &  & 0.6166 & -- & 0.0293\tabularnewline
\multirow{2}{*}{$n=500000$} &  & Unparalleled &  & -- & 0.1267 & 0.0508\tabularnewline
 &  & Parallel &  & -- & -- & 0.0374\tabularnewline
\multirow{2}{*}{$n=1000000$} &  & Unparalleled &  & -- & 0.2602 & 0.1530\tabularnewline
 &  & Parallel &  & -- & -- & 0.0574\tabularnewline
\hline 
\hline 
 &  &  &  &  &  & \tabularnewline
\end{tabular}
\par\end{centering}
{\footnotesize{Note: All running time in seconds. Parallel computation is conducted over 6 cores. $B = 1000$ when $n\leq 20000$, $B=3000$ when $n = 500000$, and $B = 5000$ when $n = 1000000$. }}{\footnotesize\par}
\end{table}

\begin{table}[t]
\begin{centering}
\caption{\label{simutab3}Comparing KMBGD and SBGD Estimators}
\begin{tabular}{ll>{\raggedright}p{1.5cm}c>{\raggedright}p{1.2cm}c>{\raggedright}p{1.2cm}>{\raggedright}p{1.2cm}}
 &  &  &  & \multicolumn{1}{c}{} &  & \multicolumn{2}{c}{}\tabularnewline
\hline 
\hline 
Distribution & Sample Size & Method &  & RMSE &  & \multicolumn{2}{c}{Running Time}\tabularnewline
\cline{1-3} \cline{2-3} \cline{3-3} \cline{5-5} \cline{7-8} \cline{8-8} 
\multirow{4}{*}{$u\sim\text{Cauchy}$} & \multirow{2}{*}{$n=500000$} & SBGD &  & 0.0620 &  & 0.8417 & 3.2841\tabularnewline
 &  & KMBGD &  & 0.0628 &  & 0.4719 & 0.1042\tabularnewline
 & \multirow{2}{*}{$n=1000000$} & SBGD &  & 0.0398 &  & 1.7304 & 13.921\tabularnewline
 &  & KMBGD &  & 0.0407 &  & 0.5002 & 0.0968\tabularnewline
\hline 
\multirow{4}{*}{$u\sim t\left(4\right)$} & \multirow{2}{*}{$n=500000$} & SBGD &  & 0.0390 &  & 0.8219 & 3.3434\tabularnewline
 &  & KMBGD &  & 0.0390 &  & 0.3954 & 0.1045\tabularnewline
 & \multirow{2}{*}{$n=1000000$} & SBGD &  & 0.0273 &  & 1.6701 & 13.893\tabularnewline
 &  & KMBGD &  & 0.0276 &  & 0.4158 & 0.4059\tabularnewline
\hline 
\multirow{4}{*}{$u\sim\chi^{2}\left(3\right)$} & \multirow{2}{*}{$n=500000$} & SBGD &  & 0.0475 &  & 0.7016 & 3.3534\tabularnewline
 &  & KMBGD &  & 0.0475 &  & 0.4098 & 0.1047\tabularnewline
 & \multirow{2}{*}{$n=1000000$} & SBGD &  & 0.0319 &  & 1.4244 & 14.196\tabularnewline
 &  & KMBGD &  & 0.0330 &  & 0.3703 & 0.3515\tabularnewline
\hline 
\multirow{4}{*}{$u\sim\mathcal{N}(0,1)$} & \multirow{2}{*}{$n=500000$} & SBGD &  & 0.0341 &  & 0.8261 & 3.3310\tabularnewline
 &  & KMBGD &  & 0.0341 &  & 0.3930 & 0.1056\tabularnewline
 & \multirow{2}{*}{$n=1000000$} & SBGD &  & 0.0216 &  & 1.6498 & 14.134\tabularnewline
 &  & KMBGD &  & 0.0218 & &0.3500   & 0.3542\tabularnewline
\hline 
\hline 
 &  &  &  &  &  &  & \tabularnewline
\end{tabular}
\par\end{centering}
{\footnotesize{}NOTE: All running time in hours.}{\footnotesize\par}
\end{table}

\begin{table}[t]
\centering{}\caption{\label{simutab4}Comparing True and Estimated Variance}
\begin{tabular}{ll>{\centering}p{1cm}>{\centering}p{1cm}>{\centering}p{1cm}>{\centering}p{1cm}>{\centering}p{1cm}>{\centering}p{1cm}>{\centering}p{1cm}>{\centering}p{1cm}>{\centering}p{1cm}}
 & \multicolumn{10}{c}{}\tabularnewline
\hline 
\hline 
 & \multicolumn{10}{c}{$u_{i}\sim\mathrm{Cauchy}$}\tabularnewline
\hline 
 &  & $\beta_{1}$ & $\beta_{2}$ & $\beta_{3}$ & $\beta_{4}$ & $\beta_{5}$ & $\beta_{6}$ & $\beta_{7}$ & $\beta_{8}$ & $\beta_{9}$\tabularnewline
\hline 
\multirow{2}{*}{$n=500000$} & True Std & 0.0173 & 0.0102 & 0.0099 & 0.0193 & 0.0442 & 0.0079 & 0.0105 & 0.0159 & 0.0411\tabularnewline
 & Est Std & 0.0173 & 0.0099 & 0.0097 & 0.0203 & 0.0444 & 0.0082 & 0.0107 & 0.0177 & 0.0409\tabularnewline
\hline 
\multirow{2}{*}{$n=1000000$} & True Std & 0.0114 & 0.0064 & 0.0069 & 0.0142 & 0.0260 & 0.0057 & 0.0083 & 0.0115 & 0.0264\tabularnewline
 & Est Std & 0.0123 & 0.0070 & 0.0068 & 0.0143 & 0.0313 & 0.0058 & 0.0075 & 0.0124 & 0.0287\tabularnewline
\hline 
 & \multicolumn{10}{c}{$u_{i}\sim t(4)$}\tabularnewline
\hline 
\multirow{2}{*}{$n=500000$} & True Std & 0.0118 & 0.0059 & 0.0063 & 0.0126 & 0.0280 & 0.0052 & 0.0074 & 0.0113 & 0.0261\tabularnewline
 & Est Std & 0.0110 & 0.0062 & 0.0060 & 0.0124 & 0.0275 & 0.0053 & 0.0068 & 0.0111 & 0.0257\tabularnewline
\hline 
\multirow{2}{*}{$n=1000000$} & True Std & 0.0071 & 0.0045 & 0.0040 & 0.0084 & 0.0196 & 0.0041 & 0.0047 & 0.0077 & 0.0180\tabularnewline
 & Est Std & 0.0078 & 0.0044 & 0.0043 & 0.0088 & 0.0194 & 0.0037 & 0.0048 & 0.0079 & 0.0182\tabularnewline
\hline 
 & \multicolumn{10}{c}{$u_{i}\sim\chi^{2}\left(3\right)$}\tabularnewline
\hline 
\multirow{2}{*}{$n=500000$} & True Std & 0.0120 & 0.0074 & 0.0066 & 0.0149 & 0.0316 & 0.0067 & 0.0093 & 0.0137 & 0.0321\tabularnewline
 & Est Std & 0.0135 & 0.0076 & 0.0071 & 0.0148 & 0.0332 & 0.0068 & 0.0089 & 0.0143 & 0.0325\tabularnewline
\hline 
\multirow{2}{*}{$n=1000000$} & True Std & 0.0092 & 0.0045 & 0.0047 & 0.0107 & 0.0226 & 0.0049 & 0.0061 & 0.0096 & 0.0214\tabularnewline
 & Est Std & 0.0096 & 0.0053 & 0.0051 & 0.0105 & 0.0235 & 0.0048 & 0.0063 & 0.0101 & 0.0230\tabularnewline
\hline 
 & \multicolumn{10}{c}{$u_{i}\sim\mathcal{N}\left(0,1\right)$}\tabularnewline
\hline 
\multirow{2}{*}{$n=500000$} & True Std & 0.0099 & 0.0053 & 0.0049 & 0.0113 & 0.0246 & 0.0048 & 0.0059 & 0.0098 & 0.0225\tabularnewline
 & Est Std & 0.0096 & 0.0054 & 0.0053 & 0.0109 & 0.0240 & 0.0046 & 0.0060 & 0.0097 & 0.0225\tabularnewline
\hline 
\multirow{2}{*}{$n=1000000$} & True Std & 0.0068 &0.0038  &0.0035  &0.0072  &0.0146  &0.0036  &0.0040  &0.0061  &0.0139 \tabularnewline
 & Est Std & 0.0068 &0.0038  &0.0037  &0.0077  &0.0170  &0.0033  &0.0042  &0.0069  & 0.0159\tabularnewline
\hline 
\hline 
 &  &  &  &  &  &  &  &  &  & \tabularnewline
\end{tabular}
\end{table}

\subsection{Computational Efficiency}\label{section5.2}

This subsection formally compares the computational efficiency of several gradient-based estimators for semiparametric montone index models. In particular, I compare  KMBGD estimator with the KBGD and SBGD estimators proposed by \citet{khan2022}. 

I first compare  the updating speed of each algorithm under different setups of sample sizes.  In particular, for each algorithm, I keep updating  100 times and  report the average running time of each single update. For kernel-based updates (KBGD and KMBGD), I consider two computation strategies: unparalleled and parallel computation. When using parallel computation,  kernel estimators are simultaneously calculated over 6 cores. I consider six   sample sizes: $n=2500, 5000, 10000, 20000, 500000,$ and $1000000$. For SBGD estimation, the sieve functions follow those used in \citet{khan2022}.  The order of sieves is chosen as $q=9$ when $n=2500$ and $5000$, $q=11$ when $n=10000$ and $20000$, and $q=31$ when $n=500000$ and $1000000$. The subsample size $B$ is chosen as $B=1000$ when $n\leq 20000$,  $B=3000$ for $n=500000$, and $B = 5000$ for $n=1000000$. The simulation results are reported in \autoref{simutab2}.

It can be seen that without parallel computation, the updating time of full-sample-based KBGD algorithm increases roughly at rate $n^2$, which is in linear with the previous discussion. In particular, when sample size is 2500, each single update requires 0.0475 seconds, which amounts to 21 updates within one second. However, such updating time increases to 0.2 seconds when sample size is 5000, which amounts to only 5 updates each second. When the sample size is 20000, without parallel computation, each single update of KBGD requires more than 3 seconds, indicating that 1000 updates may cost around 1 hour of computational time. For extremely large sample sizes $n=500000$ or $1000000$, KBGD is practically infeasible, so the computational time is not reported. It can also be seen that parallel computation may significantly decrease the updating time when $n$ is large ($n=10000, 20000$), but the updating time is still too long to be practically feasible.

I then look at the updating speed of SBGD and KMBGD. Apparently, when sample size is small or modest, SBGD exhibits excellent performance: when sample size is 2500, 5000, and 10000, each single update of SBGD requires only 0.0003, 0.0004, and 0.0006 seconds, which amounts to 3300, 2500, and 1600 updates within one second. Even when sample size is 20000, each update of SBGD requires only 0.0027 seconds, so 370 updates can be conducted within one second. This suggests that SBGD significantly outperforms KMBGD when the sample size $n$ is small or modest. However, when the sample size $n$ is extremely large, KMBGD starts dominating SBGD. In particular, when $n=500000$ and 1000000, the updating speed of KMBGD (with parallel computation) is roughly 4 and 5 times faster than that of SBGD. 

Of course, the reduction of computational time of each single update of KMBGD compared with that of SBGD may come at the cost of longer total running time or large estimation error. To study whether it is the case, I then compare the total running time of SBGD and KMBGD. I also consider four setups of random error distributions as I did in \autoref{section5.1}.  I consider two extreme sample sizes: $n=500000$ and $n=1000000$. The subsample size $B = 3000$ when $n=500000$ and $B = 5000$ when $n=1000000$. The stopping rule for SBGD is $\max_{1\leq j\leq 9}|\beta_{j,k+1} - \beta_{j,k}|<10^{-6}$ and that for KMBGD is the same as before. For both updates, the initial guess is located at Logit estimator, and the maximum number of updates is 20000. For inference, I choose subsample size $B = 3000$ when $n=500000$ 
 and $B = 6000$ when $n=1000000$. The number of subsamples is chosen as 200.  Finally, I note here that for both estimation and inference, unparalleled computation is considered. 

I report the RMSE and running time of both estimation and inference in \autoref{simutab3}.  As can be seen from the table, for all combinations of error distributions and sample sizes, the RMSE of SBGD and KMBGD are almost identical, indicating that updates based on subsamples do not result in loss of estimation accuracy. When looking at the running time, it's impressive to see that, the estimation time of KMBGD is substantially shorter compared with that of SBGD. When $n=500000$, KMBGD decreases the running time by roughly half, while when $n$ increases to 1000000, the reduction of estimation time is more significant: running time of KMBGD is only around one forth of that of SBGD. It is also interesting to see that, when the sample size increases and I use a larger subsample size, the running time of KMBGD even slightly decreases. This implies that although using a larger subsample size may make updating speed slightly slower, it makes convergence faster because the amount of noises in the update is decreased. 

I finally look at the computational burden of inference based on different methods. As can be seen from \autoref{simutab3}, the operational time of variance calculation of SBGD is over 3.2 hours without parallel computation when $n=500000$, and it rises to around 14 hours when  $n=1000000$. This implies that even SBGD may have adequate computational efficiency in terms of estimation,  it may still cost a large amount of time to conduct inference. When turning to the subsample-based infernece under KMBGD, it can be clearly seen that variance estimation only requires around 0.1 hours (10 min) when $n=500000$ and 0.4 hours (40 min) when $n=1000000$, which significantly improves the speed of inference. I also report in \autoref{simutab4} the true standard deviation and subsample-based estimator of the standard deviation of each estimator, which are close to each other. This implies that subsample-based inference improves the speed while does not suffer from much accuracy loss.

\section{Empirical Illustration}\label{section6}

In this section, I will illustrate the empirical applicability of the new subsample-based learning method by revisiting some   empirical results in  \citet{helpman2008estimating}. In their paper, \citet{helpman2008estimating} consider estimating the following model,
\begin{align}
 \text{Pr}\left(\left.T_{ij}=1\right|\text{observed variables}\right) = G\left(\gamma_0^{\star} + \xi_j^{\star} + \zeta_i^{\star} + \gamma^{\star}d_{ij} + \kappa^{\star \mathrm{T}}\phi_{ij}
\right),\label{empirical_model}
\end{align}
where $T_{ij}$ is an indicator of whether  country $j$ exports to country $i$, $\xi_j^{\star}$ is the exporter fixed effect of the $j$-th country, $\zeta_i^{\star}$ is the importer fixed effect of the $i$-th country, $d_{ij}$ is the natural logarithm  of the geographic distance between countries $i$ and $j$, and $\phi_{ij}$  is a vector of covariates that describe the variable country-pair fixed trade cost. The full sample  contains a total of 248060 observations and 338 covariates, which features both large  $n$ and $p$. The covariates contain  12 key variables including Distance, Land Border, Island, Landlock, Legal, Language, Colonial Ties, Currency Union, FTA, Religion, WTO (none) and WTO (both), and 158  exporter fixed effects, 158 importer fixed effects, and 10 year fixed effects. 

When estimating (\ref{empirical_model}) based on the full sample, \citet{helpman2008estimating} consider a parametric Probit setup, where $G$ is specified to be the CDF of standard normal distribution. In this section, I reestimate model (\ref{empirical_model}) without assuming the functional form of $G$ by applying the KMBGD algorithm. Such reestimation is well motivated because assuming normal distributed random shocks actually makes restrictive assumptions over the decreasing speed of the tails of the random shocks, which might be violated in some  empirical applications. Misspecification of distribution of random shocks may dampen the estimation results as well as the subsequent inference, as we will see in the following analysis. 

When conducting  KMBGD estimation, I need to choose one covariate and normalize its coefficient to 1. To improve the numerical performance of the method, I choose to normalize the coefficient of the continuous variable  Distance. According to \citet{khan2022}, the  covariate whose coefficient is normalized must have positive impacts on the conditional probability.  Since a larger geographic distance is generally associated with higher trading costs, the covariate Distance has negative impacts on the conditional probability of the presence of trades between two countries\footnote{When I apply Logit or Probit to model (\ref{empirical_model}), the estimated coefficient of Distance is significantly negative.}. In this case, I use the negative value of (logarithm of) Distance instead of the original variable when performing iteration. So any covarite whose coefficient is estimated to be positive can be explained to have positive impacts on the conditional probability. 

When estimating the model,  I leave out as few fixed effects as possible to ensure that my covariate matrix is nonsingular. When conducting iteration for KMBGD, I  choose learning rate $\delta_k = 1$ for all $k$ and subsample size $B = 1000$. When constructing kernel estimator, I choose sixth-order Epanechnikov kernel function, and the  bandwidth $h_n$ is chosen as $h_n = c_k\cdot h_n^{-1/10}$, where $c_k = \text{std}\left(z_{i,k}\right)$ and $z_{i,k} = X_{0,i}+\mathbf{X}_i^{\mathrm{T}}\boldsymbol{\beta}_k$. The initial guess of the parameter is fixed at the Probit estimator. I update the estimator  500000 times and use the last 50000 updated estimators to construct the AKMBGD estimator.

Apart from KMBGD estimator, I also consider the full-sample-based SBGD estimator prposed in KLTY. To construct such estimator, I choose learning rate $\delta_k=1$ for all $k$ and the order of sieves $q=25$. The basis functions  are the same as in KLTY. The initial guess is also fixed at the Probit estimator. The stopping rule is $\max_{1\leq j\leq p} \left|\beta_{j,k+1} - \beta_{j,k}\right|<10^{-6}$, where $\beta_{j,k+1}$ is the $j$-argument of $\boldsymbol{\beta}_k$ or the number of updates exceeds 500000.   To further provide some comparisons between parametric and semiparametric estimation, I also consider parametric estimation based on Logit and Probit regression.

\begin{table}
\caption{\label{emp1}Estimation Results}

\begin{tabular}{>{\raggedright}p{2.2cm}>{\raggedright}p{1.4cm}>{\raggedright}p{1.4cm}>{\raggedright}p{1.4cm}>{\raggedright}p{1.4cm}>{\raggedright}p{1.4cm}>{\raggedright}p{1.4cm}>{\raggedright}p{1.4cm}>{\raggedright}p{1.4cm}}
 &  &  &  &  &  &  &  & \tabularnewline
\hline 
\hline 
 & {\small{}Probit} & {\small{}Logit} & {\small{}KMBGD} & {\small{}SBGD} & {\small{}Probit} & {\small{}Logit} & {\small{}KMBGD} & {\small{}SBGD}\tabularnewline
\hline 
{\small{}Border} & $-0.602^{***}$

$(0.047)$ & $-0.626^{***}$

$(0.044)$ & $-0.634^{***}$

$(0.042)$ & $-0.630^{***}$

$(0.043)$ & $-0.603^{***}$

$(0.047)$ & $-0.627^{***}$

$(0.044)$ & $-0.635^{***}$

$(0.043)$ & $-0.631^{***}$

$(0.042)$\tabularnewline
{\small{}Island} & $3.600^{***}$

$(0.100)$ & $3.400^{***}$

$(0.097)$ & $3.395^{***}$

$(0.107)$ & $3.461^{***}$

$(0.010)$ & $3.296^{***}$

$(0.108)$ & $3.120^{***}$

$(0.104)$ & $3.120^{***}$

$(0.143)$ & $3.182^{***}$

$(0.106)$\tabularnewline
{\small{}Landlock} & $4.942^{***}$

$(0.134)$ & $4.731^{***}$

$(0.138)$ & $4.754^{***}$

$(0.155)$ & $4.847^{***}$

$(0.149)$ & $4.642^{***}$

$(0.140)$ & $4.455^{***}$

$(0.144)$ & $4.480^{***}$

$(0.161)$ & $4.566^{***}$

$(0.157)$\tabularnewline
Legal & $0.120^{***}$

$(0.014)$ & $0.127^{***}$

$(0.014)$ & $0.132^{***}$

$(0.014)$ & $0.133^{***}$

$(0.014)$ & $0.118^{***}$

$(0.014)$ & $0.126^{***}$

$(0.014)$ & $0.131^{***}$

$(0.014)$ & $0.132^{***}$

$(0.014)$\tabularnewline
{\small{}Language} & $0.457^{***}$

$(0.018)$ & $0.426^{***}$

$(0.018)$ & $0.423^{***}$

$(0.019)$ & $0.422^{***}$

$(0.019)$ & $0.454^{***}$

$(0.019)$ & $0.423^{***}$

$(0.018)$ & $0421^{***}$

$(0.020)$ & $0.419^{***}$

$(0.019)$\tabularnewline
{\small{}Colonial} & $0.490^{***}$

$(0.133)$ & $0.453^{***}$

$(0.134)$ & $0.467^{***}$

$(0.141)$ & $0.478^{***}$

$(0.134)$ & $0.497^{***}$

$(0.133)$ & $0.458^{***}$

$(0.134)$ & $0.471^{***}$

$(0.144)$ & $0.483^{***}$

$(0.133)$\tabularnewline
{\small{}Currency} & $0.845^{***}$

$(0.062)$ & $0.799^{***}$

$(0.059)$ & $0.799^{***}$

$(0.060)$ & $0.810^{***}$

$(0.059)$ & $0.846^{***}$

$(0.062)$ & $0.801^{***}$

$(0.059)$ & $0.802^{***}$

$(0.062)$ & $0.809^{***}$

$(0.059)$\tabularnewline
{\small{}FTA} & $3.039^{***}$

$(0.155)$ & $2.908^{***}$

$(0.154)$ & $2.930^{***}$

$(0.152)$ & $2.925^{***}$

$(0.156)$ & $3.017^{***}$

$(0.155)$ & $2.893^{***}$

$(0.154)$ & $2.919^{***}$

$(0.158)$ & $2.910^{***}$

$(0.157)$\tabularnewline
{\small{}Religion} & $0.391^{***}$

$(0.030)$ & $0.342^{***}$

$(0.028)$ & $0.334^{***}$

$(0.029)$ & $0.336^{***}$

$(0.029)$ & $0.385^{***}$

$(0.030)$ & $0.336^{***}$

$(0.028)$ & $0.330^{***}$

$(0.030)$ & $0.330^{***}$

$(0.029)$\tabularnewline
{\small{}WTO (none)} &  &  &  &  & $-0.229^{***}$

$(0.043)$ & $-0.219^{***}$

$(0.041)$ & $-0.222^{***}$

$(0.047)$ & $-0.210^{***}$

$(0.041)$\tabularnewline
{\small{}WTO (both)} &  &  &  &  & $0.376^{***}$

$(0.043)$ & $0.337^{***}$

$(0.041)$ & $0.334^{***}$

$(0.046)$ & $0.322^{***}$

$(0.041)$\tabularnewline
Running Time (Estimation) & $0.021$ & $0.018$ & $5.026$ & $8.798$ & $0.025$ & $0.019$ & $5.002$ & $8.360$\tabularnewline
Running Time (Variance) &  &  &  $0.788$& $2.302$ &  &  & $0.783$ & $2.266$\tabularnewline
\hline 
\hline  
 \tabularnewline
\end{tabular}
\small{Note: Probit and Logit estimation are conducted using MATLAB's code \texttt{fitglm.m}. For Probit and Logit estimation, running time of estimation includes the time of both parameter and covariance matrix estimation. All of the running time are in hours. $^{***}$ indicates significance at 1\%.}  
\end{table}

\begin{figure}
	\caption{\label{emp}Estimation Results under Different Methods}
	
	\includegraphics[bb=50bp 20bp 700bp 510bp,clip,scale=0.95]{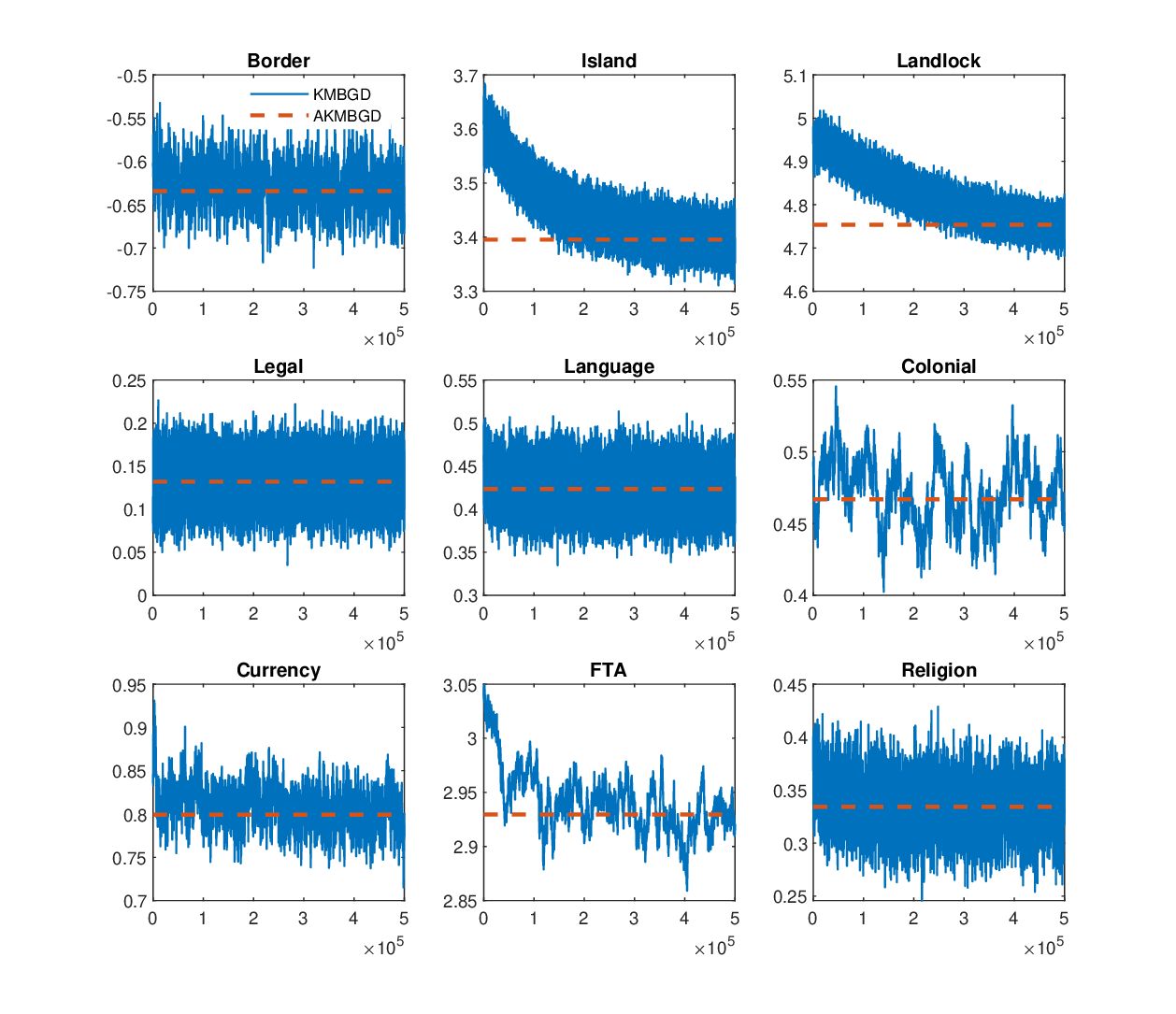}
 Note: This figure displays the estimation results without covariates WTO (both) and WTO (none).  X-axis is the number of iterations. 
\end{figure}

The estimation results are reported in \autoref{emp1}. I first compare the computational time of different methods. Obviously, parametric Probit and Logit estimation  feature fast computation, which both  take around 1 minute. On the other side,  the semiparametric estimation based on KMBGD and SBGD take 8.0--9.0 hours, which are all computationally feasible. Comparitively, the subsample-based KMBGD is slightly  faster in terms of estimation, and significantly outpeforms the SBGD method in terms of the operation time of inference. 

Next I compare the estimation results  of different estimation methods. I find that, first of all, the Logit estimator differs significantly from the Probit estimator for some coefficients. For example, the estimated coefficient of Island using Probit is 3.600 with standard deviation 0.100. So under Probit estimation, the 0.95 confidence interval for the coefficient of Island is $[3.404, 3.796]$,  which does not include the Logit estimator $3.400$. This implies that if the random shock in the binary choice model actually has a Logistic distribution instead of standard normal distribution, then there is a high probability ($\geq 50\%$) that the confidence interval based on Probit does not include the unknown true parameter. Indeed, the semiparametric estimation results strongly favor such possibility. In particular, it can be seen that the KMBGD estimator is quite close to the Logit estimator. For example, for the coefficient of Island, the Logit estimator is $3.400$ and the KMBGD estimator is $3.395$, which almost coincide with each other. Similar patterns can also be seen from the estimation results of other coefficients.  I further compare the  SBGD estimator with both Probit and Logit estimators. I also find that comparatively, the SBGD estimator is   closer to the  Logit estimator. The above  result highlights the potential of model misspecification of Probit estimation and motivates the use the semiparametric estimation.   

I finally investigate convergence of KMBGD estimator. I plot the KMBGD estimation results (without WTO (both) and WTO (none)) of the first 9 covariates produced during 500000 iterations in \autoref{emp}. It can be seen that different coefficients exhibit different converging behaviors. For example, for the coefficient of FTA, although the starting point of iteration (which is Probit estimator) deviates a lot from the final estimator, it converges very quickly and starts fluctuating around the AKMBGD estimator after roughly 100000 rounds of updates. While comparitively, the estimators of the coefficients of Island and Landlock converge slowly, which start fluctuating around the final estimators after roughly 300000 and 400000 rounds of updates, respectively.

\section{\label{section7}Concluding Remarks}

This paper investigates semiparametric estimation of monotone index models in a large-$n$ environment, where the    number of   observations is extremely large. I propose a novel subsample- and iteration-based  estimation
procedure.  Essentially, starting from an initial guess of the parameter, in each round of iteration  a subsample is randomly drawn and then used   to update the parameter based on the gradient of some well-chosen loss function, where the unknown  nonparametric component is replaced with its subsample-based kernel estimator.  The proposed algorithm essentially  generalizes the idea of  mini-batch-based algorithms to the
semiparametric setup. Compared with the KBGD algorithm proposed in KLTY, the computational speed of the new estimator substantially improves, so can be easily applied when the sample size $n$ is extremely large.  I also show that  further averaging across the estimators produced during iterations yields a $1/\sqrt{n}$ consistent and asymptotically normally distributed estimator.  

As an empirical application of the new method, I revisit the Probit estimation of the  presence of trade between countries in \citet{helpman2008estimating}. Given the large sample size and number of covariates, the computational time of estimation and inference based on KMBGD algorithm is reasonable. I also find that compared with Probit specification, the semiparametric estimation results are more in favor of the Logistic distributed random shock in the binary choice model, which highlights the use of semiparametric estimation in the empirical applications. 

Some issues in this paper remain to be addressed in the future studies. For example, similar to \citet{ichimura1993semiparametric}, I show that a particular sequence of bandwidth    satisfying some order conditions guarantees all the theorems. However, in the theorem the bandwidth is assumed to be unchanged across iterations. Obviously, as the updates proceed, the magnitude of the index value also changes, so a bandwidth adjusted to such change in index value in each round of iteration may lead to a better kernel estimator and improve the updating results. Similarly, other tuning parameters such as the learning rate $\delta $ and subsample size $B$ are all assumed to be given, while their optimal choices remain to be studied. 

Another potential future  research direction is to generalize the noval subsample-based updating techinque to the full-sample-based SBGD algorithm proposed in KLTY. Different from the kernel-based learning approach, the SBGD algorithm relies on the full sample to update the sieve coefficient in each iteration. So it is still unclear whether using subsamples to perform the update will also yield $1/\sqrt{n}$-consistent estimator. However, since the SBGD algorithm runs significantly faster than the KBGD algorithm, developing subsample-based SBGD algorithm may further improve the computational speed, which deserves further study.

\bibliographystyle{plainnat}
\bibliography{KBGD}

\begin{thebibliography}{28}
\providecommand{\natexlab}[1]{#1}
\providecommand{\url}[1]{\texttt{#1}}
\expandafter\ifx\csname urlstyle\endcsname\relax
  \providecommand{\doi}[1]{doi: #1}\else
  \providecommand{\doi}{doi: \begingroup \urlstyle{rm}\Url}\fi

\bibitem[Agarwal et~al.(2014)Agarwal, Kakade, Karampatziakis, Song, and
  Valiant]{agarwal2014least}
Alekh Agarwal, Sham Kakade, Nikos Karampatziakis, Le~Song, and Gregory Valiant.
\newblock Least squares revisited: Scalable approaches for multi-class
  prediction.
\newblock In \emph{International Conference on Machine Learning}, pages
  541--549. PMLR, 2014.

\bibitem[Ahn et~al.(2018)Ahn, Ichimura, Powell, and Ruud]{ahn2018simple}
Hyungtaik Ahn, Hidehiko Ichimura, James~L Powell, and Paul~A Ruud.
\newblock Simple estimators for invertible index models.
\newblock \emph{Journal of Business \& Economic Statistics}, 36\penalty0
  (1):\penalty0 1--10, 2018.

\bibitem[Bottou et~al.(2018)Bottou, Curtis, and
  Nocedal]{bottou2018optimization}
L{\'e}on Bottou, Frank~E Curtis, and Jorge Nocedal.
\newblock Optimization methods for large-scale machine learning.
\newblock \emph{Siam Review}, 60\penalty0 (2):\penalty0 223--311, 2018.

\bibitem[Cavanagh and Sherman(1998)]{cavanagh1998rank}
Christopher Cavanagh and Robert~P Sherman.
\newblock Rank estimators for monotonic index models.
\newblock \emph{Journal of Econometrics}, 84\penalty0 (2):\penalty0 351--381,
  1998.

\bibitem[Cosslett(1983)]{cosslett1983distribution}
Stephen~R Cosslett.
\newblock Distribution-free maximum likelihood estimator of the binary choice
  model.
\newblock \emph{Econometrica: Journal of the Econometric Society}, pages
  765--782, 1983.

\bibitem[Fan et~al.(2020)Fan, Han, Li, and Zhou]{fan2020rank}
Yanqin Fan, Fang Han, Wei Li, and Xiao-Hua Zhou.
\newblock On rank estimators in increasing dimensions.
\newblock \emph{Journal of Econometrics}, 214\penalty0 (2):\penalty0 379--412,
  2020.

\bibitem[Forneron(2022)]{forneron2022estimation}
Jean-Jacques Forneron.
\newblock Estimation and inference by stochastic optimization.
\newblock \emph{arXiv preprint arXiv:2205.03254}, 2022.

\bibitem[Hall et~al.(2004)Hall, Racine, and Li]{hall2004cross}
Peter Hall, Jeff Racine, and Qi~Li.
\newblock Cross-validation and the estimation of conditional probability
  densities.
\newblock \emph{Journal of the American Statistical Association}, 99\penalty0
  (468):\penalty0 1015--1026, 2004.

\bibitem[Han(1987)]{han1987non}
Aaron~K Han.
\newblock Non-parametric analysis of a generalized regression model: the
  maximum rank correlation estimator.
\newblock \emph{Journal of Econometrics}, 35\penalty0 (2-3):\penalty0 303--316,
  1987.

\bibitem[H{\"a}rdle et~al.(1993)H{\"a}rdle, Hall, and
  Ichimura]{hardle1993optimal}
Wolfgang H{\"a}rdle, Peter Hall, and Hidehiko Ichimura.
\newblock Optimal smoothing in single-index models.
\newblock \emph{The annals of Statistics}, 21\penalty0 (1):\penalty0 157--178,
  1993.

\bibitem[Helpman et~al.(2008)Helpman, Melitz, and
  Rubinstein]{helpman2008estimating}
Elhanan Helpman, Marc Melitz, and Yona Rubinstein.
\newblock Estimating trade flows: Trading partners and trading volumes.
\newblock \emph{The quarterly journal of economics}, 123\penalty0 (2):\penalty0
  441--487, 2008.

\bibitem[Horowitz(1992)]{horowitz1992smoothed}
Joel~L Horowitz.
\newblock A smoothed maximum score estimator for the binary response model.
\newblock \emph{Econometrica: journal of the Econometric Society}, pages
  505--531, 1992.

\bibitem[Horowitz and H{\"a}rdle(1996)]{horowitz1996direct}
Joel~L Horowitz and Wolfgang H{\"a}rdle.
\newblock Direct semiparametric estimation of single-index models with discrete
  covariates.
\newblock \emph{Journal of the American Statistical Association}, 91\penalty0
  (436):\penalty0 1632--1640, 1996.

\bibitem[Hristache et~al.(2001)Hristache, Juditsky, and
  Spokoiny]{hristache2001direct}
Marian Hristache, Anatoli Juditsky, and Vladimir Spokoiny.
\newblock Direct estimation of the index coefficient in a single-index model.
\newblock \emph{Annals of Statistics}, pages 595--623, 2001.

\bibitem[Ichimura(1993)]{ichimura1993semiparametric}
Hidehiko Ichimura.
\newblock Semiparametric least squares (sls) and weighted sls estimation of
  single-index models.
\newblock \emph{Journal of econometrics}, 58\penalty0 (1-2):\penalty0 71--120,
  1993.

\bibitem[Khan et~al.(2023)Khan, Lan, and Tamer]{khan2022}
Shakeeb Khan, Xiaoying Lan, and Elie Tamer.
\newblock Estimating high dimensional monotone index models by iterative convex
  optimization1.
\newblock \emph{arXiv preprint arXiv:2110.04388}, 2023.

\bibitem[Klein and Spady(1993)]{klein1993efficient}
Roger~W Klein and Richard~H Spady.
\newblock An efficient semiparametric estimator for binary response models.
\newblock \emph{Econometrica: Journal of the Econometric Society}, pages
  387--421, 1993.

\bibitem[Lewbel(2000)]{lewbel2000semiparametric}
Arthur Lewbel.
\newblock Semiparametric qualitative response model estimation with unknown
  heteroscedasticity or instrumental variables.
\newblock \emph{Journal of econometrics}, 97\penalty0 (1):\penalty0 145--177,
  2000.

\bibitem[Manski(1975)]{manski1975maximum}
Charles~F Manski.
\newblock Maximum score estimation of the stochastic utility model of choice.
\newblock \emph{Journal of econometrics}, 3\penalty0 (3):\penalty0 205--228,
  1975.

\bibitem[Manski(1985)]{manski1985semiparametric}
Charles~F Manski.
\newblock Semiparametric analysis of discrete response: Asymptotic properties
  of the maximum score estimator.
\newblock \emph{Journal of econometrics}, 27\penalty0 (3):\penalty0 313--333,
  1985.

\bibitem[Murty and Kabadi(1987)]{murty1987some}
Katta~G Murty and Santosh~N Kabadi.
\newblock Some np-complete problems in quadratic and nonlinear programming.
\newblock \emph{Mathematical Programming}, 39:\penalty0 117--129, 1987.

\bibitem[Ouyang and Yang(2023)]{ouyang2023}
Fu~Ouyang and Thomas~Tao Yang.
\newblock High dimensional binary choice model with unknown heteroskedasticity
  or instrumental variables.
\newblock 2023.

\bibitem[Polyak and Juditsky(1992)]{polyak1992acceleration}
Boris~T Polyak and Anatoli~B Juditsky.
\newblock Acceleration of stochastic approximation by averaging.
\newblock \emph{SIAM journal on control and optimization}, 30\penalty0
  (4):\penalty0 838--855, 1992.

\bibitem[Powell et~al.(1989)Powell, Stock, and
  Stoker]{powell1989semiparametric}
James~L Powell, James~H Stock, and Thomas~M Stoker.
\newblock Semiparametric estimation of index coefficients.
\newblock \emph{Econometrica: Journal of the Econometric Society}, pages
  1403--1430, 1989.

\bibitem[Ruder(2016)]{ruder2016overview}
Sebastian Ruder.
\newblock An overview of gradient descent optimization algorithms.
\newblock \emph{arXiv preprint arXiv:1609.04747}, 2016.

\bibitem[Sherman(1993)]{sherman1993limiting}
Robert~P Sherman.
\newblock The limiting distribution of the maximum rank correlation estimator.
\newblock \emph{Econometrica: Journal of the Econometric Society}, pages
  123--137, 1993.

\bibitem[Stoker(1986)]{stoker1986consistent}
Thomas~M Stoker.
\newblock Consistent estimation of scaled coefficients.
\newblock \emph{Econometrica: Journal of the Econometric Society}, pages
  1461--1481, 1986.

\bibitem[Toulis and Airoldi(2017)]{toulis2017asymptotic}
Panos Toulis and Edoardo~M Airoldi.
\newblock Asymptotic and finite-sample properties of estimators based on
  stochastic gradients.
\newblock \emph{The Annals of Statistics}, 45\penalty0 (4):\penalty0
  1694--1727, 2017.

\end{thebibliography}

\newpage

\section*{Appendix}

\begin{lemma}
	\label{thm:3.3}Suppose that  \autoref{assu1}--\autoref{assu:5}  
	hold with $D\geq 4$.
	Suppose moreover that 
	$\delta_{k}=\delta<\min\left\{ 1/\left(2\underline{\lambda}_{\varLambda}\right),1/\left(4p^{2}\left\Vert G^{\prime}\right\Vert _{\infty}\right)\right\} $, $\phi<\delta \underline{\lambda}_{\varLambda}/\left(16p^{2}\left\Vert G^{\prime}\right\Vert _{\infty}\zeta\right)$, 
	$h_{n}$ is chosen such that $h_nn^{1/2D}\rightarrow0$ and $h_{n}n^{1/6}/\log^{1/3} \left(n\right)\rightarrow\infty$. If 
	  $\boldsymbol{\beta}_k$ is updated under (\ref{subsample_NW}) and (\ref{KcMBGD estimator}) with $\mathfrak{I}_{B,k} = {1,\cdots, n}$, then 
	
	(i) There exists some positive integer $k_{KBGD}$ such that 
\[
	\sup_{k\geq k_{KBGD}}\left\Vert \Delta\boldsymbol{\beta}_{k}\right\Vert =O_{p}\left(n^{-1/2}\right);
\]

	(ii) Define 
$
	\boldsymbol{\xi}_{n}^{\phi}=\frac{1}{n}\sum_{i=1}^{n}(\widehat{G}\left(\left.z_{i}^{\star}\right|\boldsymbol{\beta}^{\star}\right)-y_{i})\mathbf{X}_{i}^{\phi}$, where $z_i^{\star} = z\left(\mathbf{X}_{e,i}, \boldsymbol{\beta}^{\star}\right)$. There holds 
	\[
	\Delta\boldsymbol{\beta}_{k+1} = \left(I_p - \delta \varLambda_{\phi}\left(\boldsymbol{\beta}^{\star}\right)\right)\Delta\boldsymbol{\beta}_k - \delta \boldsymbol{\xi}_n^{\phi} + \delta\widetilde{\varOmega}_k^{\phi},
	\]
	where $\sup_{k\geq k_{KBGD}}\Vert \widetilde{\varOmega}_k^{\phi}\Vert = o_p\left(n^{-1/2}\right)$. Define $\widehat{\boldsymbol{\beta}}=\boldsymbol{\beta}_{k}$
	for any $k$ such that $k -k_{KBGD} \rightarrow\infty$. There holds  $\Delta\widehat{\boldsymbol{\beta}} = -\varLambda_{\phi}^{-1}\left(\boldsymbol{\beta}^{\star}\right)\boldsymbol{\xi}_{n}^{\phi} + o_p (n^{-1/2})$, and 
	\[	\sqrt{n}\Delta\widehat{\boldsymbol{\beta}}\rightarrow_d N\left(0,\Sigma_{\boldsymbol{\beta}}^{\phi}\right),
\]
	where $\Sigma_{\boldsymbol{\beta}}^{\phi}=\varLambda_{\phi}^{-1}\left(\boldsymbol{\beta}^{\star}\right)\Sigma_{\boldsymbol{\xi}}^{\phi}\left(\varLambda_{\phi}^{-1}\left(\boldsymbol{\beta}^{\star}\right)\right)^{\mathrm{T}}$  and 
	\[
	\Sigma_{\boldsymbol{\xi}}^{\phi}=\mathbb{E}\left[\left(1-G\left(z_{i}^{\star}\right)\right)G\left(z_{i}^{\star}\right)\left(\mathbf{X}_{i}^{\phi}-\mathbb{E}\left(\left.\mathbf{X}_{i}^{\phi}\right|z_{i}^{\star}\right)\right)\left(\mathbf{X}_{i}^{\phi}-\mathbb{E}\left(\left.\mathbf{X}_{i}^{\phi}\right|z_{i}^{\star}\right)\right)^{\mathrm{T}}\right].
	\]
\end{lemma}
\begin{proof}[Proof of \autoref{thm:3.3}]
	See \citet{khan2022}.
\end{proof}

\subsection*{Proof of \autoref{lem:4.2}}

\begin{proof}
	We start with the proof of the first result. Define $\psi(n, h_n, D) = \sqrt{\log(n)/nh_n} + h_n^D$.  \citet{khan2022} show that 
	\[
	\sup_{z\in\mathcal{Z}^{\phi},\boldsymbol{\beta}\in\mathcal{B}}\left|\widehat{G}\left(\left.z\right|\boldsymbol{\beta}\right)-\mathbb{E}\left(\left.y\right|X_{0}+\mathbf{X}^{T}\boldsymbol{\beta}=z\right)\right|=O_{p}\left(\psi\left(n,h_{n},D\right)\right).
	\]
	Define event 
	\[
	e_{1,n}=\left\{ \sup_{z\in\mathcal{Z}^{\phi},\boldsymbol{\beta}\in\mathcal{B}}\left|\widehat{G}\left(\left.z\right|\boldsymbol{\beta}\right)\right|\leq2\right\} ,
	\]
	then $P\left(e_{1,n}\right)\rightarrow1$ since $\psi\left(n,h_{n},D\right)\rightarrow0$ according to the choice of $h_n$.
	Over event $e_{1,n}$, we have that 
	\[
	\mathbb{E}_{k}^{*}\left\Vert \frac{1}{B}\sum_{i\in\mathfrak{I}_{B,k}}\left(\widehat{G}\left(\left.X_{0}+\mathbf{X}_{i}^{\mathrm{T}}\boldsymbol{\beta}_{k}\right|\boldsymbol{\beta}_{k}\right)-y_{i}\right)\mathbf{X}_{i}^{\phi}-\frac{1}{n}\sum_{i=1}^{n}\left(\widehat{G}\left(\left.X_{0}+\mathbf{X}_{i}^{\mathrm{T}}\boldsymbol{\beta}_{k}\right|\boldsymbol{\beta}_{k}\right)-y_{i}\right)\mathbf{X}_{i}^{\phi}\right\Vert \leq\frac{C}{B}.
	\]
	
	Now we prove the second result. Recall that $A_{n,y}\left(z,\boldsymbol{\beta}\right)=\frac{1}{n}\sum_{i=1}^{n}K_{h_{n}}\left(z-X_{0,i}-\mathbf{X}_{i}^{\mathrm{T}}\boldsymbol{\beta}\right)y_{i}$,
	$A_{n,1}\left(z,\boldsymbol{\beta}\right)=\frac{1}{n}\sum_{i=1}^{n}K_{h_{n}}\left(z-X_{0,i}-\mathbf{X}_{i}^{\mathrm{T}}\boldsymbol{\beta}\right)$,
	$A_{n,y}\left(\left.z,\boldsymbol{\beta}\right|\mathfrak{I}_{B,k}\right)=\frac{1}{B}\sum_{i\in\mathfrak{I}_{B,k}}K_{h_{n}}\left(z-X_{0,i}-\mathbf{X}_{i}^{\mathrm{T}}\boldsymbol{\beta}\right)y_{i}$,
	and $A_{n,1}\left(\left.z,\boldsymbol{\beta}\right|\mathfrak{I}_{B,k}\right)=\frac{1}{B}\sum_{i\in\mathfrak{I}_{B,k}}K_{h_{n}}\left(z-X_{0,i}-\mathbf{X}_{i}^{\mathrm{T}}\boldsymbol{\beta}\right)$.
	According to \citet{khan2022}, 
	\[
	\sup_{z\in\mathcal{Z}^{\phi},\boldsymbol{\beta}\in\mathcal{B}}\left|A_{n,1}\left(z,\boldsymbol{\beta}\right)-f_{Z}\left(\left.z\right|\boldsymbol{\beta}\right)\right|=O_{p}\left(\psi\left(n,h_{n},D\right)\right).
	\]
	Note that $\inf_{z\in\mathcal{Z}^{\phi},\boldsymbol{\beta}\in\mathcal{B}}f_{Z}\left(\left.z\right|\boldsymbol{\beta}\right)\geq3\underline{c}_{f}$
	and $\sup_{z\in\mathcal{Z}^{\phi},\boldsymbol{\beta}\in\mathcal{B}}f_{Z}\left(\left.z\right|\boldsymbol{\beta}\right)\leq\overline{c}_{f}$, where $\overline{c}_{f}$ is some sufficiently large positive constant, 
	define event 
	\[
	e_{2,n}=\left\{ 2\underline{c}_{f}\leq\inf_{z\in\mathcal{Z}^{\phi},\boldsymbol{\beta}\in\mathcal{B}}A_{n,1}\left(z,\boldsymbol{\beta}\right)\leq\sup_{z\in\mathcal{Z}^{\phi},\boldsymbol{\beta}\in\mathcal{B}}A_{n,1}\left(z,\boldsymbol{\beta}\right)\leq2\overline{c}_{f}\right\} .
	\]
	Since $\psi\left(n,h_{n},D\right)\rightarrow0$, we have that $P\left(e_{2,n}\right)\rightarrow1$.
	Moreover, $P\left(e_{1,n}\cap e_{2,n}\right)\rightarrow1$ and over
	$e_{1,n}\cap e_{2,n}$, we have that 
	\begin{align*}
		\sup_{z\in\mathcal{Z}^{\phi},\boldsymbol{\beta}\in\mathcal{B}}\left|A_{n,y}\left(z,\boldsymbol{\beta}\right)\right| & \leq\sup_{z\in\mathcal{Z}^{\phi},\boldsymbol{\beta}\in\mathcal{B}}\left|A_{n,1}\left(z,\boldsymbol{\beta}\right)\right|\cdot\sup_{z\in\mathcal{Z}^{\phi},\boldsymbol{\beta}\in\mathcal{B}}\left|\widehat{G}\left(\left.z\right|\boldsymbol{\beta}\right)\right|\\
		& \leq4\overline{c}_{f}.
	\end{align*}
	
	Define 
	\[
	e_{3,n,k}^{\epsilon}=\left\{ \sup_{z\in\mathcal{Z}^{\phi}}\left|A_{n,y}\left(\left.z,\boldsymbol{\beta}_{k}\right|\mathfrak{I}_{B,k}\right)-A_{n,y}\left(z,\boldsymbol{\beta}_{k}\right)\right|<\epsilon\right\} 
	\]
	and 
	\[
	e_{4,n,k}^{\epsilon}=\left\{ \sup_{z\in\mathcal{Z}^{\phi}}\left|A_{n,1}\left(\left.z,\boldsymbol{\beta}_{k}\right|\mathfrak{I}_{B,k}\right)-A_{n,1}\left(z,\boldsymbol{\beta}_{k}\right)\right|<\epsilon\right\} .
	\]
	For $\epsilon=\epsilon\left(\zeta\right)=2\underline{c}_{f}/\zeta$
	with $\zeta>2$, we have that over $e_{1,n}\cap e_{2,n}\cap e_{3,n,t}^{\epsilon}\cap e_{4,n,t}^{\epsilon}$,
	there holds
	\begin{align*}
		& \sup_{z\in\mathcal{Z}^{\phi}}\left|\frac{A_{n,y}\left(\left.z,\boldsymbol{\beta}_{k}\right|\mathfrak{I}_{B,k}\right)}{A_{n,1}\left(\left.z,\boldsymbol{\beta}_{k}\right|\mathfrak{I}_{B,k}\right)}-\frac{A_{n,y}\left(z,\boldsymbol{\beta}_{k}\right)}{A_{n,1}\left(z,\boldsymbol{\beta}_{k}\right)}\right|\\
		& \leq\sup_{z\in\mathcal{Z}^{\phi}}\left|\frac{A_{n,y}\left(\left.z,\boldsymbol{\beta}_{k}\right|\mathfrak{I}_{B,k}\right)-A_{n,y}\left(z,\boldsymbol{\beta}_{k}\right)}{A_{n,1}\left(z,\boldsymbol{\beta}_{k}\right)}\right|+\sup_{z\in\mathcal{Z}^{\phi}}\left|\frac{A_{n,y}\left(\left.z,\boldsymbol{\beta}_{k}\right|\mathfrak{I}_{B,k}\right)\left(A_{n,1}\left(\left.z,\boldsymbol{\beta}_{k}\right|\mathfrak{I}_{B,k}\right)-A_{n,1}\left(z,\boldsymbol{\beta}_{k}\right)\right)}{A_{n,1}\left(\left.z,\boldsymbol{\beta}_{k}\right|\mathfrak{I}_{B,k}\right)A_{n,1}\left(z,\boldsymbol{\beta}_{k}\right)}\right|\\
		& \leq\frac{1}{2\underline{c}_{f}}\sup_{z\in\mathcal{Z}^{\phi}}\left|A_{n,y}\left(\left.z,\boldsymbol{\beta}_{k}\right|\mathfrak{I}_{B,k}\right)-A_{n,y}\left(z,\boldsymbol{\beta}_{k}\right)\right|+\frac{4\overline{c}_{f}+2\underline{c}_{f}/\zeta}{\left(2\underline{c}_{f}\right)\left(2\underline{c}_{f}-2\underline{c}_{f}/\zeta\right)}\sup_{z\in\mathcal{Z}^{\phi}}\left|A_{n,1}\left(\left.z,\boldsymbol{\beta}_{k}\right|\mathfrak{I}_{B,k}\right)-A_{n,1}\left(z,\boldsymbol{\beta}_{k}\right)\right|\\
		& \leq c_{1}\left(\zeta\right)\epsilon,
	\end{align*}
	where
	\[
	c_{1}\left(\zeta\right)=\frac{1}{2\underline{c}_{f}}+\frac{4\overline{c}_{f}\zeta+2\underline{c}_{f}}{4\underline{c}_{f}^{2}\left(\zeta-1\right)}\leq c_{1}^{\infty},
	\]
	and $c_{1}^{\infty}$ is a positive constant depending only on $\overline{c}_{f}$
	and $\underline{c}_{f}$. Moreover, when $\epsilon=\underline{c}_{f}/\zeta$
	is chosen such that $\zeta>2$, there holds $2\underline{c}_{f}/\zeta<\underline{c}_{f}$,
	so over $e_{1,n}\cap e_{2,n}\cap e_{3,n,k}^{\epsilon}\cap e_{4,n,k}^{\epsilon}$,
	there holds $\inf_{z\in\mathcal{Z}^{\phi}}A_{n,1}\left(\left.z,\boldsymbol{\beta}_{k}\right|\mathfrak{I}_{B,k}\right)\geq\underline{c}_{f}$,
	and $\widehat{G}\left(\left.z\right|\boldsymbol{\beta}_{k},\mathfrak{I}_{B,k},\underline{c}_{f}\right)=A_{n,y}\left(\left.z,\boldsymbol{\beta}_{k}\right|\mathfrak{I}_{B,k}\right)/A_{n,1}\left(\left.z,\boldsymbol{\beta}_{k}\right|\mathfrak{I}_{B,k}\right)$. 
	
	Since $\left|K_{h_{n}}\left(z-X_{0,i}-\mathbf{X}_{i}^{\mathrm{T}}\boldsymbol{\beta}_{k}\right)\right|\leq Ch_{n}^{-1}$,
	we have that for any fixed $z$ and $\epsilon$, 
	\[
	\mathbb{P}_{k}^{*}\left(\left|A_{n,1}\left(\left.z,\boldsymbol{\beta}_{k}\right|\mathfrak{I}_{B,k}\right)-A_{n,1}\left(z,\boldsymbol{\beta}_{k}\right)\right|>\epsilon\right)\leq2\exp\left(-CBh_{n}^{2}\epsilon^{2}/2\right),
	\]
	and
	\[
	\mathbb{P}_{k}^{*}\left(\left|A_{n,y}\left(\left.z,\boldsymbol{\beta}_{k}\right|\mathfrak{I}_{B,k}\right)-A_{n,y}\left(z,\boldsymbol{\beta}_{k}\right)\right|>\epsilon\right)\leq2\exp\left(-CBh_{n}^{2}\epsilon^{2}/2\right),
	\]
	Also note that 
	\begin{align*}
		& \sup_{z\in\mathcal{Z}^{\phi}}\left|A_{n,1}\left(\left.z,\boldsymbol{\beta}_{k}\right|\mathfrak{I}_{B,k}\right)-A_{n,1}\left(z,\boldsymbol{\beta}_{k}\right)\right|\\
		& \leq\max_{1\leq s\leq S}\left|A_{n,1}\left(\left.z_{s},\boldsymbol{\beta}_{k}\right|\mathfrak{I}_{B,k}\right)-A_{n,1}\left(z_{s},\boldsymbol{\beta}_{k}\right)\right|+Ch_{n}^{-2}/S,
	\end{align*}
	for any positive integer $S$ and a set of well-chosen points $z_{1},\cdots,z_{S}$
	in $\mathcal{Z}^{\phi}$, where the positive constant $C$ does not depend on $\boldsymbol{\beta}_k$, the index set $\mathfrak{I}_{B,k}$,  $S$, and the choice of $z_1, \cdots, z_S$. Let $S$ be such that $Ch_{n}^{-2}/S<\epsilon$,
	we have that 
	\begin{align}
		& \mathbb{P}_{k}^{*}\left(\sup_{z\in\mathcal{Z}^{\phi}}\left|A_{n,1}\left(\left.z,\boldsymbol{\beta}_{k}\right|\mathfrak{I}_{B,k}\right)-A_{n,1}\left(z,\boldsymbol{\beta}_{k}\right)\right|>\epsilon\right)\nonumber \\
		& \leq\sum_{s=1}^{S}\mathbb{P}_{k}^{*}\left(\left|A_{n,1}\left(\left.z_{s},\boldsymbol{\beta}_{k}\right|\mathfrak{I}_{B,k}\right)-A_{n,1}\left(z_{s},\boldsymbol{\beta}_{k}\right)\right|>\epsilon-Ch_{n}^{-2}/S\right)\nonumber \\
		& \leq2\exp\left(\log S-Bh_{n}^{2}\left(\epsilon-Ch_{n}^{-2}/S\right)^{2}/2\right).\label{concentration_y}
	\end{align}
	Using similar method, we can show that 
	\begin{align}
		& \mathbb{P}_{k}^{*}\left(\sup_{z\in\mathcal{Z}^{\phi}}\left|A_{n,y}\left(\left.z,\boldsymbol{\beta}_{k}\right|\mathfrak{I}_{B,k}\right)-A_{n,y}\left(z,\boldsymbol{\beta}_{k}\right)\right|>\epsilon\right)\nonumber \\
		& \leq2\exp\left(\log S-Bh_{n}^{2}\left(\varepsilon-Ch_{n}^{-2}/S\right)^{2}/2\right).\label{concentration_1}
	\end{align}
	
	Now consider $\mathbb{E}_{k}^{*}\left\Vert \pi_{2,n,k}\right\Vert ^{2}$
	when $e_{1,n}\cap e_{2,n}$ occurs. We first have that
	\begin{align*}
		\mathbb{E}_{k}^{*}\left\Vert \pi_{2,n,k}\right\Vert ^{2} & =\mathbb{E}_{k}^{*}\left(\left.\left\Vert \pi_{2,n,k}\right\Vert ^{2}\right|e_{3,n,k}^{\epsilon}\cap e_{4,n,k}^{\epsilon}\right)\mathbb{P}_{k}^{*}\left(e_{3,n,k}^{\epsilon}\cap e_{4,n,k}^{\epsilon}\right)\\
		& +\mathbb{E}_{k}^{*}\left(\left.\left\Vert \pi_{2,n,k}\right\Vert ^{2}\right|\left(e_{3,n,k}^{\epsilon}\cap e_{4,n,k}^{\epsilon}\right)^{C}\right)\mathbb{P}_{k}^{*}\left(\left(e_{3,n,k}^{\epsilon}\cap e_{4,n,k}^{\epsilon}\right)^{C}\right).
	\end{align*}
	For $\epsilon<2\underline{c}_{f}/\zeta$ with $\zeta>2$, we have
	that 
	\[
	\mathbb{E}_{k}^{*}\left(\left.\left\Vert \pi_{2,n,k}\right\Vert ^{2}\right|e_{3,n,k}^{\epsilon}\cap e_{4,n,k}^{\epsilon}\right)\leq c_{1}^{\infty2}\left\Vert \mathbf{X}^{\phi}\right\Vert _{\infty}^{2}\epsilon^{2}=C\epsilon^{2}.
	\]
	On the other side, according to (\ref{concentration_y}) and (\ref{concentration_1}),
	we have that 
	\begin{align*}
		& \mathbb{E}_{k}^{*}\left(\left.\left\Vert \pi_{2,n,k}\right\Vert ^{2}\right|\left(e_{3,n,k}^{\epsilon}\cap e_{4,n,k}^{\epsilon}\right)^{C}\right)\mathbb{P}_{k}^{*}\left(\left(e_{3,n,k}^{\epsilon}\cap e_{4,n,k}^{\epsilon}\right)^{C}\right)\\
		& \leq Ch_{n}^{-2}\mathbb{P}_{k}^{*}\left(\left(e_{3,n,k}^{\epsilon}\cap e_{4,n,k}^{\epsilon}\right)^{C}\right)\leq Ch_{n}^{-2}\exp\left(C\log S- CBh_{n}^{2}\left(\epsilon-Ch_{n}^{-2}/S\right)^{2}/2\right).
	\end{align*}
	Together we have that over $e_{1,n}\cap e_{2,n}$, there holds
	\[
	\mathbb{E}_{k}^{*}\left\Vert \pi_{2,n,k}\right\Vert ^{2}\leq C\left(\epsilon^{2}+h_{n}^{-2}\exp\left(C\log S-CBh_{n}^{2}\left(\epsilon-Ch_{n}^{-2}/S\right)^{2}/2\right)\right).
	\]
	If we choose 
	\[
	S=2C\sqrt{\frac{Bh_{n}^{-2}}{\log\left(Bh_{n}^{-2}\right)}},\ \epsilon=\sqrt{\frac{8\left(\log\left(h_{n}^{-2}\right)+\log\left(4C^{2}Bh_{n}^{-2}\right)+\log\left(8Bh_{n}^{2}\right)\right)}{Bh_{n}^{2}}},
	\]
	we have that $Ch_{n}^{-2}/S\leq\epsilon/2$ and $\epsilon<2\underline{c}_{f}$
	for $n$ sufficiently large, and 
	\[
	\mathbb{E}_{k}^{*}\left\Vert \pi_{2,n,k}\right\Vert ^{2}\leq C\frac{\log\left(Bh_{n}^{-2}\right)}{Bh_{n}^{2}}.
	\]
	Since $\sup_{k\geq1}\mathbb{E}_{k}^{*}\left\Vert \pi_{2,n,k}\right\Vert ^{2}\leq C$
	implies that $\sup_{k\geq1}\mathbb{E}^{*}\left\Vert \pi_{2,n,k}\right\Vert ^{2}\leq C$,
	we have that 
	\begin{align*}
		P\left(\sup_{k\geq1}\mathbb{E}^{*}\left\Vert \pi_{2,n,k}\right\Vert ^{2}\leq C\frac{\log\left(Bh_{n}^{-2}\right)}{Bh_{n}^{2}}\right) & \geq P\left(\sup_{k\geq1}\mathbb{E}_{k}^{*}\left\Vert \pi_{2,n,k}\right\Vert ^{2}\leq C\frac{\log\left(Bh_{n}^{-2}\right)}{Bh_{n}^{2}}\right)\\
		& \geq P\left(e_{1,n}\cap e_{2,n}\right)\rightarrow1.
	\end{align*}
	This proves the result.
\end{proof}

\subsection*{Proof of \autoref{thm7}}
\begin{proof}
	Note that 
	\begin{align*}
		\left\Vert \Delta\boldsymbol{\beta}_{k+1}\right\Vert  & \leq\sup_{\boldsymbol{\beta}\in\mathcal{B}}\overline{\sigma}\left(I_{p}-\delta\varLambda_{\phi}\left(\boldsymbol{\beta}\right)\right)\left\Vert \Delta\boldsymbol{\beta}_{k}\right\Vert +\delta\left(\sup_{\boldsymbol{\beta}\in\mathcal{B}}\left\Vert \eta_{1,n}\left(\boldsymbol{\beta}\right)\right\Vert +\left\Vert \eta_{2,n}\right\Vert +\left\Vert \pi_{1,n,k}\right\Vert +\left\Vert \pi_{2,n,k}\right\Vert \right)\\
		& \leq\left(1-\delta\underline{\lambda}_{\varLambda}/16\right)\left\Vert \Delta\boldsymbol{\beta}_{k}\right\Vert +\delta\left(\sup_{\boldsymbol{\beta}\in\mathcal{B}}\left\Vert \eta_{1,n}\left(\boldsymbol{\beta}\right)\right\Vert +\left\Vert \eta_{2,n}\right\Vert +\left\Vert \pi_{1,n,k}\right\Vert +\left\Vert \pi_{2,n,k}\right\Vert \right),
	\end{align*}
	where \[
	\eta_{1,n}\left(\boldsymbol{\beta}\right)=\frac{1}{n}\sum_{i=1}^{n}\widehat{G}\left(\left.z\left(\mathbf{X}_{e,i},\boldsymbol{\beta}\right)\right|\boldsymbol{\beta}\right)\boldsymbol{\mathbf{X}}_{i}-\mathbb{E}\left[L\left(z\left(\boldsymbol{\mathbf{X}}_{e,i},\boldsymbol{\beta}\right),\boldsymbol{\beta}\right)\boldsymbol{\mathbf{X}}_{i}\right],
	\]
	\[
	\eta_{2,n}=\left(\frac{1}{n}\sum_{i=1}^{n}G\left(z_{i}^{\star}\right)\boldsymbol{\mathbf{X}}_{i}-\mathbb{E}\left[G\left(z_{i}^{\star}\right)\boldsymbol{\mathbf{X}}_{i}\right]\right)+\frac{1}{n}\sum_{i=1}^{n}\varepsilon_{i}\cdot\boldsymbol{\mathbf{X}}_{i}.
	\] 
	Using Minkovski inequality, we have that
	\begin{align*}
		\left(\mathbb{E}^{*}\left\Vert \Delta\boldsymbol{\beta}_{k+1}\right\Vert ^{2}\right)^{1/2} & \leq\left(1-\delta\underline{\lambda}_{\varLambda}/16\right)\left(\mathbb{E}^{*}\left\Vert \Delta\boldsymbol{\beta}_{k}\right\Vert ^{2}\right)^{1/2}+\delta\sup_{\boldsymbol{\beta}\in\mathcal{B}}\left\Vert \eta_{1,n}\left(\boldsymbol{\beta}\right)\right\Vert +\delta\left\Vert \eta_{2,n}\right\Vert \\
		& +\delta\left(\mathbb{E}^{*}\left\Vert \pi_{1,n,k}\right\Vert ^{2}\right)^{1/2}+\delta\left(\mathbb{E}^{*}\left\Vert \pi_{2,n,k}\right\Vert ^{2}\right)^{1/2}\\
		& \leq\left(1-\delta\underline{\lambda}_{\varLambda}/16\right)\left(\mathbb{E}^{*}\left\Vert \Delta\boldsymbol{\beta}_{k}\right\Vert ^{2}\right)^{1/2}+\delta\sup_{\boldsymbol{\beta}\in\mathcal{B}}\left\Vert \eta_{1,n}\left(\boldsymbol{\beta}\right)\right\Vert +\delta\left\Vert \eta_{2,n}\right\Vert \\
		& +CB^{-1/2}+C\left(\frac{\log\left(Bh_{n}^{-2}\right)}{Bh_{n}^{2}}\right)^{1/2}.
	\end{align*}
	This implies that
	\begin{align*}
\left(\mathbb{E}^{*}\left\Vert \Delta\boldsymbol{\beta}_{k+1}\right\Vert ^{2}\right)^{1/2}	&\leq\left(1-\delta\underline{\lambda}_{\varLambda}/16\right)^{k}\left(\mathbb{E}^{*}\left\Vert \Delta\boldsymbol{\beta}_{1}\right\Vert ^{2}\right)^{1/2}\\
	&+C\left(\sup_{\boldsymbol{\beta}\in\mathcal{B}}\left\Vert \eta_{1,n}\left(\boldsymbol{\beta}\right)\right\Vert +\left\Vert \eta_{2,n}\right\Vert +\left(\frac{\log\left(Bh_{n}^{-2}\right)}{Bh_{n}^{2}}\right)^{1/2}\right).
\end{align*}
	Then when $k\geq k_n+1$, we have that 
	\[
	\left(1-\delta\underline{\lambda}_{\varLambda}/16\right)^{k}\left(\mathbb{E}^{*}\left\Vert \Delta\boldsymbol{\beta}_{1}\right\Vert ^{2}\right)^{1/2}\leq\sup_{\boldsymbol{\beta}\in\mathcal{B}}\left\Vert \eta_{1,n}\left(\boldsymbol{\beta}\right)\right\Vert +\left\Vert \eta_{2,n}\right\Vert +\left(\log\left(Bh_{n}^{-2}\right)/Bh_{n}^{2}\right)^{1/2},
	\]
	implying that $\left(\mathbb{E}^{*}\left\Vert \Delta\boldsymbol{\beta}_{k+1}\right\Vert ^{2}\right)^{1/2}=O_{p}\left(\sup_{\boldsymbol{\beta}\in\mathcal{B}}\left\Vert \eta_{1,n}\left(\boldsymbol{\beta}\right)\right\Vert +\left\Vert \eta_{2,n}\right\Vert +\left(\log\left(Bh_{n}^{-2}\right)/Bh_{n}^{2}\right)^{1/2}\right)$.
	Finally, \citet{khan2022} show that $\sup_{\boldsymbol{\beta}\in\mathcal{B}}\left\Vert \eta_{1,n}\left(\boldsymbol{\beta}\right)\right\Vert +\left\Vert \eta_{2,n}\right\Vert = O_p (\psi(n, h_n, D))$. Since $B\leq n$, we have that 
	\[
	\mathbb{E}^{*}\left\Vert \Delta\boldsymbol{\beta}_{k+1}\right\Vert ^{2}=O_{p}\left(h_{n}^{2D}+\frac{\log\left(Bh_{n}^{-2}\right)}{Bh_{n}^{2}}\right).
	\]
\end{proof}

\subsection*{Proof of  \autoref{lem:4.3} } 
\begin{proof} Note that 
	\begin{align*}
		\Delta\boldsymbol{\beta}_{k+1} & =\int_{0}^{1}\left(I_{p}-\delta\varLambda_{\phi}\left(\boldsymbol{\beta}^{\star}+\tau\Delta\boldsymbol{\beta}_{k}\right)\right)d\tau\Delta\boldsymbol{\beta}_{k}-\delta\boldsymbol{\xi}_{n}^{\phi}\\
		& -\delta\int_{0}^{1}\left(\frac{1}{n}\sum_{i=1}^{n}\left.\mathbf{X}_{i}^{\phi}\frac{\partial\widehat{G}\left(\left.X_{0,i}+\mathbf{X}_{i}^{\mathrm{T}}\boldsymbol{\beta}\right|\boldsymbol{\beta}\right)}{\partial\boldsymbol{\beta}^{\mathrm{T}}}\right|_{\boldsymbol{\beta}=\boldsymbol{\beta}^{\star}+\tau\Delta\boldsymbol{\beta}_{k}}-\varLambda_{\phi}\left(\boldsymbol{\beta}^{\star}+\tau\Delta\boldsymbol{\beta}_{k}\right)\right)d\tau\Delta\boldsymbol{\beta}_{k}(i)\\
		& -\delta\left(\frac{1}{B}\sum_{i\in\mathfrak{I}_{B,k}}\left(\widehat{G}\left(\left.z_{i,k}\right|\boldsymbol{\beta}_{k}\right)-y_{i}\right)\mathbf{X}_{i}^{\phi}-\frac{1}{n}\sum_{i=1}^{n}\left(\widehat{G}\left(\left.z_{i,k}\right|\boldsymbol{\beta}_{k}\right)-y_{i}\right)\mathbf{X}_{i}^{\phi}\right)(ii)\\
		& -\delta\left(\frac{1}{B}\sum_{i\in\mathfrak{I}_{B,k}}\left(\widehat{G}\left(\left.z_{i,k}\right|\boldsymbol{\beta}_{k},\mathfrak{I}_{B,k},\underline{c}_{f}\right)-\widehat{G}\left(\left.z_{i,k}\right|\boldsymbol{\beta}_{k}\right)\right)\mathbf{X}_{i}^{\phi}\right)(iii).
	\end{align*}
	For (i), we have that 
	\begin{align*}
		\sup_{k\geq k_{n}+1}\mathbb{E}_{k}^{*}\left\Vert (i)\right\Vert  & =O_{p}\left(\left(h_{n}^{-2}\sqrt{\frac{\log\left(n\right)}{n}}+h_{n}^{D}\right)\left(h_{n}^{D}+\sqrt{\frac{\log\left(n\right)}{Bh_{n}^{2}}}\right)+h_{n}^{2D}+\frac{\log\left(Bh_{n}^{-2}\right)}{Bh_{n}^{2}}\right)\\
		& =O_{p}\left(\sqrt{\frac{\log^{2}\left(n\right)}{nBh_{n}^{6}}}+h_{n}^{D-2}\sqrt{\frac{\log\left(n\right)}{n}}+\frac{\log\left(Bh_{n}^{-2}\right)}{Bh_{n}^{2}}+h_{n}^{2D}\right).
	\end{align*}
	This implies that given the choice of $B$ and $h_n$,  $\mathbb{E}^{*}\left\Vert (i)\right\Vert$ is  $o_p(n^{-1/2})$ uniformly with respect to $k$. 
	
	Now we look at (iii). To further simplify our notations, we denote
	$A_{n,y}\left(z_{i,k},\boldsymbol{\beta}_{k}\right)=A_{n,y,i,k}$,
	$A_{n,1}\left(z_{i,k},\boldsymbol{\beta}_{k}\right)=A_{n,1,i,k}$,
	$A_{n,y}\left(\left.z_{i,k},\boldsymbol{\beta}_{k}\right|\mathfrak{I}_{B,k}\right)=A_{n,y,i,k}^{\mathfrak{I}}$,
	$A_{n,1}\left(\left.z_{i,k},\boldsymbol{\beta}_{k}\right|\mathfrak{I}_{B,k}\right)=A_{n,1,i,k}^{\mathfrak{I}}$.
	We have that 
	\begin{align*}
		(iii) & =\frac{1}{B}\sum_{i\in\mathfrak{I}_{B,k}}\left(\frac{A_{n,y,i,k}^{\mathfrak{I}}}{A_{n,1,i,k}^{\mathfrak{I}}\wedge\underline{c}_{f}}-\frac{A_{n,y,i,k}}{A_{n,1,i,k}}\right)\mathbf{X}_{i}^{\phi}\\
		& =\frac{1}{B}\sum_{i\in\mathfrak{I}_{B,k}}\frac{\mathbf{X}_{i}^{\phi}}{A_{n,1,i,k}}\cdot\left(A_{n,y,i,k}^{\mathfrak{I}}-A_{n,y,i,k}\right)(iv)\\
		& -\frac{1}{B}\sum_{i\in\mathfrak{I}_{B,k}}\frac{A_{n,y,i,k}\mathbf{X}_{i}^{\phi}}{A_{n,1,i,k}^{2}}\left(A_{n,1,i,k}^{\mathfrak{I}}\wedge\underline{c}_{f}-A_{n,1,i,k}^{\mathfrak{I}}\right)(v)-\frac{1}{B}\sum_{i\in\mathfrak{I}_{B,k}}\frac{A_{n,y,i,k}\mathbf{X}_{i}^{\phi}}{A_{n,1,i,k}^{2}}\left(A_{n,1,i,k}^{\mathfrak{I}}-A_{n,1,i,k}\right)(vi)\\
		& -\frac{1}{B}\sum_{i\in\mathfrak{I}_{B,k}}\frac{\mathbf{X}_{i}^{\phi}}{\widetilde{A}_{n,1,i,k}^{2}}\left(A_{n,y,i,k}^{\mathfrak{I}}-A_{n,y,i,k}\right)\left(A_{n,1,i,k}^{\mathfrak{I}}\wedge\underline{c}_{f}-A_{n,1,i,k}^{\mathfrak{I}}\right)(vii)\\
		& -\frac{1}{B}\sum_{i\in\mathfrak{I}_{B,k}}\frac{\mathbf{X}_{i}^{\phi}}{\widetilde{A}_{n,1,i,k}^{2}}\left(A_{n,y,i,k}^{\mathfrak{I}}-A_{n,y,i,k}\right)\left(A_{n,y,i,k}^{\mathfrak{I}}-A_{n,y,i,k}\right)(viii)\\
		& +\frac{2}{B}\sum_{i\in\mathfrak{I}_{B,k}}\frac{A_{n,1,i,k}\mathbf{X}_{i}^{\phi}}{\widetilde{\widetilde{A}}_{n,1,i,k}^{3}}\left(A_{n,y,i,k}^{\mathfrak{I}}-A_{n,y,i,k}\right)^{2}(ix),
	\end{align*}
	where $\widetilde{A}_{n,1,i,k}^{2}$ and $\widetilde{\widetilde{A}}_{n,1,i,k}^{3}$
	both lie between $A_{n,1,i,k}^{\mathfrak{I}}\wedge\underline{c}_{f}$
	and $A_{n,1,i,k}$. Define $\ mathbb{E}_k^*\{|j\}$ as the conditional expectation with respect to $\mathbb{P}_k^*$ holding the $j$-th index $i_{k,j}$ fixed. Note that for any $1\leq j\leq B$ and $k$, 
	\begin{align*}
		& \mathbb{E}_{k}^{*}\left\{ \left.\left(A_{n,y,i_{k,j},k}^{\mathfrak{I}}-A_{n,y,i_{k,j},k}\right)^{2}\right|j\right\}\\
		& =\mathbb{E}_{k}^{*}\left\{ \left.\left(\frac{1}{B}\sum_{b=1}^{B}K_{h_{n}}\left(z_{i_{k,j},k}-z_{i_{k,b},k}\right)y_{i_{b}}-\frac{1}{n}\sum_{b=1}^{n}K_{h_{n}}\left(z_{i_{k,j},k}-z_{b,k}\right)y_{k,b}\right)^{2}\right|j\right\} \\
		& \leq C\left\{ \left(\frac{y_{i_{k,j}}}{Bh_{n}}-\frac{1}{Bn}\sum_{b=1}^{n}K_{h_{n}}\left(z_{i_{k,j},k}-z_{b,k}\right)y_{b}\right)^{2}+\frac{B-1}{B^{2}}\frac{1}{n}\sum_{b=1}^{n}K_{h_{n}}^{2}\left(z_{i_{k,j},k}-z_{b,k}\right)y_{b}^{2}\right\} \leq\frac{C}{Bh_{n}^{2}},
	\end{align*}
	for some positive constant $C$ that does not depend on $k$ and $j$.
	Similarly, we have that for all $1\leq j\leq B$ and $k$, 
	\[
	\mathbb{E}_{k}^{*}\left\{ \left.\left(A_{n,1,i_{k,j},k}^{\mathfrak{I}}-A_{n,1,i_{k,j},k}\right)^{2}\right|j\right\} \leq\frac{C}{Bh_{n}^{2}}.
	\]
	So with probability going to 1, for all $k$
	\begin{align*}
		\mathbb{E}_{k}^{*}\left\Vert (viii)\right\Vert  & \leq\frac{C}{B}\mathbb{E}_{k}^{*}\left(\sum_{i\in\mathfrak{I}_{B,k}}\left|\left(A_{n,y,i,k}^{\mathfrak{I}}-A_{n,y,i,k}\right)\left(A_{n,1,i,k}^{\mathfrak{I}}-A_{n,1,i,k}\right)\right|\right)\\
		& \leq\frac{C}{B}\mathbb{E}_{k}^{*}\left(\sum_{j=1}^{B}\mathbb{E}_{k}^{*}\left(\left.\left|\left(A_{n,y,i_{k,j},k}^{\mathfrak{I}}-A_{n,y,i_{k,j},k}\right)\left(A_{n,1,i_{k,j},k}^{\mathfrak{I}}-A_{n,1,i_{k,j},k}\right)\right|\right|j\right)\right)\\
		& \leq\frac{C}{B}\mathbb{E}_{k}^{*}\left(\sum_{j=1}^{B}\sqrt{\mathbb{E}_{k}^{*}\left\{ \left.\left(A_{n,y,i_{k,j},k}^{\mathfrak{I}}-A_{n,y,i_{k,j},k}\right)^{2}\right|j\right\} }\sqrt{\mathbb{E}_{k}^{*}\left\{ \left.\left(A_{n,1,i_{k,j},k}^{\mathfrak{I}}-A_{n,1,i_{k,j},k}\right)^{2}\right|j\right\} }\right)\\
		& \leq\frac{C}{B}\mathbb{E}_{k}^{*}\left(\sum_{j=1}^{B}\frac{C}{Bh_{n}^{2}}\right)\leq\frac{C}{Bh_{n}^{2}}.
	\end{align*}
	Similarly, we have that $\mathbb{E}_{k}^{*}\left\Vert (ix)\right\Vert \leq C/Bh_{n}^{2}$
	for all $k$ with probability going to 1. Due to the choice of $B$ and $h_n$, we have that $\mathbb{E}^{*}\left\Vert (viii)\right\Vert$ and $\mathbb{E}^{*}\left\Vert (ix)\right\Vert$ are both $o_p(n^{-1/2})$ uniformly with respect to $k$. On the other side, note
	that 
	\begin{align*}
		\mathbb{E}_{k}^{*}\left\Vert (vii)\right\Vert  & \leq C\mathbb{E}_{k}^{*}\left(\frac{1}{B}\sum_{j=1}^{B}\sqrt{\mathbb{E}_{k}^{*}\left\{ \left.\left(A_{n,y,i_{k,j},k}^{\mathfrak{I}}-A_{n,y,i_{k,j},k}\right)^{2}\right|j\right\} }\sqrt{\mathbb{E}_{k}^{*}\left\{ \left.\left(A_{n,1,i_{k,j},k}^{\mathfrak{I}}\wedge\underline{c}_{f}-A_{n,1,i_{k,j},k}\right)^{2}\right|j\right\} }\right)\\
		& \leq C\mathbb{E}_{k}^{*}\left(\frac{1}{B}\sum_{j=1}^{B}\left(\frac{C}{\sqrt{Bh_{n}^{2}}}\right)\sqrt{\mathbb{E}_{k}^{*}\left\{ \left.\left(A_{n,1,i_{k,j},k}^{\mathfrak{I}}\wedge\underline{c}_{f}-A_{n,1,i_{k,j},k}\right)^{2}\right|j\right\} }\right).
	\end{align*}
	Note that 
	\[
	\mathbb{E}_{k}^{*}\left\{ \left.\left(A_{n,1,i_{k,j},k}^{\mathfrak{I}}\wedge\underline{c}_{f}-A_{n,1,i_{k,j},k}\right)^{2}\right|j\right\} \leq Ch_{n}^{-2}\mathbb{P}_{k}^{*}\left(\left.A_{n,1,i_{j},k}^{\mathfrak{I}}<\underline{c}_{f}\right|j\right).
	\]
	Now consider $\mathbb{P}_{k}^{*}\left(\left.A_{n,1,i_{k,j},k}^{\mathfrak{I}}<\underline{c}_{f}\right|j\right)$.
	Note that 
	\begin{align*}
		A_{n,1,i_{k,j},k}^{\mathfrak{I}}<\underline{c}_{f}  \Longrightarrow &\frac{1}{B}\sum_{b=1}^{B}K_{h_{n}}\left(z_{i_{k,j},k}-z_{i_{b}}\right)y_{i_{b}}-\frac{1}{n}\sum_{i=1}^{n}K_{h_{n}}\left(z_{i_{k,j},k}-z_{i,k}\right)y_{i}\\
		&<\underline{c}_{f}-\frac{1}{n}\sum_{i=1}^{n}K_{h_{n}}\left(z_{i_{k,j},k}-z_{i,k}\right)y_{i}\\
		 \Longrightarrow&\frac{1}{B}\sum_{b\neq j}^{B}K_{h_{n}}\left(z_{i_{k,j},k}-z_{i_{k,b}}\right)y_{i_{k,b}}-\frac{1}{n}\sum_{i=1}^{n}K_{h_{n}}\left(z_{i_{k,j},k}-z_{i,k}\right)y_{i}<-\underline{c}_{f}-\frac{y_{i_{k,j}}}{Bh_{n}}\\
		 \Longrightarrow&\sup_{z\in\mathcal{Z}^{\phi}}\left|\frac{1}{B}\sum_{b\neq j}^{B}K_{h_{n}}\left(z_{i_{k,j},k}-z_{i_{k,b}}\right)y_{i_{k,b}}-\frac{B-1}{B}\frac{1}{n}\sum_{i=1}^{n}K_{h_{n}}\left(z_{i_{k,j},k}-z_{i,k}\right)y_{i}\right|>\underline{c}_{f}+\frac{C}{Bh_{n}}.
	\end{align*}
	This implies that 
	\begin{align*}
		& \mathbb{P}_{k}^{*}\left(\left.A_{n,1,i_{j},k}^{\mathfrak{I}}<\underline{c}_{f}\right|j\right)\\
		& \leq\mathbb{P}_{k}^{*}\left(\left.\sup_{z\in\mathcal{Z}}\left|\frac{1}{B}\sum_{b\neq j}^{B}K_{h_{n}}\left(z_{i_{k,j},k}-z_{i_{k,b}}\right)y_{i_{k,b}}-\frac{B-1}{B}\frac{1}{n}\sum_{i=1}^{n}K_{h_{n}}\left(z_{i_{k,j},k}-z_{i,k}\right)y_{i}\right|>\underline{c}_{f}+\frac{C}{Bh_{n}}\right|j\right)\\
		& \leq2\exp\left(\log S-Bh_{n}^{2}\left(\underline{c}_{f}+\frac{C}{Bh_{n}}-Ch_{n}^{-2}/S\right)^{2}/2\right)
	\end{align*}
	for any sufficiently large positive integer $S$. Let $S=Bh_{n}^{-1}$, we have that
	for $n$ sufficiently large, we have that 
	\[
	\exp\left(\log S-Bh_{n}^{2}\left(\underline{c}_{f}-\frac{C}{Bh_{n}}+\frac{C}{h_{n}^{2}S}\right)^{2}/2\right)\leq C\exp\left(C\left(\log\left(Bh_{n}^{-1}\right)-Bh_{n}^{2}\right)\right),
	\]
	implying that 
	\[
	\mathbb{E}_{k}^{*}\left\{ \left.\left(A_{n,1,i_{k,j},k}^{\mathfrak{I}}\wedge\underline{c}_{f}-A_{n,1,i_{k,j},k}\right)^{2}\right|j\right\} \leq Ch_{n}^{-2}\exp\left(C\left(\log\left(Bh_{n}^{-1}\right)-Bh_{n}^{2}\right)\right).
	\]
	So uniformly with respect to  $k$, there holds 
	\[
	\mathbb{E}_{k}^{*}\left\Vert (vii)\right\Vert \leq\frac{C\exp\left(C\left(\log\left(Bh_{n}^{-1}\right)-Bh_{n}^{2}\right)\right)}{\sqrt{Bh_{n}^{4}}}.
	\]
	Similarly, we have that $\mathbb{E}_{k}^{*}\left\Vert (v)\right\Vert \leq Ch_{n}^{-1}\exp\left(C\left(\log\left(Bh_{n}^{-1}\right)-Bh_{n}^{2}\right)\right)$
	for all $k$. Given the choice of $B$ and $h_n$, we have that $\mathbb{E}^{*}\left\Vert (vii)\right\Vert$ and $\mathbb{E}^{*}\left\Vert (v)\right\Vert $ are both $o_p(n^{-1/2})$ uniformly with respect to $k$.
	
	We finally note that uniformly for all $k$, 
	\[
	\mathbb{E}^{*}\left\Vert \left(\int_{0}^{1}\varLambda_{\phi}\left(\boldsymbol{\beta}^{\star}+\tau\Delta\boldsymbol{\beta}_{k}\right)d\tau-\varLambda_{\phi}\left(\boldsymbol{\beta}^{\star}\right)\right)\Delta\boldsymbol{\beta}_{k}\right\Vert \leq C\mathbb{E}^{*}\left\Vert \Delta\boldsymbol{\beta}_{k}\right\Vert ^{2}=O_{p}\left(h_{n}^{2D}+\frac{\log\left(Bh_{n}^{-2}\right)}{Bh_{n}^{2}}\right).
	\]
	This finishes the proof. 
\end{proof}

\subsection*{Proof of \autoref{thm8}}

\begin{proof}
	Define 
	\[
	\boldsymbol{\Xi}_{1,k}^{\phi}=\frac{1}{B}\sum_{i\in\mathfrak{I}_{B,k}}\left(\widehat{G}\left(\left.z_{i,k}\right|\boldsymbol{\beta}_{k}\right)-y_{i}\right)\mathbf{X}_{i}^{\phi}-\frac{1}{n}\sum_{i=1}^{n}\left(\widehat{G}\left(\left.z_{i,k}\right|\boldsymbol{\beta}_{k}\right)-y_{i}\right)\mathbf{X}_{i}^{\phi},
	\]
	\[
	\boldsymbol{\Xi}_{2,k}^{\phi}=\frac{1}{B}\sum_{i\in\mathfrak{I}_{B,k}}\frac{\mathbf{X}_{i}^{\phi}}{A_{n,1}\left(z_{i,k},\boldsymbol{\beta}_{k}\right)}\left(A_{n,y}\left(\left.z_{i,k},\boldsymbol{\beta}_{k}\right|\mathfrak{I}_{B,k}\right)-A_{n,y}\left(z_{i,k},\boldsymbol{\beta}_{k}\right)\right),
	\]
	and 
	\[
	\boldsymbol{\Xi}_{3,k}^{\phi}=\frac{1}{B}\sum_{i\in\mathfrak{I}_{B,k}}\frac{A_{n,y}\left(z_{i,k},\boldsymbol{\beta}_{k}\right)\mathbf{X}_{i}^{\phi}}{A_{n,1}^{2}\left(z_{i,k},\boldsymbol{\beta}_{k}\right)}\left(A_{n,1}\left(\left.z_{i,k},\boldsymbol{\beta}_{k}\right|\mathfrak{I}_{B,k}\right)-A_{n,1}\left(z_{i,k},\boldsymbol{\beta}_{k}\right)\right).
	\]
	We obviously have that $\sup_{k}\mathbb{E}_{k}^{*}\left\Vert \boldsymbol{\Xi}_{1,k}^{\phi}\right\Vert ^{2}\leq C/B$,
	so $\sup_{k}\mathbb{E}^{*}\left\Vert \boldsymbol{\Xi}_{1,k}^{\phi}\right\Vert ^{2}\leq C/B$
	holds. Moreover, $\mathbb{E}_{k}^{*}\left(\boldsymbol{\Xi}_{1,k}^{\phi}\boldsymbol{\Xi}_{1,k^{\prime}}^{\phi\mathrm{T}}\right)=0$
	for all $k\neq k^{\prime}$, so $\mathbb{E}^{*}\left(\boldsymbol{\Xi}_{1,k}^{\phi}\boldsymbol{\Xi}_{1,k^{\prime}}^{\phi\mathrm{T}}\right)=0$
	for all $k\neq k^{\prime}$. We then show that
	\[
	\sup_{k\geq k_{n}+1}\mathbb{E}^{*}\left\Vert \boldsymbol{\boldsymbol{\Xi}}_{2,k}^{\phi}\right\Vert ^{2}=O_{p}\left(\frac{1}{Bh_{n}^{2}}\right),\text{ }\sup_{k\geq k_{n}+1}\mathbb{E}^{*}\left\Vert \boldsymbol{\boldsymbol{\Xi}}_{3,k}^{\phi}\right\Vert ^{2}=O_{p}\left(\frac{1}{Bh_{n}^{2}}\right)
	\]
	and 
	\[
	\sup_{k,k^{\prime}\geq k_{n}+1,k\neq k^{\prime}}\left\Vert \mathbb{E}^{*}\boldsymbol{\Xi}_{2,k}^{\phi}\boldsymbol{\boldsymbol{\Xi}}_{2,k^{\prime}}^{\phi\mathrm{T}}\right\Vert =O_{p}\left(\frac{\sqrt{\log n}}{B^{2}h_{n}^{2}}\right),\text{ }\sup_{k,k^{\prime}\geq k_{n}+1,k\neq k^{\prime}}\left\Vert \mathbb{E}^{*}\boldsymbol{\boldsymbol{\Xi}}_{3,k}^{\phi}\boldsymbol{\boldsymbol{\Xi}}_{3,k^{\prime}}^{\phi\mathrm{T}}\right\Vert =O_{p}\left(\frac{\sqrt{\log n}}{B^{2}h_{n}^{2}}\right).
	\]
	We will only show the results for $\boldsymbol{\boldsymbol{\Xi}}_{2,k}^{\phi}$.
	The results for $\boldsymbol{\boldsymbol{\Xi}}_{3,k}^{\phi}$ can
	be similarly proved. For the first result, according to the proof
	of Lemma 2, we note that with probability going to 1, 
	\begin{align*}
		\mathbb{E}^{*}\left\Vert \boldsymbol{\boldsymbol{\Xi}}_{2,k}^{\phi}\right\Vert ^{2} & \leq\frac{1}{B^{2}}\sum_{j=1}^{B}\sum_{l\neq j}^{B}\mathbb{E}^{*}\left(\left\Vert \frac{\mathbf{X}_{i_{k,j}}^{\phi}\mathbf{X}_{i_{k,l}}^{\phi\mathrm{T}}}{A_{n,1,i_{k,j},k}A_{n,1,i_{k,l},k}}\right\Vert \left|\left(A_{n,1,i_{k,j},k}^{\mathfrak{I}}-A_{n,1,i_{k,j},k}\right)\left(A_{n,1,i_{k,l},k}^{\mathfrak{I}}-A_{n,1,i_{k,l},k}\right)\right|\right)\\
		& +\frac{1}{B^{2}}\sum_{j=1}^{B}\mathbb{E}^{*}\left(\left\Vert \frac{\mathbf{X}_{i_{k,j}}^{\phi}\mathbf{X}_{i_{k,j}}^{\phi\mathrm{T}}}{A_{n,1,i_{k,j},k}^{2}}\right\Vert \left(A_{n,1,i_{k,j},k}^{\mathfrak{I}}-A_{n,1,i_{k,j},k}\right)^{2}\right)\\
		& \leq\frac{C}{B^{2}}\sum_{j=1}^{B-1}\sum_{l=j+1}^{B}\frac{1}{Bh_{n}^{2}}+\frac{C}{B^{2}}\sum_{j=1}^{B}\frac{1}{Bh_{n}^{2}}\leq\frac{C}{Bh_{n}^{2}}.
	\end{align*}
	The derivation of the second result is more complicated. Without loss
	of generality, we assume that $\boldsymbol{\Xi}_{2,k}^{\phi}$ is
	one-dimensional and $k<k^{\prime}$. Then $\mathbb{E}^{*}\boldsymbol{\Xi}_{2,k}^{\phi}\boldsymbol{\boldsymbol{\Xi}}_{2,k^{\prime}}^{\phi}=\mathbb{E}^{*}\left(\mathbb{E}_{k}^{*}\boldsymbol{\Xi}_{2,k}^{\phi}\left(\mathbb{E}_{k^{\prime}}^{*}\boldsymbol{\boldsymbol{\Xi}}_{2,k^{\prime}}^{\phi}\right)\right)$.
	We first look at $\mathbb{E}_{k}^{*}\boldsymbol{\boldsymbol{\Xi}}_{2,k}^{\phi}$
	for general $k$. We have that 
	\begin{align*}
		& \mathbb{E}_{k}^{*}\boldsymbol{\boldsymbol{\Xi}}_{2,k}^{\phi}\\
		& =\frac{1}{B}\sum_{j=1}^{B}\mathbb{E}_{k}^{*}\left[\frac{\mathbf{X}_{i_{k,j}}^{\phi}}{A_{n,1}\left(z_{i_{k,j},k},\boldsymbol{\beta}_{k}\right)}\mathbb{E}_{k}^{*}\left\{ \left.A_{n,y}\left(\left.z_{i_{k,j},k},\boldsymbol{\beta}_{k}\right|\mathfrak{I}_{B,k}\right)-A_{n,y}\left(z_{i_{k,j},k},\boldsymbol{\beta}_{k}\right)\right|j\right\} \right]\\
		& =\frac{1}{B}\sum_{j=1}^{B}\mathbb{E}_{k}^{*}\left[\frac{\mathbf{X}_{i_{k,j}}^{\phi}}{A_{n,1}\left(z_{i_{k,j},k},\boldsymbol{\beta}_{k}\right)}\left\{ \mathbb{E}_{k}^{*}\left\{ \left.\frac{1}{B}\sum_{l=1}^{B}K_{h_{n}}\left(z_{i_{k,j},k}-z_{i_{k,l},k}\right)y_{i_{k,l}}-\frac{1}{n}\sum_{l=1}^{n}K_{h_{n}}\left(z_{i_{k,j},k}-z_{l,k}\right)y_{l}\right|j\right\} \right\} \right],
	\end{align*}
	Obviously, for $l\neq j$, we have that $\mathbb{E}_{k}^{*}\left\{ \left.K_{h_{n}}\left(z_{i_{k,j},k}-z_{i_{k,l},k}\right)y_{i_{k,l}}\right|j\right\} =\frac{1}{n}\sum_{l=1}^{n}K_{h_{n}}\left(z_{i_{k,j},k}-z_{l,k}\right).$
	So 
	\begin{align*}
		& \mathbb{E}_{k}^{*}\left\{ \left.\frac{1}{B}\sum_{l=1}^{B}K_{h_{n}}\left(z_{i_{k,j},k}-z_{i_{k,l},k}\right)y_{i_{k,l}}-\frac{1}{n}\sum_{l=1}^{n}K_{h_{n}}\left(z_{i_{k,j},k}-z_{l,k}\right)y_{l}\right|j\right\} \\
		& =\frac{1}{B}\left(K\left(0\right)y_{i_{k,j}}-\frac{1}{n}\sum_{l=1}^{n}K_{h_{n}}\left(z_{i_{k,j},k}-z_{l,k}\right)y_{l}\right).
	\end{align*}
	So 
	\begin{align*}
		\mathbb{E}_{k}^{*}\boldsymbol{\boldsymbol{\Xi}}_{2,k}^{\phi} & =\frac{1}{B}\sum_{j=1}^{B}\mathbb{E}_{k}^{*}\left(\frac{1}{B}\frac{\mathbf{X}_{i_{k,j}}^{\phi}\left(K\left(0\right)y_{i_{k,j}}-\frac{1}{n}\sum_{l=1}^{n}K_{h_{n}}\left(z_{i_{k,j},k}-z_{l,k}\right)y_{l}\right)}{A_{n,1}\left(z_{i_{k,j},k},\boldsymbol{\beta}_{k}\right)}\right)
	\end{align*}
	Now define $z_{i}^{\star}=X_{0,i}+\mathbf{X}_{i}^{\mathrm{T}}\boldsymbol{\beta}^{\star}$,
	we have that with probability going to 1, there holds 
	\begin{align*}
		& \left|\frac{\mathbf{X}_{i_{k,j}}^{\phi}\left(K\left(0\right)y_{i_{k,j}}-\frac{1}{n}\sum_{l=1}^{n}K_{h_{n}}\left(z_{i_{k,j}}^{\star}-z_{l}^{\star}\right)y_{l}\right)}{A_{n,1}\left(z_{i_{k,j},k},\boldsymbol{\beta}_{k}\right)}-\frac{\mathbf{X}_{i_{k,j}}^{\phi}\left(K\left(0\right)y_{i_{k,j}}-\frac{1}{n}\sum_{l=1}^{n}K_{h_{n}}\left(z_{i_{k,j}}^{\star}-z_{l}^{\star}\right)y_{l}\right)}{A_{n,1}\left(z_{i_{k,j}}^{\star},\boldsymbol{\beta}^{\star}\right)}\right|\\
		& \leq C\left\Vert \Delta\boldsymbol{\beta}_{k}\right\Vert ,
	\end{align*}
	Then 
	\[
	\left|\mathbb{E}_{k}^{*}\boldsymbol{\boldsymbol{\Xi}}_{2,k}^{\phi}-\frac{1}{B}\sum_{j=1}^{B}\mathbb{E}_{k}^{*}\left(\frac{1}{B}\frac{\mathbf{X}_{i_{k,j}}^{\phi}\left(K\left(0\right)y_{i_{k,j}}-\frac{1}{n}\sum_{l=1}^{n}K_{h_{n}}\left(z_{i_{k,j}}^{\star}-z_{l}^{\star}\right)y_{l}\right)}{A_{n,1}\left(z_{i_{k,j}}^{\star},\boldsymbol{\beta}^{\star}\right)}\right)\right|\leq\frac{C\left\Vert \Delta\boldsymbol{\beta}_{k}\right\Vert }{B},
	\]
	which is equivalent to 
	\[
	\left|\mathbb{E}_{k}^{*}\boldsymbol{\boldsymbol{\Xi}}_{2,k}^{\phi}-\frac{1}{nB}\sum_{i=1}^{n}\left(\frac{\mathbf{X}_{i}^{\phi}\left(K\left(0\right)y_{i}-\frac{1}{n}\sum_{l=1}^{n}K_{h_{n}}\left(z_{i}^{\star}-z_{l}^{\star}\right)y_{l}\right)}{A_{n,1}\left(z_{i}^{\star},\boldsymbol{\beta}^{\star}\right)}\right)\right|\leq\frac{C\left\Vert \Delta\boldsymbol{\beta}_{k}\right\Vert }{B},
	\]
	Based on such result, we have that 
	\begin{align*}
		& \left|\mathbb{E}_{k}^{*}\left(\boldsymbol{\Xi}_{2,k}^{\phi}\left(\mathbb{E}_{k^{\prime}}^{*}\boldsymbol{\boldsymbol{\Xi}}_{2,k^{\prime}}^{\phi}\right)\right)-\mathbb{E}_{k}^{*}\left(\boldsymbol{\Xi}_{2,k}^{\phi}\frac{1}{nB}\sum_{i=1}^{n}\left(\frac{\mathbf{X}_{i}^{\phi}\left(K\left(0\right)y_{i}-\frac{1}{n}\sum_{l=1}^{n}K_{h_{n}}\left(z_{i}^{\star}-z_{l}^{\star}\right)y_{l}\right)}{A_{n,1}\left(z_{i}^{\star},\boldsymbol{\beta}^{\star}\right)}\right)\right)\right|\\
		& \leq C\mathbb{E}_{k}^{*}\left(\left|\boldsymbol{\Xi}_{2,k}^{\phi}\right|\left\Vert \Delta\boldsymbol{\beta}_{k^{\prime}}\right\Vert \right)/B\leq C\sqrt{\mathbb{E}_{k}^{*}\left|\boldsymbol{\Xi}_{2,k}^{\phi}\right|^{2}}\sqrt{\mathbb{E}_{k}^{*}\left\Vert \Delta\boldsymbol{\beta}_{k^{\prime}}\right\Vert ^{2}}/B\leq\frac{C\sqrt{\log n}}{B^{2}h_{n}^{2}}
	\end{align*}
	uniformly for all $k$ when $k\geq k_{n}+1$. On the other side, 
	\begin{align*}
		& \mathbb{E}_{k}^{*}\left(\boldsymbol{\Xi}_{2,k}^{\phi}\frac{1}{nB}\sum_{i=1}^{n}\left(\frac{\mathbf{X}_{i}^{\phi}\left(K\left(0\right)y_{i}-\frac{1}{n}\sum_{l=1}^{n}K_{h_{n}}\left(z_{i}^{\star}-z_{l}^{\star}\right)y_{l}\right)}{A_{n,1}\left(z_{i}^{\star},\boldsymbol{\beta}^{\star}\right)}\right)\right)\\
		& =\frac{1}{nB}\sum_{i=1}^{n}\left(\frac{\mathbf{X}_{i}^{\phi}\left(K\left(0\right)y_{i}-\frac{1}{n}\sum_{l=1}^{n}K_{h_{n}}\left(z_{i}^{\star}-z_{l}^{\star}\right)y_{l}\right)}{A_{n,1}\left(z_{i}^{\star},\boldsymbol{\beta}^{\star}\right)}\right)\mathbb{E}_{k}^{*}\left(\frac{1}{B}\left(K\left(0\right)y_{i_{k,j}}-\frac{1}{n}\sum_{l=1}^{n}K_{h_{n}}\left(z_{i_{k,j},k}-z_{l,k}\right)y_{l}\right)\right)\\
		& =O_{p}\left(\frac{1}{B^{2}}\right)
	\end{align*}
	uniformly for all $k$. This proves the desired result. 
	
	Now denote $\widetilde{k}=\left[-\log\left(n\right)/\log\left(1-\delta\underline{\lambda}_{\varLambda}/8\right)\right]$,
	so $k^{*}=k_{n}+\widetilde{k}$. We have that 
	\begin{align*}
		& \Delta\boldsymbol{\beta}_{k^{*}+1+t}\\
		& =\left(I-\delta\varLambda_{\phi}\left(\boldsymbol{\beta}^{\star}\right)\right)^{t+\widetilde{k}}\Delta\boldsymbol{\beta}_{k_{n}+1}+\delta\sum_{k=0}^{t+\widetilde{k}-1}\left(I-\delta\varLambda_{\phi}\left(\boldsymbol{\beta}^{\star}\right)\right)^{t+\widetilde{k}-1-k}\varOmega_{k_{n}+1+k}^{\phi}\\
		& -\delta\sum_{k=0}^{t+\widetilde{k}-1}\left(I-\delta\varLambda_{\phi}\left(\boldsymbol{\beta}^{\star}\right)\right)^{t+\widetilde{k}-1-k}\boldsymbol{\xi}_{n}^{\phi}-\delta\sum_{k=0}^{t+\widetilde{k}-1}\left(I-\delta\varLambda_{\phi}\left(\boldsymbol{\beta}^{\star}\right)\right)^{t+\widetilde{k}-1-k}\left(\boldsymbol{\mathbf{\Xi}}_{1,k_{n}+1+k}^{\phi}+\boldsymbol{\mathbf{\Xi}}_{2,k_{n}+1+k}^{\phi}-\boldsymbol{\mathbf{\Xi}}_{3,k_{n}+1+k}^{\phi}\right).
	\end{align*}
	So 
	\begin{align*}
		\frac{1}{T}\sum_{t=1}^{T}\Delta\boldsymbol{\beta}_{k^{*}+1+t} & =\frac{1}{T}\sum_{t=1}^{T}\left(I-\delta\varLambda_{\phi}\left(\boldsymbol{\beta}^{\star}\right)\right)^{t+\widetilde{k}}\Delta\boldsymbol{\beta}_{k_{n}+1}+\frac{\delta}{T}\sum_{t=1}^{T}\sum_{k=0}^{t+\widetilde{k}-1}\left(I-\delta\varLambda_{\phi}\left(\boldsymbol{\beta}^{\star}\right)\right)^{t+\widetilde{k}-1-k}\varOmega_{k_{n}+1+k}^{\phi}\\
		& -\varLambda_{\phi}^{-1}\left(\boldsymbol{\beta}^{\star}\right)\boldsymbol{\xi}_{n}^{\phi}-\frac{1}{T}\sum_{t=1}^{T}\left(\delta\sum_{k=0}^{t+\widetilde{k}-1}\left(I-\delta\varLambda_{\phi}\left(\boldsymbol{\beta}^{\star}\right)\right)^{k}-\varLambda_{\phi}^{-1}\left(\boldsymbol{\beta}^{\star}\right)\right)\boldsymbol{\xi}_{n}^{\phi}\\
		& -\frac{\delta}{T}\sum_{t=1}^{T}\sum_{k=0}^{t+\widetilde{k}-1}\left(I-\delta\varLambda_{\phi}\left(\boldsymbol{\beta}^{\star}\right)\right)^{t+\widetilde{k}-1-k}\left(\boldsymbol{\mathbf{\Xi}}_{1,k_{n}+1+k}^{\phi}+\boldsymbol{\mathbf{\Xi}}_{2,k_{n}+1+k}^{\phi}-\boldsymbol{\mathbf{\Xi}}_{3,k_{n}+1+k}^{\phi}\right).
	\end{align*}
	We look at the above terms separately. We have that 
	\begin{align*}
		\mathbb{E}^{*}\left\Vert \frac{1}{T}\sum_{t=1}^{T}\left(I-\delta\varLambda_{\phi}\left(\boldsymbol{\beta}^{\star}\right)\right)^{t+\widetilde{k}}\Delta\boldsymbol{\beta}_{k_{n}+1}\right\Vert  & \leq\left(1-\delta\underline{\lambda}_{\varLambda}/8\right)^{\widetilde{k}}\frac{1}{T}\sum_{t=1}^{T}\left(1-\delta\underline{\lambda}_{\varLambda}/8\right)^{t}\mathbb{E}^{*}\left\Vert \Delta\boldsymbol{\beta}_{k_{n}+1}\right\Vert \\
		& \leq C\left(1-\delta\underline{\lambda}_{\varLambda}/8\right)^{\widetilde{k}}\mathbb{E}^{*}\left\Vert \Delta\boldsymbol{\beta}_{k_{n}+1}\right\Vert =O_{p}\left(n^{-1}\right),
	\end{align*}
	\begin{align*}
		\mathbb{E}^{*}\left\Vert \frac{\delta}{T}\sum_{t=1}^{T}\sum_{k=0}^{t+\widetilde{k}-1}\left(I-\delta\varLambda_{\phi}\left(\boldsymbol{\beta}^{\star}\right)\right)^{t+\widetilde{k}-1-k}\varOmega_{k_{n}+1+k}^{\phi}\right\Vert  & \leq\frac{\delta}{T}\sum_{t=1}^{T}\sum_{k=0}^{\infty}\left(1-\delta\underline{\lambda}_{\varLambda}/8\right)^{k}\mathbb{E}^{*}\left\Vert \varOmega_{k_{n}+1+k}^{\phi}\right\Vert \\
		& \leq C\sup_{k\geq k_{n}+1}\mathbb{E}^{*}\left(\left\Vert \varOmega_{k}^{\phi}\right\Vert \right)=o_{p}\left(n^{-1/2}\right),
	\end{align*}
	\begin{align*}
		\left\Vert \frac{1}{T}\sum_{t=1}^{T}\left(\delta\sum_{k=0}^{t+\widetilde{k}-1}\left(I-\delta\varLambda_{\phi}\left(\boldsymbol{\beta}^{\star}\right)\right)^{k}-\varLambda_{\phi}^{-1}\left(\boldsymbol{\beta}^{\star}\right)\right)\boldsymbol{\xi}_{n}^{\phi}\right\Vert  & =\left\Vert \frac{1}{T}\sum_{t=1}^{T}\left(\delta\sum_{k=t+\widetilde{k}}^{\infty}\left(I-\delta\varLambda_{\phi}\left(\boldsymbol{\beta}^{\star}\right)\right)^{k}\right)\boldsymbol{\xi}_{n}^{\phi}\right\Vert \\
		& \leq C\left(1-\delta\underline{\lambda}_{\varLambda}/8\right)^{\widetilde{k}+1}\left\Vert \boldsymbol{\xi}_{n}^{\phi}\right\Vert =o_{p}\left(n^{-1/2}\right).
	\end{align*}
	We finally look at the last term. We will focus on $\frac{\delta}{T}\sum_{t=1}^{T}\sum_{k=0}^{t+\widetilde{k}-1}\left(I-\delta\varLambda_{\phi}\left(\boldsymbol{\beta}^{\star}\right)\right)^{t+\widetilde{k}-1-k}\boldsymbol{\mathbf{\Xi}}_{2,k_{n}+1+k}^{\phi}$
	only, because verifying the remaining terms can be done similarly.
	Without loss of generality, we again assume that $\boldsymbol{\mathbf{\Xi}}_{2,k_{n}+1+k}^{\phi}$
	is one-dimensional. We note that 
	\begin{align*}
		& \frac{1}{T}\sum_{t=1}^{T}\sum_{k=0}^{t+\widetilde{k}-1}\left(I-\delta\varLambda_{\phi}\left(\boldsymbol{\beta}^{\star}\right)\right)^{t+\widetilde{k}-1-k}\boldsymbol{\mathbf{\Xi}}_{2,k_{n}+1+k}^{\phi}\\
		& =\frac{1}{T}\sum_{t=1}^{T}\sum_{t^{\prime}=1}^{t}\left(I-\delta\varLambda_{\phi}\left(\boldsymbol{\beta}^{\star}\right)\right)^{t^{\prime}-1}\boldsymbol{\mathbf{\Xi}}_{2,k_{n}+\widetilde{k}+T-t}^{\phi}+\frac{1}{T}\sum_{l=1}^{\widetilde{k}-1}\sum_{t=1}^{T}\left(I-\delta\varLambda_{\phi}\left(\boldsymbol{\beta}^{\star}\right)\right)^{t+l-1}\boldsymbol{\Xi}_{2,k_{n}+\widetilde{k}-l}^{\phi}.
	\end{align*}
	We have that 
	\begin{align*}
		& \mathbb{E}^{*}\left(\frac{1}{T}\sum_{l=1}^{\widetilde{k}-1}\sum_{t=1}^{T}\left(I-\delta\varLambda_{\phi}\left(\boldsymbol{\beta}^{\star}\right)\right)^{t+l-1}\boldsymbol{\Xi}_{2,k_{n}+\widetilde{k}-l}^{\phi}\right)^{2}\\
		& =\mathbb{E}^{*}\left(\frac{1}{T}\sum_{l=1}^{\widetilde{k}-1}\left(I-\delta\varLambda_{\phi}\left(\boldsymbol{\beta}^{\star}\right)\right)^{l}\sum_{t=1}^{T}\left(I-\delta\varLambda_{\phi}\left(\boldsymbol{\beta}^{\star}\right)\right)^{t-1}\boldsymbol{\Xi}_{2,k_{n}+\widetilde{k}-l}^{\phi}\right)^{2}\\
		& \leq\frac{1}{T^{2}}\sum_{t=1}^{T}\sum_{l=1}^{T}\sum_{t^{\prime}=1}^{t}\left(1-\delta\underline{\lambda}_{\varLambda}/8\right)^{t^{\prime}-1}\sum_{l^{\prime}=1}^{l}\left(1-\delta\underline{\lambda}_{\varLambda}/8\right)^{l^{\prime}-1}\mathbb{E}^{*}\left(\boldsymbol{\mathbf{\Xi}}_{2,k_{n}+\widetilde{k}+T-t}^{\phi}\boldsymbol{\mathbf{\Xi}}_{2,k_{n}+\widetilde{k}+T-l}^{\phi}\right)\\
		& =O_{p}\left(\frac{1}{TBh_{n}^{2}}+\frac{\sqrt{\log n}}{B^{2}h_{n}^{2}}\right).
	\end{align*}
	On the other side, we have that 
	\begin{align*}
		& \mathbb{E}^{*}\left(\frac{1}{T}\sum_{l=1}^{\widetilde{k}-1}\sum_{t=1}^{T}\left(I-\delta\varLambda_{\phi}\left(\boldsymbol{\beta}^{\star}\right)\right)^{t+l-1}\boldsymbol{\Xi}_{2,k_{n}+\widetilde{k}-l}^{\phi}\right)^{2}\\
		& =\frac{1}{T^{2}}\sum_{l=1}^{\widetilde{k}-1}\sum_{l^{\prime}=1}^{\widetilde{k}-1}\sum_{t=1}^{T}\sum_{t^{\prime}=1}^{T}\left(I-\delta\varLambda_{\phi}\left(\boldsymbol{\beta}^{\star}\right)\right)^{t+t^{\prime}+l+l^{\prime}-2}\mathbb{E}^{*}\left(\boldsymbol{\Xi}_{2,k_{n}+\widetilde{k}-l}^{\phi}\boldsymbol{\Xi}_{2,k_{n}+\widetilde{k}-l}^{\phi}\right)\\
		& \leq\frac{C}{T^{2}}\left(\sum_{l=1}^{\infty}\left(1-\delta\underline{\lambda}_{\varLambda}/8\right)^{l}\right)^{4}\sup_{k,k^{\prime}}\left|\mathbb{E}^{*}\left(\boldsymbol{\Xi}_{2,k_{n}+k}^{\phi}\boldsymbol{\Xi}_{2,k_{n}+k^{\prime}}^{\phi}\right)\right|\\
		& =O_{p}\left(\frac{1}{T^{2}Bh_{n}^{2}}\right)
	\end{align*}
	This implies that 
	\[
	\frac{1}{T}\sum_{t=1}^{T}\sum_{k=0}^{t+\widetilde{k}-1}\left(I-\delta\varLambda_{\phi}\left(\boldsymbol{\beta}^{\star}\right)\right)^{t+\widetilde{k}-1-k}\boldsymbol{\mathbf{\Xi}}_{2,k_{n}+1+k}^{\phi}=O_{\mathbf{P}}\left(\frac{1}{\sqrt{TBh_{n}^{2}}}+\frac{\log^{1/4}\left(n\right)}{Bh_{n}}\right).
	\]
	This proves the result.
	\end{proof}
\subsection*{Proof of \autoref{thm9}}
\begin{proof}
    To prove the result, it remains to show that 
    \[
    P\left( \mathbb{P}^*\lim_{R\rightarrow \infty} \widetilde{\Sigma}_{\boldsymbol{\beta}}^{\phi} = \widehat{\Sigma}_{\boldsymbol{\beta}}^{\phi} \right)\rightarrow 1,
    \]
    where $\widehat{\Sigma}_{\boldsymbol{\beta}}^{\phi}$ is the full-sample-based covariance matrix estimator prposed in \citet{khan2022}. In particular, define 
	\[\widehat{\Sigma}_{\boldsymbol{\xi}}^{\phi}=\frac{1}{n}\sum_{i=1}^{n}\left(\widehat{G}_{i}\left(1-\widehat{G}_{i}\right)\left(\mathbf{X}_{i}^{\phi}-\widehat{\mathbb{E}}\left(\left.\mathbf{X}_{i}^{\phi}\right|\widehat{z}_{i}\right)\right)\left(\mathbf{X}_{i}^{\phi}-\widehat{\mathbb{E}}\left(\left.\mathbf{X}_{i}^{\phi}\right|\widehat{z}_{i}\right)\right)^{\mathrm{T}}\right),
	\]
	and 
	\[
	\widehat{\varLambda}_{\phi}\left(\widehat{\boldsymbol{\beta}}\right)=\frac{1}{n}\sum_{i=1}^{n}\mathbf{X}_{i}^{\phi}\frac{\partial\widehat{G}\left(\left.z\left(\mathbf{X}_{e,i},\overline{\boldsymbol{\beta}}\right)\right|\overline{\boldsymbol{\beta}}\right)}{\partial\boldsymbol{\beta}^{\mathrm{T}}},
	\]
	where 
	\[
	\widehat{G}_{i}=\frac{\sum_{j=1}^{n}K_{h_{n}}\left(\widehat{z}_{i}-\widehat{z}_{j}\right)y_{j}}{\sum_{j=1}^{n}K_{h_{n}}\left(\widehat{z}_{i}-\widehat{z}_{j}\right)},\ \widehat{\mathbb{E}}\left(\left.\mathbf{X}_{i}^{\phi}\right|\widehat{z}_{i}\right)=\frac{\sum_{j=1}^{n}K_{h_{n}}\left(\widehat{z}_{i}-\widehat{z}_{j}\right)\mathbf{X}_{j}^{\phi}}{\sum_{j=1}^{n}K_{h_{n}}\left(\widehat{z}_{i}-\widehat{z}_{j}\right)},
	\]
	and $\widehat{z}_{i}=X_{0,i}+\mathbf{X}_{i}^{\mathrm{T}}\overline{\boldsymbol{\beta}}$.
	Then $\widehat{\Sigma}_{\boldsymbol{\beta}}^{\phi}$ is defined by $\widehat{\Sigma}_{\boldsymbol{\beta}}^{\phi}=\widehat{\varLambda}_{\phi}^{-1}\left(\overline{\boldsymbol{\beta}}\right)\widehat{\Sigma}_{\boldsymbol{\xi}}^{\phi}\left(\widehat{\varLambda}_{\phi}^{-1}\left(\overline{\boldsymbol{\beta}}\right)\right)^{\mathrm{T}}$. So we only need to show that, with probability going to 1, 
 \[
 \frac{1}{R}\sum_{r=1}^R\widehat{\varLambda}^r_{\phi}\left(\overline{\boldsymbol{\beta}}\right) \rightarrow_{\mathbb{P}^*} \widehat{\varLambda}_{\phi}\left(\overline{\boldsymbol{\beta}}\right) 
 \] and 
 \[
 \frac{1}{R}\sum_{r=1}^R\widehat{\Sigma}_{\boldsymbol{\xi}}^{\phi,r}\rightarrow_{\mathbb{P}^*}\widehat{\Sigma}_{\boldsymbol{\xi}}^{\phi}
 \]
 as $R$ increases to infinity. This can be easily done using the previous proof method.
	
\end{proof}

\end{document}